%% file: notes.tex
\documentclass[11pt,headsepline,openany]{scrbook}

\usepackage{a4wide}
\usepackage{scrpage2}
\usepackage{subeqn}
\usepackage{axodraw}
\usepackage{lscape}
\usepackage[dvips]{color}
\usepackage{epsfig}
\usepackage{wrapfig}
\usepackage{url}

\usepackage{pstricks-add}
\usepackage{pstricks,pst-node,pst-text,pst-3d}

\usepackage{fancybox}
\usepackage{rotating}

\include{defs}

\begin{document}

\title{Lecture Notes on High Energy Cosmic Rays\\
{\small prepared for the 17.th Jyv\"askly\"a Summer School, August 2007}}
\author{M.~Kachelrie\ss
\\
{\it\small Institutt for fysikk, NTNU, Trondheim, Norway}
}
\date{\small
\begin{minipage}{12cm}
Abstract: I give a concise introduction into high energy cosmic ray physics,
including also few related aspects of high energy gamma-ray and neutrino 
astrophysics. The main emphasis is placed on astrophysical questions, 
and the level of the presentation is kept basic. 
\end{minipage}
}

\maketitle
\tableofcontents

\include{intro}    
\include{galprop}

\include{accel}

\include{gamma}

\include{exgal}

\include{he_nu}

\include{newphys}


\addcontentsline{toc}{chapter}{Epilogue}
\chapter*{Epilogue}

Congratulation to the readers who worked through these notes until the end.
Certainly many questions have been left unanswered and some new ones
have popped up. Still I hope that these notes could provide a first 
introduction into high energy cosmic ray physics.
For further studies I recommend the following two excellent references,
\begin{enumerate}
\item[{[E1]}]
V.~S.~Berezinsky, S.~V.~Bulanov, V.~A.~Dogiel, V.~L.~Ginzburg, V.~S.~Ptu\-skin,
{\it Astrophysics of Cosmic Rays\/} (North-Holland, Amsterdam 1990),
\item[{[E2]}]
T.~K.~Gaisser,
{\it Cosmic Rays and Particle Physics\/} 
(Cambridge University Press, Cambridge 1990).
\end{enumerate}
As the titles indicate, the focus of the first book is on
astrophysical aspects of cosmic rays, while the second discusses
additionally the interactions of cosmic rays in the atmosphere
and the formation of extensive air showers. 
The presentation of several topics I have given was heavily influenced
by these books. I omitted generally references to the original literature,
apart from references to the sources  of the reproduced figures, and
refer also in this respect to the books of Berezinsky {\it et al.\/} and
Gaisser. Additionally, the following review articles may be helpful:
\begin{enumerate}
\item
  V.~Berezinsky,
  ``Ultra high energy neutrino astronomy,''
  Nucl.\ Phys.\ Proc.\ Suppl.\  {\bf 151}, 260 (2006)
  [arXiv:astro-ph/0505220].
\sitem
P.~Lipari,
  ``Perspectives of high energy neutrino astronomy,''
  Nucl.\ Instrum.\ Meth.\  A {\bf 567}, 405 (2006)
  [arXiv:astro-ph/0605535].
\sitem
V.~Berezinsky,
  ``Transition from galactic to extragalactic cosmic rays,''
to appear in the proceedings of 30th ICRC 2007,
  arXiv:0710.2750 [astro-ph].
\sitem
M.~Kachelrie\ss,
  ``Status of particle physics solutions to the UHECR puzzle,''
  Comptes Rendus Physique {\bf 5}, 441 (2004)
  [arXiv:hep-ph/0406174].
\sitem
  G.~Jungman, M.~Kamionkowski and K.~Griest,
  ``Supersymmetric dark matter,''
  Phys.\ Rept.\  {\bf 267}, 195 (1996)
  [arXiv:hep-ph/9506380].
\end{enumerate}

It is a pleasure to thank all my collaborators in the field of cosmic ray 
physics, but in particular Roberto Aloisio, Venya Berezinsky, 
Sergey Ostapchenko, Dima Semikoz, Pasquale Serpico, and Ricard Tom\`as,
as well as G\"unter Sigl for many illuminating discussions. I am grateful
to Sergey Ostapchenko and Pasquale Serpico for reading (parts of) the 
manuscript and pointing out several errors and misunderstandings 
contained in the text.

Last but not least, I would like to thank Kimmo Kainulainen for inviting me
to lecture at the Jyv\"askly\"a Summer School and for the excellent working 
conditions provided there.

\end{document}

%% file: defs.tex
\def\be{\begin{equation}}
\def\ee{\end{equation}}
\def\ba{\begin{eqnarray}}
\def\ea{\end{eqnarray}}

\def\lsim{\raise0.3ex\hbox{$\;<$\kern-0.75em\raise-1.1ex\hbox{$\sim\;$}}}
\def\gsim{\raise0.3ex\hbox{$\;>$\kern-0.75em\raise-1.1ex\hbox{$\sim\;$}}}

\def\eps{\varepsilon}
\def\theta{\vartheta}

\def\d{{\rm d}}
\def\e{{\rm e}}

\def\vx{{\bf x}}

\def\vr{{\bf r}}
\def\vp{{\bf p}}

\def\vv{{\bf v}}

\def\vj{{\bf j}}
\def\vB{{\bf B}}
\def\vF{{\bf F}}

\def\F{{\cal F}}

\def\Ecr{E_{\rm cr}}

\def\ap{\approx}
\def\const{{\rm const.}}

\def\He4{\:^4{\rm He}}
\def\Fe56{\:^{56}{\rm Fe}}

\def\sitem{\vspace*{-0.2cm}\item}

\global\newcount\paperno \global\paperno=1
\def \paper #1 {\vskip 10pt
\hangindent=1cm \noindent \hbox to 0.85cm
{\hfil[\the\paperno]\hbox to .5cm{}}#1\global\advance\paperno by 1}

\def\ex#1{\bigskip\hrule\smallskip\noindent 
{\bf Ex.:} {\small { #1}}\smallskip\hrule\smallskip\smallskip}


%% file: intro.tex
\chapter{Introduction}

\section{Preludes}

\subsection*{What do we want to discuss?}

\begin{figure}
\epsfig{file=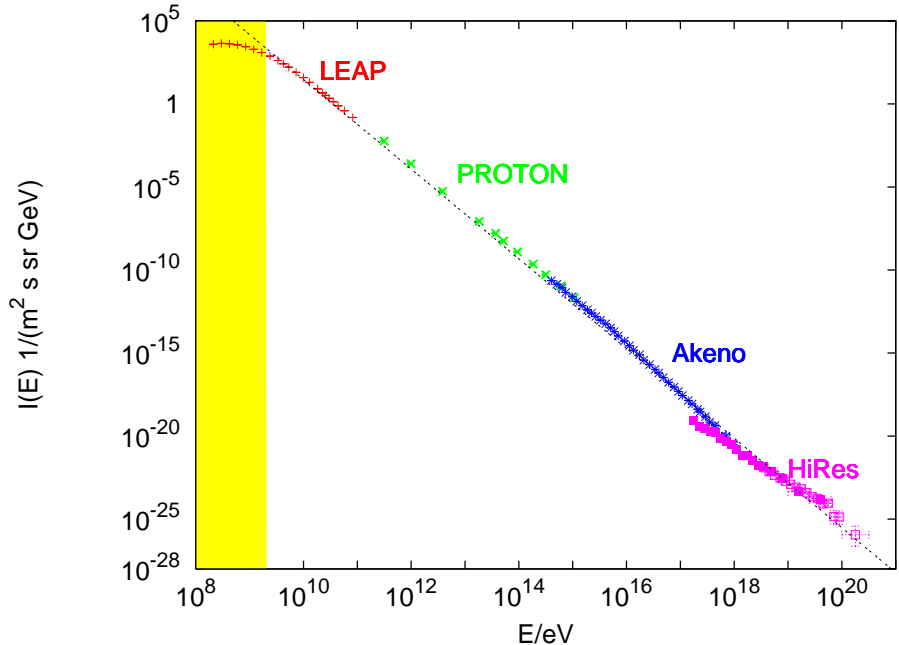,width=12.5cm}  
\vspace{-0.3cm}
\caption{\label{cosmic ray}
The cosmic ray spectrum $I(E)$ as function of kinetic energy $E$, 
compiled using results 
from the LEAP, proton, Akeno, and HiRes experiments~\cite{exp}. 
The energy region 
influenced by the Sun is marked yellow and an $1/E^{2.7}$ power-law is
also shown.}
\end{figure}

The term cosmic rays may be defined either as all radiation consisting of
relativistic particles impinging on the Earth's atmosphere from outer
space or, more narrowly, including only charged particles. We shall
follow generally the latter definition, but discuss also high energy gamma rays
and neutrinos, since $(i)$ they can be created as 
secondaries of high energy cosmic
rays and $(ii)$ they may help us to learn more about the charged
component and their sources (``multi-messenger approach'').   
The qualification high energy (or relativistic) means that
we do not consider cosmic rays below a few GeV that may be produced or
are influenced by the Sun and its wind. The cosmic ray spectrum, a
nearly featureless power-law extending over eleven decades in energy up to
a ${\rm few}\times 10^{20}\:$eV, is shown in Fig.~\ref{cosmic ray}.

We cover mainly astrophysical aspects (What are the sources? How do they
accelerate cosmic rays? What happens during the journey of cosmic rays
to the Earth?),  
but discuss also briefly their interactions in the atmosphere and the 
experimental methods used to detect them. In the final chapter we
review connections between cosmic ray physics and searches for
physics beyond the standard model.

The level of the presentation is as basic as possible, preferring
``back on the envelope calculations'', and the material covered
corresponds roughly to 20\,hours of lectures.

\subsection*{Where are we?}

Our standard length unit is the parsec, the distance  from the Earth to
a star that has a parallax\footnote{Relatively nearby stars are seen
  at slightly different positions on the celestial sphere (i.e. the
  background of stars that are ``infinitely'' far away) as the Earth moves
  around the Sun. Half of this angular difference is called the
  parallax angle or simply the parallax $p$. Thus $p\ap\tan p={\rm
    AU}/{d}$, where 1~AU denotes the mean distance between the
  Earth and the Sun.} of one arcsecond. Since one arcsecond is
$1/(360\times 60\times 60)=1/206 265$ 
fraction of $2\pi$, a parsec corresponds to $206,265\,{\rm AU} 
=3.086\times 10^{18}\:{\rm cm}= 3.26\,{\rm lyr} = 1\,{\rm pc}$.

\begin{figure}
\vspace*{0.5cm}
\hspace*{2.cm}
\epsfig{width=.65\textwidth,angle=0,file=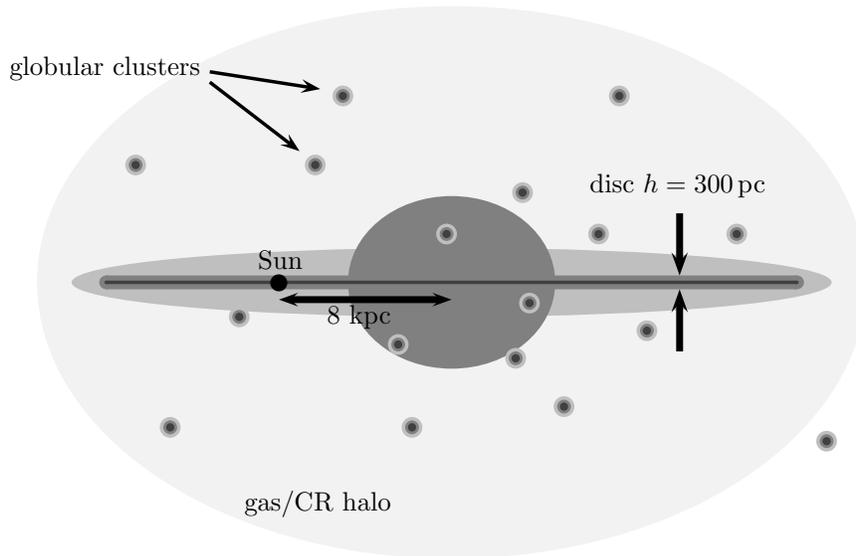}
\caption{\label{Milky}
Schematic picture of the Milky Way with a gas and dust disc, an extended
halo of gas and cosmic rays, surrounded by globular clusters. Everything
is immersed in a halo of dark matter. } 
\end{figure}

A schematic picture of our home galaxy, the Milky Way, is shown in 
Fig.~\ref{Milky}. Most 
stars are concentrated in the galactic disc of height $h\ap 300\:$pc in 
the form of spiral arms. The disc is filled with warm
atomic gas that consists to $90\%$ of H and to $10\%$~ of He and has
an average density $n\sim  1/$cm$^3$. It contains also an ordered
magnetic field with strength $B\sim 3\mu$G. The energy when the Larmor
radius 
\be
 R_L =\frac{cp}{ZeB} \ap 
 100\,{\rm pc} \;\frac{3\mu{\rm G}}{B}\: \frac{E}{Z\times 10^{18}\rm eV}
\ee
of a particle with charge $Ze$ and momentum $p$ equals the height of
the Galactic disc marks approximately the transition between 
diffusive and rectilinear propagation of cosmic rays. We can hope to 
perform ``charged-particle astronomy'' only for energies well above this 
transition. This is one of the reasons why we are especially interested 
in  cosmic rays of the highest energies.

A halo extends with $n\sim 0.01/$cm$^3$ and a turbulent magnetic field 
probably up to distances $\sim (10-15)\,$kpc above the disc. 
The average strength of this turbulent
magnetic field is not well restricted and may reach up to 
$B\lsim 10\,\mu$G. The whole visible part of the galaxy
is embedded in a much larger dark matter halo, which comprises 90\% of
the total mass of the Milky Way. The center of the Milky Way contains,
as probably most other galaxies, a supermassive black hole (SMBH) with mass 
$\sim 10^6M_\odot$.

The Milky Way is a member of the Local Group, a rather small cluster
of galaxies with the Andromeda galaxy as the other prominent member. Its
diameter is $\sim 2\,$Mpc. 
The distance to the nearest galaxy cluster, the Virgo cluster, is
18\,Mpc, i.e.\ approximately equal to the mean free path of cosmic rays with
the highest observed energies, $E\sim 10^{20}\,$eV.

A naive estimate for the size $R$ of the observable Universe can be
obtained from the observed recession velocity $v=H_0d$ of galaxies at the
distance $d$. Using for the Hubble parameter
$H_0=70\,$km\,s$^{-1}$\,Mpc$^{-1}$, 
it follows $R=c/H_0\sim 4300\,$Mpc. For comparison, the cosmic ray horizon 
reaches Gpc scale at $E=10^{19}\:$eV.  

We shall use mostly natural units with $\hbar=c=k_B=1$, but keep often
explicitly the speed of light $c$ in formulas within a more astrophysical 
context.

\section{Historical remarks}

\begin{itemize}
\sitem[1912:]
Victor Hess discovered on a balloon flight that ionizing radiation
increases with altitude. As he wrote ``The results of the present
observations are most easily 
  explained by the assumption that radiation with very high
  penetrating power enters the atmosphere from above; even in its
  lower layers, this radiation produces part of the ionization
  observed in closed vessels\ldots Since there was neither a decrease
  at night or during solar eclipse, the Sun can hardly be considered as
  the source\ldots''~\cite{hess1912}. 
\sitem[1929:] 
Skobelzyn observed first cosmic rays with a cloud chamber. Bothe and
Kolh\"orster showed that the tracks are curved by a magnetic field. 
This proved that the observed cosmic rays on
ground are charged particles---now we know that these are mainly muons
produced as secondaries in cosmic ray interactions in the higher atmosphere.  
\sitem[28/29:] 
Clay observed the ``latitude effect'': The cosmic ray intensity depends on 
the (geomagnetic) latitude. Bothe and Kolh\"orster provided first the correct
interpretation of this effect as an anisotropy induced by
the magnetic field of the Earth, providing in turn evidence that 
(the primary) cosmic rays are charged. 
\sitem[1932:]
Anderson discovered the positron in cosmic rays. This was just the start
for a series of new particles detected in cosmic rays: The muon 1936 again
by Andersen, charged pions in 1947, and strange particles 1947--50. 
\sitem[1932:]
Raged debate in the US about sources and primary type of the new
radiation. Millikan and Compton
favored gamma rays and coined therefore the name "cosmic rays." 
\sitem[1934:] The sign of the east-west asymmetry showed that the
cosmic ray primaries are positively charged particles. 
\sitem[34/38:]
Rossi and independently  Auger discovered through coincidence measurements
"extensive air showers," showers of secondary particles
caused by the collision of high energy cosmic rays with air nuclei.  
\sitem[1934:]
Bethe and Heitler developed the electromagnetic cascade theory; the
observed particles on ground are secondaries. 
\sitem[1947:] Zatsepin discovered the scaling of hadronic 
interactions studying the evolution of extensive air showers.
\sitem[1949:] 
Fermi proposed that cosmic rays are accelerated by bouncing
off moving magnetic clouds in the Galaxy.
\sitem[52-54:] The first human accelerators reaching $p\gsim 1\:$GeV were 
built. 
As a consequence, cosmic ray and high energy physics started to decouple, and
cosmic ray physicists focused with time more on astrophysical questions.
\sitem[1954:]
 First measurements of high energy cosmic rays via sampling of
 extensive air showers done at the Harvard College Observatory. 
\sitem[1972:] The launch of the SAS-2 satellite marked the start of
high energy gamma astronomy. 
\sitem[1976:] Start of the first prototype of a large-scale underwater
detector for high energy neutrino astronomy, DUMAND in Hawaii. 
\sitem[1998:] The Superkamiokande experiment found the first convincing
evidence that neutrinos are massive observing flavor oscillations of 
atmospheric neutrinos.
\sitem[2007:] Completion of the Pierre Auger Observatory in Argentina,
the first combination of a ground array and fluorescence telescopes in
the same experiment. With its size of $A\ap 3000\,$km$^2$, it is a
factor 30 larger than previous experiments.
\end{itemize}

\chapter{Basic notations of particle physics}

\section{Relativistic kinematic and cross sections}

\subsection{Kinematics}

The description of scattering reactions is simplified, if Lorentz invariant
quantities are used. Consider for instance the squared center-of-mass energy 
$s=(p_a+p_b)^2=(p_c+p_d)^2$ with four-momenta $p_i=(E_i,\vp_i$) in the 
$2\to 2$ scattering $a+b\to c+d$,
\be
 s=(p_a+p_b)^2=m_a^2+m_b^2 + 2(E_aE_b - \vp_a\vp_b) = 
 m_a^2+m_b^2 + 2E_aE_b(1 - \beta_a\beta_b\cos\theta) \,,
\ee
where $\beta_i=v_i=p_i/E_i$. We are interested often in the threshold
energy $s_{\min}^{1/2}$ of a certain process. An important example
for cosmic ray physics is the reaction $p+\gamma\to p+\pi^0$. If in this
process the proton is at rest, then  
\be
 s= m_p^2+2E_\gamma m_p \geq (m_p+m_\pi)^2=m_p^2+2m_pm_\pi+m_\pi^2
\ee
or
\be
 E_\gamma \geq m_\pi+\frac{m_\pi^2}{2m_p} \ap 145\: {\rm MeV}\,.
\ee
Thus the photo-production of pions on protons at rest is only possible
for $E_\gamma\geq 145\:$MeV. Photons with lower energy do not
interact via this reaction, while higher energetic ones do. We will
consider this process later again, but then as the scattering of
ultrahigh energy protons on low energy photons, that is an important 
example for the transition from a transparent Universe at low energies
to an opaque one at high energies. 

\ex{
Calculate the minimal energy of a proton able to produce
anti-protons scattering on another proton at rest.
\\
{\small 
Because of baryon number conservation, anti-proton production is first
possible in $pp\to ppp\bar p$ with $s=2m_p^2+2E_pm_p\geq 16m_p^2$ or
$E_p\geq 7m_p$. Furthermore, the cross section of this reaction is
small close to the threshold. Hence the anti-proton flux at small
energies should be strongly suppressed, if anti-protons are only produced as
secondaries in cosmic ray interactions. 
}
}

The second important quantity characterizing a scattering process is the
four-momentum transfer $t=(p_a-p_c)^2=(p_b-p_d)^2$. As an example, we
consider electron-proton scattering $e^-+p\to e^-+p$. Then 
\be
 t = (p_e-p_e')^2 = 
 2m_e^2 - 2E_eE_e' (1- \beta_e\beta_e^\prime\cos\theta)  \,.
\ee
For high energies, $\beta_e,\beta_e^\prime\to 1$, and 
\be
 t\ap -  2E_eE_e' (1-\cos\theta) = -4E_eE_e^\prime \sin^2\theta/2 \,.
\ee
Because of four-momentum conservation, the variable $t$ corresponds to
the squared momentum $Q$ of the exchanged virtual photon,
$t=Q^2=(p_e-p_e^\prime)^2<0$.  
Thus a virtual particle does not fulfill the relativistic energy-momentum 
relation. The energy-time uncertainty relation $\Delta E\Delta t\gsim 1$
allows such a violation, if the virtual particle is exchanged only
during a short enough time. 
Note also that the angular dependence of Rutherford scattering,
$\d\sigma/\d\Omega\propto 1/\sin^4\theta/2$ is obtained if
$\d\sigma/\d\Omega\propto 1/t^2$.

In the reaction $e^-+p\to e^-+p$ the scattering happens most likely at
small angles $\theta$ and small four-momentum transfer $t$. (Note that
this does not imply small energy transfer in the lab system.) In a 
collider experiment, this kinematical region is difficult to observe
because of the beam pipe. Moreover, the search for physics beyond the
standard model as one of the most important problems in high energy
physics is often performed as search for new, heavy
particles. The production of particles with mass $M$ requires
typically $|t|\gsim M^2$ and is therefore rare. On the
other hand, for the description of cosmic ray interactions in the
atmosphere we are interested mainly in the most common
interactions. This is one of the reasons why high energy particle
physics with accelerators and with cosmic rays are focusing
partly on different aspects of high energy interactions.

\subsection{Cross section, interaction depth, and  slant depth}

Consider  a cylinder of length $l$ and area $A$ filled with $N$
scattering centers each with cross section $\sigma$. Their number
density $n$ is $n=N/(Al)$. Let us assume first that $N\sigma\ll A$.
Then the fraction of incoming particles absorbed in the cylinder is
simply 
\be
 \frac{N\sigma}{A}=nl\sigma\equiv\tau
\ee
and defines the optical or interaction depth $\tau\equiv nl\sigma$.  
Our assumption $N\sigma\ll A$ corresponds to $\tau\ll 1$, an ``(optical)
thin'' or transparent source in contrast to an ``(optical) thick''
source with $\tau\gg 1$. In general, we have to calculate how much
radiation is absorbed passing the infinitesimal distance $\d l$, 
\be \label{rad1}
 \d I= -I \d\tau= -I n\sigma \d l \,.
\ee
Integrating gives
\be
 I=I_0\exp(-\tau) \qquad {\rm or}\qquad
 I(l)=I_0\exp\left(-\int_{0}^l \d l \: n\sigma\right)\,.
\ee

In cosmic ray physics, one introduces often the interaction length
$\lambda_i=m/\sigma_i$ measured in g/cm$^2$ and replaces the path length 
$l$ by the slant depth $X$, 
\be \label{slant}
 X(l) = \int_0^l \d l\: \rho(l') \qquad{\rm or}\qquad
 X(h) = \int_h^\infty\d h\: \rho(h') \,.
\ee
Hence, $X$ measures the weight per area of the material crossed. The
version on the RHS of Eq.~(\ref{slant}) is appropriate in the case of 
the Earth atmosphere, where $X=0$ corresponds to $h\to\infty$,
while $X$ measured from the zenith is numerically equal to 
1030\,g/cm$^2$ at sea level. The advantage of $X$ is obviously to hide 
the integration in $X$,
\be
 \tau = \int_0^X \frac{\d X'}{\lambda}= \frac{X}{\lambda} \,.
\ee

\section{Particles and interactions}

\subsection{Our particle inventory}

The particles  known around 1950 are shown in Table~\ref{part50} together
with their main decay mode and their life-time $t_{1/2}$ and range
$ct_{1/2}$. 
\begin{table}
\begin{tabular}{c|c|c|c}
particle  & main decay mode & life-time $t_{1/2}$ & range $ct_{1/2}$
\\[0.5ex] \hline
$\gamma$ & -- & $\infty$ & $\infty$ \\
$e^-$ & -- & $\infty$ & $\infty$ \\
p & -- & $\infty$ & $\infty$  \\
n & $n\to p+e^-  + \bar\nu_e$ & 886s & $2.65\times 10^{13}\:$cm  \\
$\mu^-$ & $\mu^-\to e^-+\bar\nu_e+\nu_\mu$ & $t_{1/2}\sim 2.20  \times10^{-6}$s &
$659\:$m 
\\
$\pi^-$ & $\pi^-\to\mu^-+\bar\nu_\mu$ & $t_{1/2}\sim 2.60\times  10^{-8}\:$s & 
$780\:$cm  \\
$\pi^0$ & $\pi^0\to 2\gamma$ & $t_{1/2}\sim 8.4\times  10^{-17}\:$s  &
$25.1\:$nm  \\
$\nu$ & -- & $\infty$  & $\infty$   
\end{tabular}
\caption{\label{part50} 
The known particles around 1950 together with their main decay mode,
their life-time and range.}
\end{table}
Some immediate consequences of the life-times given are:
\begin{itemize}
\item 
Neutrons with $\Gamma=E/m_n\gsim 10^9$ or $E\gsim 10^{18}\:$eV are
stable on galactic scales ($\sim 10\,$kpc). A source of neutrons with $E\gsim
10^{18}\:$eV in our galaxy would provide  neutral, strong interacting
primaries, while at lower energies it would be visible as a $\bar\nu_e$
source from neutron decay on the flight.
\sitem
In the atmosphere, only low energy charged pions will decay and
produce neutrinos, whilst at high energies charged pions mainly
scatter. Similarly, most high energy muons do not decay within the
extension of the atmosphere, $\sim 15\,$km. Thus the neutrino flux
produced by cosmic rays in the atmosphere should be a steeper function of
energy than the cosmic ray flux. This opens the possibility to perform high
energy neutrino astronomy at energies $E\gg$~TeV, for which the
background of atmospheric neutrinos becomes negligible.
\sitem
The long muon range can be used to enlarge the effective volume of a
neutrino detector, observing also muons produced outside the proper
detector  volume.
\end{itemize}

\subsection{Comparison of electromagnetic, weak and strong interactions}

The energy-time uncertainty relation $\Delta E\Delta t\gsim 1$
restricts the emission of a heavy particle with mass $M$ by a lighter
one to times $\Delta t\lsim 1/M$. Thus the range $\Delta t$ of an
interaction over which a massive particle can be exchanged should be
limited by $1/M$. 

This idea leads to the generalization of the Coulomb potential to the
Yukawa potential,
\be
 V(r)= g^2 \: \frac{\exp(-Mr)}{r}\to \frac{g^2}{r} 
 \qquad{\rm for}\quad 
 M\to 0 \,.
\ee

In non-relativistic quantum mechanics, the connection between the
cross section $\sigma$ and the
potential $V$ is given via the scattering amplitude $f(\theta)$,
\be
 \sigma = \int\d\Omega\: \frac{\d\sigma}{\d\Omega} 
        = \int\d\Omega\: |f(\theta)|^2 \,.
\ee
In the so-called Born approximation, this scattering amplitude $f(\theta)$
is the Fourier transform of the potential,
\be
 f(q) = -\frac{m}{2q\hbar^2}\int\d^3q \:  \exp(-i({\bf q r})) V(r) \,,
\ee
where ${\bf q=p-p'}$ or $q=2p\sin\theta/2$.

Performing the integral for the Yukawa potential one obtains
\be
 \frac{\d\sigma}{\d\Omega} \propto \frac{g^4}{(q^2+M^2)^2}  \,.
\ee
This result explains the difference in strength between
electromagnetic and weak interactions, although their coupling
constants have roughly the same size, $g\ap e$. 
Photons are massless and thus 
\be
 \frac{\d\sigma}{\d\Omega} \propto \frac{e^4}{q^4}  \,,
\ee
while the bosons exchanged in weak interactions are heavy. If $s\ll
M^2$, then also $|t|\ll M^2$ and $q^2\ll M^2$, and thus the exchange of 
massive gauge bosons results in
\be
 \frac{\d\sigma}{\d\Omega} \propto \frac{g^4}{M^4}  \propto G_F^2\,, 
\quad{\rm with}\quad G_F\equiv \frac{\sqrt{2}g^2}{8M_W^2}\,.
\ee
The masses of the $W$ and $Z$ bosons that intermediate weak
interactions have been measured at
accelerators as $M_Z=91.2\,$GeV and $M_W=80.4\,$GeV.

The total cross section $\sigma$ is obtained by integrating over
$\d\Omega\propto \d(\cos\theta)\propto \d t$. We expect from
dimensional analysis that $\sigma\sim G_F^2 ({\rm energy})^2$ and
often the result can be estimated. For instance, for the scattering 
of a high-energy neutrino on a
nucleon at rest the integration gives an additional factor
$s\ap 2m_NE_\nu$, and the cross section is numerically for $s\ll m_W^2$
\be
 \sigma_{\nu N}\sim G_F^2 s \ap 
 10^{-37} {\rm cm}^2 \:\left(\frac{E_\nu}{\rm 100~GeV}\right) \,.
\ee

Finally, we consider strong interactions. The largest contribution to the 
proton-proton cross section should result from the exchange of the lightest, 
strongly interacting particle, the pion. Integrating  gives
\be
 \sigma\sim \frac{g_\pi^4}{(4\pi)^2}\int_{-s}^0 \frac{\d t}{(t-m_\pi^2)^2} =
 \alpha_\pi^2 \:\frac{s}{m_\pi^2(s-m_\pi^2)} \to
 \frac{\alpha_\pi^2}{m_\pi^2} \ap  20\alpha_s^2\:{\rm mbarn}
\ee
compared to the measured value $\sigma\ap 50\:$mbarn. 
Hence the proton-pion coupling $\alpha_\pi=g_\pi^2/(4\pi)$ is of order
one and perturbation theory for strong interactions seems to make no sense. 
However, coupling constants are---despite the name---in generally not 
constant but depend on the scale $Q^2$ probed. In the case of strong 
interactions, the coupling becomes smaller
for large $Q^2$, a phenomenon called ``asymptotic freedom.'' 
As a consequence, perturbation theory can be used in QCD, however only 
in a restricted kinematical range, $Q^2\gsim {\rm few}\times {\rm GeV}^2$. 
Unfortunately, for cosmic ray interactions  most important is
the regime of small momentum-transfer that is not (yet) accessible to
predictions from first principles and requires therefore the use of 
phenomenological models.

\begin{figure}
\epsfig{width=.55\textwidth,angle=0,file=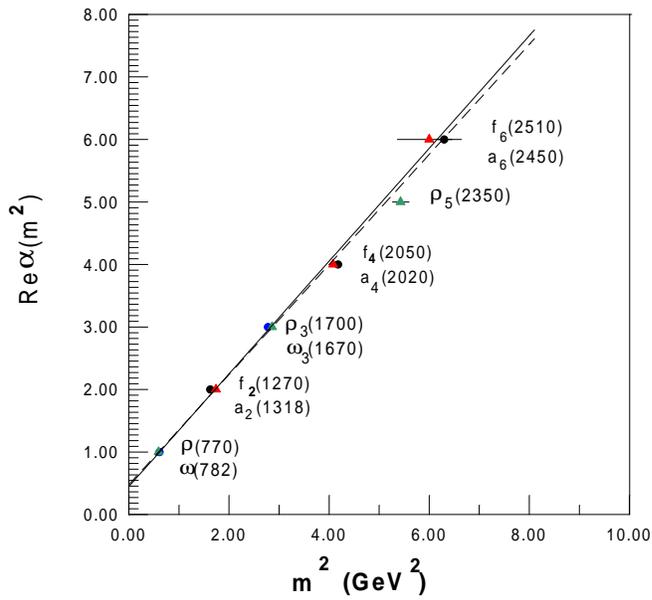}
\caption{\label{chew-frautschi}
Chew-Frautschi plot, i.e.\ the spin against the squared mass, for the 
trajectories 
of the $\omega$, $\rho$, $f_2$, and $a_2$ mesons, from Ref.~\cite{ped}.}
\end{figure}

We have argued that the exchange of the lightest (relevant) particle
dominates the total cross section. But what happens, when an infinite
number of particles can be transmitted, with increasing multiplicity
for increasing mass? This is the case of hadrons for which a zoo of
resonances exists, cf. the Chew-Frautschi plot in Fig.~\ref{chew-frautschi}.  
For instance, the vector mesons $A_2,\rho_3,\ldots$
can be viewed as $J=2,3,\ldots$ excitations of the $\rho$ meson with
spin $J=1$. 
After summing up the infinite number of exchanged hadrons per trajectory,
one can replace them by a single ``effective particle''
(called ``reggeons'').
The high energy behavior of hadronic cross sections is modeled as an
exchange of these reggeons, with their properties read off from
Chew-Frautschi plots as in Fig.~\ref{chew-frautschi}.   

\paragraph{Quark model and hadronization}
For $q\gsim 1/m_p$, the scattering amplitude for e.g. electron-proton
scattering should reflect the proton size $10^{-13}\:$cm. However, a
form factor typical for scattering on point-like particles was found
experimentally.  
This led to the parton model: Protons (as all hadrons) consist of point-like
constituents, quarks and gluons. The valence quark content of some hadrons is
given in Tab.~\ref{quark}. At high momentum transfer, a probe scatters
on partons independently (``spectator model''). Hence, the parton model
predicts e.g. for the ratio of pion-proton and proton-proton cross sections 
\be
 \frac{\sigma_{\pi p}}{\sigma_{pp}}= \frac{2}{3} \,,
\ee
in good agreement with experiment. 

Which kind of hadrons are mainly produced in the process $e^+e^-\to
q\bar q\to\,$hadrons? Mainly mesons ($q\bar q$) with only $\sim 5\%$ of nucleons
($qqq$), because it is easier to combine two quarks into a colorless,
light meson than three quarks into a colorless, rather heavy baryon.
Out of the mesons, 90\% are pions while the remaining ones are mainly
kaons. While 1/3 of pions are neutral ones and produce photons, the
other 2/3 are charged and produce neutrinos. Hence photon and
neutrino fluxes produced by hadronization are closely connected.

\begin{table}
\begin{tabular}{c|c||c|c}
hadron  & valence quarks & hadron  & valence quarks 
\\[0.5ex] \hline
p & uud & $\Lambda$& uds\\
n & udd & $K^+$ & u$\bar s$\\
$\pi^+$ & $u\bar d$ & $K^0$& $d\bar s$\\
$\pi^0$ & $(u\bar u +d\bar d)/\sqrt{2}$ & & 
\end{tabular}
\caption{\label{quark}Valence quark content of hadrons}
\end{table}

\section{Exercises}

\begin{enumerate}
\item
Derive the connection between $\d\sigma/\d\Omega$ and $\d\sigma/\d t$ 
and the integration limits in $t$ for a general $2\to 2$ scattering process.
\item
Find the minimal energy $E_{\rm th}$ of a proton scattering on a photon with
the typical energy of the cosmic microwave background ($T\ap 2.7\:$K)
for the process $p+\gamma\to p+\pi^0$. Guess the cross section of this
reaction, check it against the curve at \url{http://pdg.lbl.gov}, and
estimate the mean free path of a proton with $E\gg E_{\rm th}$.
\item
Consider the 2-particle decay $\pi^0\to 2 \gamma$. What are the
minimal and maximal photon energies, if the pion moves with velocity
$v$? What is the shape of the photon spectrum $\d N/\d\!E$?
\end{enumerate}

%% file: galprop.tex
\chapter{Galactic cosmic rays}

\section{Basic observations}
\subsection{Cosmic ray intensity and composition}

The integral intensity $I(>E)$ of cosmic rays is defined as the number
of particles with energy $>E$ crossing an unit area per unit time and
unit solid angle. Thus its units are 
$[I]={\rm cm}^{-2}{\rm s}^{-1}{\rm sr}^{-1}$. 
The (differential) intensity $I(E)$  and the integral intensity are
connected by 
\be
 I(>E)=\int_E^\infty \d E' \: I(E') \,.
\ee
The particle flux $\F$ from one hemisphere through a planar detector is
\be \label{fluxplanar}
 \F(E) = \int\!\! \d\Omega \, I(E)\cos\theta = 
 I(E) \int_0^{2\pi}\!\!\d\phi \int_0^{\pi/2} \!\!
 \d\theta\sin\theta\cos\theta = 
 \pi I(E) \int_0^{\pi/2} \!\! \d\theta\sin2\theta = \pi I(E)\,,
\ee
where we have assumed that the intensity is isotropic. The 
(differential) number density of cosmic rays with velocity $v$ is  
\be
 n(E)=\frac{4\pi}{v} I(E) \,.
\ee
\begin{wrapfigure}{r}{.45\textwidth}
\epsfig{width=.45\textwidth,angle=0,file=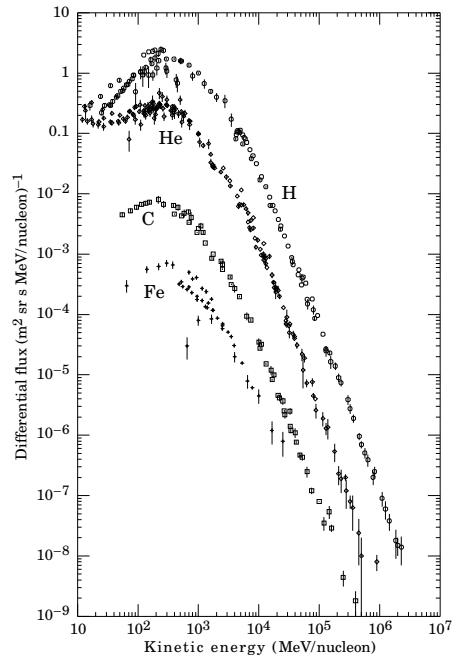}
\caption{\label{cr_comp_spec}
Cosmic rays intensity for several elements, from Ref.~\cite{PDG}.}
\end{wrapfigure}
More generally, the intensity $I$ may depend both on the position
$\vx$ of the detector and on its orientation $\phi,\theta$. We can 
connect $I$ to the phase space distribution $f(\vx,\vp)$ of cosmic rays by
comparing $\d N=f(\vx,\vp)\d^3 x\d^3 p$ with the definition
$I=\d N/(\d\!A\:\d t\:\d\Omega\:\d\!E)$,
\be \label{IF}
 I(\vx,p,\theta,\phi)= vp^2\,\frac{\d p}{\d\!E}\: f(\vx,\vp)
 = p^2 f(\vx,\vp) \,.
\ee

Figure~\ref{cr_comp_spec} shows $I(E)$ separately for some important
elements. First, one recognizes that the main 
component of cosmic rays are protons, with additionally around 10\%  of 
helium and an even smaller admixture of heavier elements.

\begin{wrapfigure}{r}{.45\textwidth}
\hskip0.2cm\epsfig{width=.35\textwidth,angle=270,file=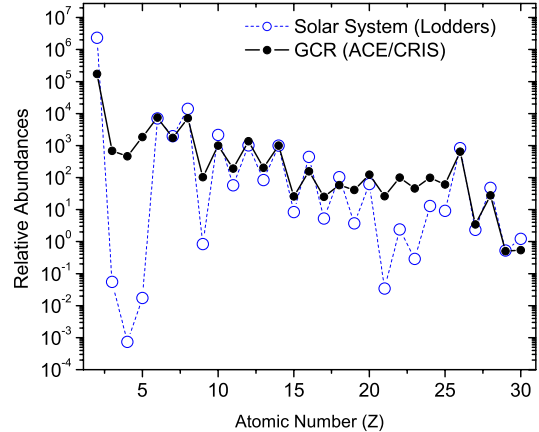}
\caption{\label{cr_aband}
Abundance of elements measured in cosmic rays compared to the solar
system abundance, from Ref.~\cite{aband}.}
\end{wrapfigure}

The relative abundance of elements measured in cosmic rays (dark,
filled circles) is compared to the one in solar system (blue, open circles)
in Fig.~\ref{cr_aband}. Both curves show the odd even effect,
i.e.\ the tighter bound nuclei with an even numbers of protons
and neutrons are more
abundant.  The main difference of the two curves is that the Li-Be-B
group ($Z=3-5$) and the Sc-Ti-V-Cr-Mn ($Z=21-25$) group are much more
abundant in cosmic rays than in the solar system. We explain this later
as a propagation effect: The elements from the Li-Be-B group are
produced as secondaries in cosmic rays interactions in the Galaxy.

Second, the spectra shown in Fig.~\ref{cr_comp_spec} are above a few
GeV power-laws, practically without any spectral features. The total
cosmic ray spectrum is 
\be
 I(E) \sim 1.8 E^{-\alpha} \:\frac{\rm particles}{\rm cm^2\: s\: st\: GeV}
\ee
in the energy range from a few~GeV to 100~TeV with
$\alpha\ap 2.7$. Around $10^{15}\:$eV (the ``knee''), the slope
steepens from $\alpha\ap 2.7$ to $\alpha\ap 3.0$. The spectrum
above $10^{18}\,$eV will be discussed in Chapter~6.

The power-law form of the cosmic ray spectrum indicates that they are
produced via non-thermal processes, in contrast to all other
radiation sources like e.g. stars or (super-) novae known until the
1950's. 

Third, small differences in the exponent $\alpha$ of the power-law for
different elements are visible: The relative contribution of heavy
elements increases with energy.

The kinetic energy density of cosmic rays is
\be
 \rho_{\rm CR} = \int\d E \:E_k n(E) 
               = 4\pi \int\d E \:\frac{E_k}{v}\: I(E) \,.
\ee
Extrapolated outside the reach of the solar wind, it is
\be
 \rho_{\rm CR} \ap 0.8 \, {\rm eV}/{\rm cm^3}
\ee
compared to the average energy density 
$\rho_{\rm b} \ap 100\,{\rm eV}/{\rm cm^3}$ of baryons in the Universe,
of star light $\rho_{\rm light}\ap 5\,$eV/cm$^3$ in the disc, and
in magnetic fields $\rho_{\rm mag} = 0.5\,{\rm eV}/{\rm cm^3}$ for
$B=6\,\mu$G.
If the local value of $\rho_{\rm CR}$ would be representative for the
Universe, 1\% of the energy of all baryons would be in the form of
relativistic particles. This is rather unlikely and suggests that
cosmic rays are accumulated in the Galaxy.

\paragraph{Solar modulations}

When cosmic rays enter our Solar System, they must overcome the
outward-flowing solar wind. This wind impedes and slows the incoming
cosmic rays, reducing their energy and preventing the lowest energy ones from
reaching the Earth. This effect is known as solar modulation. 
The Sun has an 11-year activity cycle which is reflected in the 
ability of the solar wind to modulate cosmic rays. 
As a result, the cosmic ray intensity at Earth is
anti-correlated with the level of solar activity, i.e., when solar
activity is high and there are lots of sunspots, the cosmic ray intensity at
Earth is low, and vice versa. 

\begin{figure}
\hspace*{1.cm}
\epsfig{width=.45\textwidth,angle=270,file=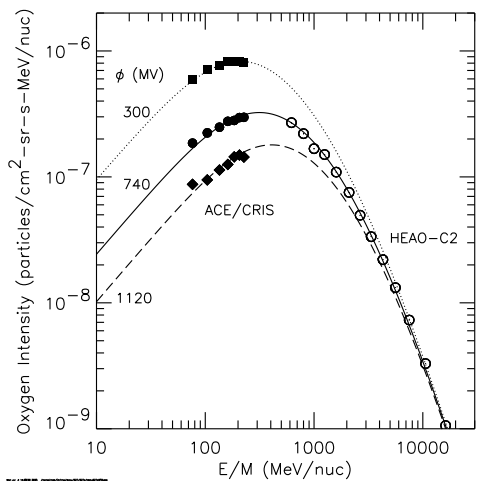}
\caption{\label{cr_solar_mod}
Oxygen cosmic ray intensity during three different periods: Sept. 1997
(squares), Feb 2000 (circles), Jan. 2001 (diamonds), from Ref.~\cite{wied}.} 
\end{figure}

Since the number of cosmic rays  increases with decreasing energy,
most cosmic rays are not visible to us. This suppression effect at
energies below a few GeV is clearly visible in Fig.~\ref{cr_solar_mod}, 
where the intensity of oxygen is shown for three different periods.

\subsection{Anisotropies and deflections in regular magnetic fields}

\begin{wrapfigure}{r}{.45\textwidth}
\epsfig{width=.45\textwidth,angle=0,file=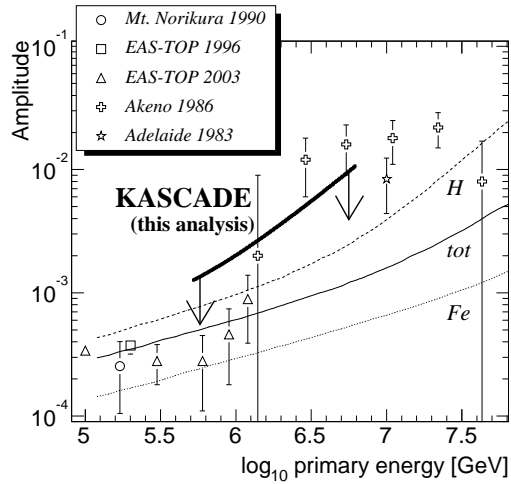}
\caption{\label{aniso}
Experimental limits on the anisotropy of cosmic rays, from
Ref.~\cite{Antoni:2003jm}.}
\end{wrapfigure}

\paragraph{Experimental results on dipole anisotropies}
Above 100\,GeV, when the influence of the Sun becomes negligible, the
cosmic ray flux is consistent with isotropy. At higher energies, when
cosmic rays drift slowly out the galaxy, one expects to detect first
anisotropies on large angular scales. Such anisotropies can be
characterized in first approximation by a pure dipole,  
$I=I_0+I_1\cos\theta$,
\be
 \delta = \frac{I_{\max}-I_{\min}}{I_{\max}+I_{\min}}
        = \frac{I_1}{I_0} \,.
\ee
Most experimental detections or limits for $\delta$ at energies 
$E\lsim 10^{15}$~eV are in the range $10^{-4}\lsim\delta\lsim 10^{-3}$, 
cf. Fig.~\ref{aniso}.
Many experimental searches are using that the experimental exposure of
a cosmic ray experiment taking continuously data is uniform in right ascension%
\footnote{In the equatorial coordinate system one projects the
usual geographical latitude and longitude on the celestial sphere.
The right ascension measures the angle east from the vernal equinox
point to an object projected on the celestial equator.}. Then one
performs an one-dimensional harmonic analysis: One sums  
\be
 a_k = \frac{2}{n} \sum_{a=1}^n \cos(k\phi_i) \quad\mbox{and}\quad
 b_k = \frac{2}{n} \sum_{a=1}^n \sin(k\phi_i)
\ee
over the $n$ data points.
The amplitude $r_k$ and and phase $\phi_k$ of the $k$.th harmonic are
given by 
\be
 r_k = \sqrt{a_k^2+b_k^2} \quad\mbox{and}\quad
 \phi_k=\arctan(b_k/a_k) 
\ee
with the chance probability
\be \label{rayleigh}
 p_{\rm ch}=\exp\left( -nr_k^2/4 \right) 
\ee
to observe a larger value of $r_k$ in an isotropic distribution.

\paragraph{Generalized Liouville theorem}
We measure cosmic rays after they traversed the magnetic field of the
Earth and of the Milky Way, and possibly extragalactic magnetic
fields. It is therefore important to separate between genuine
anisotropies and those eventually induced by the cosmic ray
propagation in magnetic fields.  

We want to show now that the intensity  $I$ is constant along any
possible cosmic ray trajectory. Let us consider the evolution of the phase
space distribution $f(\vx,\vp)$ from the time $t$ to $t+\d t$.
The number $\d N$ of particles around the point $\vx',\vp'$ at $
t'=t+\d t$ is
\be
 f(\vx',\vp') \d^3x' \d^3p' \,.
\ee
This number remains constant, if the Jacobian of the transformation 
$\vx,\vp\to\vx',\vp'$ is one,
\be
 J=\frac{\partial(\vx',\vp')}{\partial(\vx,\vp)} =1\,.
\ee
It is sufficient to show that the time-derivative $\d J/\d t=0$, i.e.\
that in the expansion of $J$ the first-order terms $\d t$ vanish.
From $\vx'=\vx+\vv \d t$ and $\vp'=\vp+\vF \d t$,
the diagonal of $J$ follows as
\be
 1,\;1,\;1,\;
 1+\frac{\partial F_x}{\partial p_x}\:\d t,\;
 1+\frac{\partial F_y}{\partial p_y}\:\d t,\;
1+\frac{\partial F_z}{\partial p_z}\:\d t\,,
\ee
while the off-diagonal elements are of order ${\cal O}(\d t)$. 
The expansion of $\d J$ to first-order in $\d t$ is thus
\be
 J=1+\left( \frac{\partial F_x}{\partial p_x}+
  \frac{\partial F_y}{\partial p_y}+
  \frac{\partial F_z}{\partial p_z} \right)\d t +\ldots
\ee
Thus the phase space distribution $f(\vx,\vp)$ is constant along a
trajectory, if $\nabla_p \vF=0$, which is fulfilled in an
electromagnetic field. Moreover, $p=|\vp|$ is constant in a pure
magnetic field and hence also $p^2 f(\vx,p)$. With Eq.~(\ref{IF}) it
follows that also the intensity $I$ is constant along any 
possible cosmic ray trajectory.  

Finally, we remark that the Liouville theorem does not imply that
magnetic fields can not enhance existing anisotropies. In fact, they
can act as magnetic lenses, enhancing or decreasing the flux received
from {\em point\/} sources similar as gravitational lenses.

\begin{figure}
\hspace*{1.cm}
\epsfig{width=.45\textwidth,angle=0,file=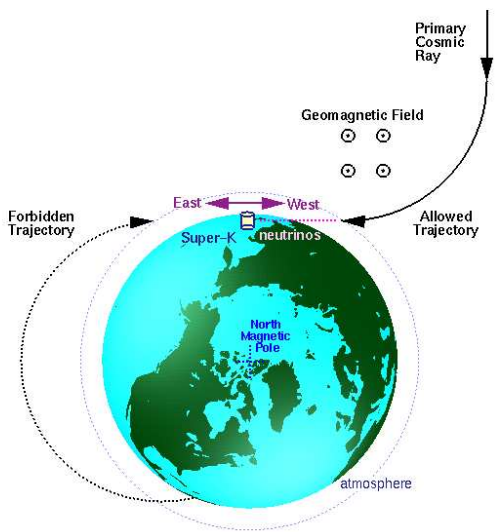}
\caption{\label{east-west}
A sketch of the east-west effect~\cite{SK}.}
\end{figure}

\paragraph{Latitude and east-west effect}
An isotropic cosmic ray flux remains isotropic propagating through a
magnetic field as long as the phase space is
simply connected. In other words, a necessary condition is that all
trajectories starting from the point considered on Earth (after
reversing the charge of the particle) reach $r=\infty$. At low enough
energies, this condition may be violated, because trajectories can be
deflected back to the Earth or stay within a finite distance $r$,
cf.~Fig.~\ref{east-west}.  In this case, the magnetic field does
induce anisotropies in the observed flux. 

Consider a particle of charge $Ze$ with orbit in the
equatorial plane of a dipole with magnetic moment $M$ (as a good model
for the geomagnetic field). Equating the centrifugal and the Lorentz
force gives
\be
 Ze|{\vv \times \vB}| =\frac{mv^2}{r} 
\ee
with $B=\mu_0/(4\pi) M/r^3$. The radius of the orbit is
\be
 r=  \left( \frac{\mu_0}{4\pi}\:\frac{ZeM}{p} \right)^{1/2} \,.
\ee
Setting $r=R_\oplus$ and using $M=8\times 10^{22}\,$Am 
as magnetic moment of the Earth, it follows
\be
\frac{p}{Z}= \frac{\mu_0}{4\pi}\:\frac{eM}{R_\oplus^2} \ap 
 59.6\:{\rm GeV} \,.
\ee
This is the minimal momentum of a proton able to reach the Earth from the
east, if its orbit is exactly  in the (magnetic) equatorial plane. 
The sign of this east-west asymmetry was used by Rossi and others to
show that the cosmic ray primaries are positively charged.
Towards the poles, the influence of the dipole field becomes weaker
($\vv\times\vB$), and the cutoff momentum becomes thus smaller. Thus
the integrated cosmic ray intensity increases with latitude for
charged particles (``latitude effect''). 

\paragraph{Compton-Getting effect}
\label{Compton-Getting}
\def\bu{{\mathbf u}}
\def\bp{{\mathbf p}}
\def\br{{\mathbf r}}
\def\bpp{{\mathbf p}^\prime}
\def\brp{{\mathbf r}^\prime}
\def\pp{p^\prime}
\def\fp{f^\prime}
\def\Ep{E^\prime}
\def\Ip{I^\prime}

Compton and Getting first discussed that a relative motion of observer
and cosmic ray sources results in an anisotropic cosmic ray flux,
using this effect 
as signature for the Galactic origin of cosmic rays with $E\gsim 0.1\:$GeV.

Lorentz invariance\footnote{The differentials $\d^3x$ and $\d^3p$
  transform opposite under Lorentz transformations, while the particle
  number $\d N$ is obviously a scalar.}  requires that the phase space
distribution function
$f$ in the frame of the observer, $f'(\brp,\bpp)$, equals the one in
the frame in which the cosmic ray flux is isotropic,
$f(\br,\bp)$. Expanding in the small parameter $\bp-\bpp\ap -p\,\bu$,
it follows 
\be
 \fp(\bpp) = f(\bpp)-p\,\bu\cdot\frac{\partial f(\bpp)}{\partial\bpp} +
   {\mathcal O}(u^2) 
  = 
 f(\bpp)\left(1-\frac{\bu \cdot \bp}{p}\frac{\d \ln f}{\d\ln\pp}\right) \,.
\ee
Since $u\equiv|\bu|\ll 1$, the anisotropy induced by the Compton-Getting 
effect is dominated by the lowest moment, i.e. its dipole
moment. Changing to the differential intensity $I(E)= p^2f(p)$,
one obtains  
\begin{equation}
\Ip(\Ep)\simeq I(E)
\left[1+\left(2-\frac{\d\ln I}{\d\ln \Ep}\right)\frac{\bu \cdot \bp}{p}\right].
\end{equation}
Thus the dipole anisotropy due to the Compton-Getting effect has the amplitude
\begin{equation}
  \delta_{\rm CG}\equiv \frac{I_{\rm max}-I_{\rm min}}{I_{\rm max}+I_{\rm
min}}=\left(2-\frac{\d\ln I}{\d\ln E}\right)\,u \,.
\end{equation}
The Sun moves with $u_\odot=220\,$km/s around the center of the Milky Way.
Most likely, the local ``cosmic ray rest frame'' is co-rotating with
the nearby stars and the  relevant velocity $u$ for the CG effect is
therefore much smaller. 
Taking into account the observed spectrum $I(E)\propto E^{-2.7}$ of
cosmic rays below the knee, the Compton-Getting effect should results
in a dipole anisotropy which amplitude is bounded by 
$\delta_{\rm CG}=(2+2.7)\,u\lsim 0.4\%$.

\section{Propagation of Galactic cosmic rays}
\label{GCRpropagation}
\paragraph{Cascade equation}
We want to explain the large over-abundance of the group Li-Be-B in
cosmic rays compared  to the Solar system. We consider two species,
primaries with number density $n_p$ and secondaries with number
density $n_s$. If the two species are coupled by the spallation
process $p\to s+X$, then  
\begin{subequations}
\label{spallation}
\ba 
 \frac{\d n_p}{\d X} &=& -\frac{n_p}{\lambda_p}\,, \\ 
 \frac{\d n_s}{\d X} &=& -\frac{n_s}{\lambda_s} +\frac{p_{\rm
     sp}n_p}{\lambda_p } \,,
\ea
\end{subequations}
where $X=\int\d l \:\rho(l)$  measures the amount of traversed  matter,
$\lambda_i=m/\sigma_i$ are the interaction lengths (in gr/cm$^2$),
and $p_{\rm sp}=\sigma_{\rm sp}/\sigma_{\rm tot}$ is the spallation
probability.  

\begin{figure}
\hspace*{2cm}
\epsfig{width=.45\textwidth,angle=0,file=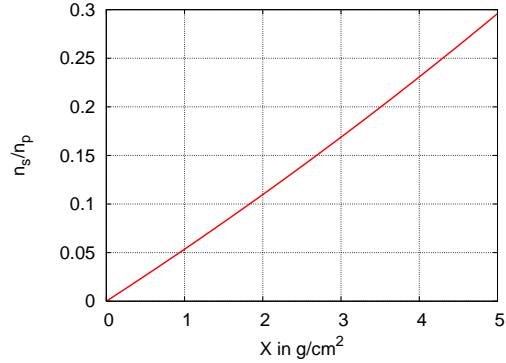}
\caption{\label{LiBeB}
The ratio $n_s/n_p$ as function of the traversed amount of matter $X$.}
\end{figure}

The system Eqs.~(\ref{spallation}) can be easily solved (Exercise 3.1) and 
using as initial condition $n_s(0)=0$ we obtain as ratio
\be
 \frac{n_s}{n_p} = \frac{p_{\rm sp}\lambda_s}{\lambda_s-\lambda_p } 
 \left[ \exp\left( \frac{X}{\lambda_p}-\frac{X}{\lambda_s} \right) 
        -1 \right] \,.
\ee 
If we consider as secondaries a group like Li-Be-B that has a much smaller
abundance in the solar system than in cosmic rays, most of them have to be
produced by spallation from heavier 
elements like the C-N-O group. With $\lambda_{\rm CNO}\ap 6.7\,$g/cm$^2$,
$\lambda_{\rm LiBeB}\ap 10\,$g/cm$^2$, and $p_{\rm sp}\ap 0.35$
measured at accelerators, the observed ratio 0.25 is reproduced for
$X\ap 4.3\,$g/cm$^2$, see Fig.~\ref{LiBeB}.

With $h=300\:{\rm pc}\ap 10^{21}\:$cm as thickness of the Galactic
disc, $n_H\ap 1/$cm$^3$ as density of the interstellar medium, a
cosmic ray following a straight line perpendicular the disc crosses
only $X=m_H n_H h\ap 10^{-3}$g/cm$^2$. 
The residence time of cosmic rays in the galaxy follows as
$t\sim (4.3/10^{-3})(h/c)\sim 1.4\times 10^{14}{\rm s}\sim 5\times 10^6\,$yr.
This result can only be explained, if the
propagation of cosmic rays resembles a random-walk.
Moreover, it suggests that acceleration and propagation 
can be treated separately. 

\paragraph{Random walks}
After $N$ steps ${\bf l_i}$ of the same size $|{\bf l_i}|=l$ a
particle that started at zero is at the position ${\bf d}=\sum_{i=1}^N
{\bf l}_i$. We assume that the direction of each step ${\bf l_i}$ is
chosen randomly. Then the scalar product of ${\bf d}$ with itself is
\be
{\bf d}\cdot{\bf d} = \sum_{i=1}^N\sum_{j=1}^N {\bf l}_i\cdot{\bf l}_j \,,
\ee
and splitting the sum into the diagonal and the off-diagonal
terms, we obtain
\be
 d^2 = 
 Nl^2 + 2l^2 \sum_{i=1}^N\sum_{j<i}^N  \cos\theta_{ij}
 \ap  Nl^2 \,.
\ee
By assumption, the angles $\theta_{ij}$ between ${\bf l}_i$ and
${\bf l}_j$ are chosen randomly and thus the off-diagonal terms cancel
against each other.

One might think of two different reasons behind the random walk:
Scattering with isotropic re-emission as in the case of photons in the Sun or
diffusion in turbulent magnetic fields. 
The first possibility is excluded by the small scattering probability
$\tau=h\sigma n_H\sim 10^{21}\times 10^{-25}\times 1\ap 10^{-4}$ and
by the fact that at high energies scattered particle are concentrated
in the forward direction within a cone of opening angle $\theta\sim 1/\Gamma$.
Hence we are lead to examine the second possibility, the scattering of
cosmic rays on turbulent magnetic fields in the Galactic disc.

\paragraph{Diffusion equation}
The continuity equation for the  number density $n$ and its current
${\bf j}$ (that corresponds to $\F$ in our usual notation),
\be
 \nabla {\bf j} + \frac{\partial n}{\partial t} = 0 \,,
\ee
leads together with Fick's law for an isotropic medium,
\be
 {\bf j} = - D \nabla n \,,
\ee
to the diffusion equation
\be
\frac{\partial n}{\partial t} -\nabla (D \nabla n) = Q \,,
\ee
where we added additionally a source term $Q=Q(E,{\bf x},t)$.
Note that the diffusion equation can be transformed for a diffusion
coefficient $D$ that is independent on $\vx$ 
into the free Schr\"odinger equation substituting $D\leftrightarrow
\hbar^2/(2m)$ and $t\leftrightarrow -it$. Hence we can borrow the free
propagator for a non-relativistic particle as Green's function  $G(r)$
for the diffusion equation with $D=\const$ and obtain with the mentioned
substitutions,  
\be
 G(r)= \frac{1}{(4\pi Dt)^{3/2}} \exp[-r^2/(4Dt)] \,.
\ee
Thus the mean distance traveled outward is $\propto \sqrt{Dt}$, as in
a random walk with $\langle r^2\rangle \sim N l_0^2$. Connecting the
two pictures, we obtain $D\sim N l_0^2/t\sim v l_0$ with $ v=Nl_0/t$. 
Therefore, the diffusion coefficient $D$ can be estimated as the product
of the cosmic ray velocity $v\ap c$ and its mean free path $l_0$. 
A more precise analysis gives $D=l_0v/3$, where the factor three reflects 
the number of spatial dimensions.

\paragraph{Diffusion coefficient} We estimate now the
energy-dependence of $D(E)$ and its numerical value for a cosmic ray
propagating in the Galactic disc. We start picturing its propagation as a
random-walk caused by scatterings  
on magnetic clouds of size $r_0$. Then one can distinguish two different
regimes: 
\begin{itemize}
\sitem
At low energies, i.e. when the Larmor radius $R_L=p/(ZeB)$ is smaller 
than the size $r_0$ of magnetic clouds with density $n$, the angles
between the entrance and the exit directions are isotropically
distributed. Since the direction is on average changed considerably  
in each scattering process, the mean free path $l_0$ is simply the
distance between clouds, $l_0=1/(\sigma n)\sim 1/(r_0^2 n)$ and thus 
\be \label{D_lowB}
  D_0=\frac{1}{3} l_0v \sim \frac{1}{3}\frac{c}{r_0^2 n} \sim\const
\ee
\sitem
At high energies, cosmic rays are deflected in each cloud only by a small angle
$\delta\sim r_0/R_L$. The directions are uncorrelated and thus the mean
deflection is zero, $\langle\delta\rangle=0$, and the variance
$\langle \delta^2\rangle$ is given again by the result for a random-walk,
$\langle \delta^2\rangle \sim N (r_0/R_L)^2$. The effective free mean
path $l_0$ is the distance after which $\langle\delta^2\rangle\sim 1$.
Hence the energy-dependence of the diffusion coefficient is
\be
 D(E) = \left( \frac{R_L}{r_0} \right)^2 D_0 \propto E^2 \,.
\ee
\sitem
The transition between these two regimes happens when $R_L(\Ecr)=r_0$. 
Numerically,  this energy is given by $\Ecr\ap 10^{15}{\rm eV}
(B/\mu{\rm G}) (r_0/{\rm pc})$.
\end{itemize}
Obviously, the picture of magnetic clouds or domains with an unique
size $r_0$ is an oversimplification. In a more realistic picture, there is a
distribution of magnetic field fluctuations that can be easiest
characterized by the spectrum of its Fourier components, 
$\langle B^2(k)\rangle\propto  k^{-\alpha}$. 
Charged particles scatter mainly at field fluctuations which wave
numbers $k$ matches their Larmor radius, $k\sim 1/R_L$. If the
amplitude of this resonant magnetic field fluctuation is 
$\delta B_{\rm res}$, then  $D\ap (\delta B_{\rm res}/B)^{-2} vR_L/3$
instead of our over-simplified estimate~(\ref{D_lowB}).
Thus the energy dependence of $D$ below $E_{\rm cr}$ is determined
by the power-spectrum of magnetic field fluctuations.
The size $r_0$ of magnetic field domains is in this picture
replaced by the correlation length $l_c$, i.e.\ the length scale below
the field is smooth. 

\paragraph{Diffusion and cosmic ray anisotropies}
In the diffusion picture, the resulting net cosmic ray flux is connected via
Fick's law with the diffusion tensor. We assume a small, pure dipole
anisotropy, $I=I_0+I_1\cos\theta$ and $I_1\ll I_0$, and choose the $z$
axis along the dipole axis. Only the dipole term contributes
integrating $I(E,\theta)$,   
\be  \label{diff1}
 \F_z(E) = 2\pi \int_0^\pi  \d\theta\sin\theta \, I(E,\theta)\cos\theta = 
  \frac{4\pi}{3}\, I_1(E)  \,.
\ee
On the other hand, for diffusion Fick's law is valid, 
\be
 \F_z(E) = -D_{zz} \partial n(E)/\partial z \,.
\ee
Comparing the two equations we obtain
\be
 I_1 = -\frac{3}{4\pi}D_{zz}\: \frac{\partial n}{\partial z} \,.
\ee
Then the anisotropy $\delta$ is
\be
 \delta = \frac{I_{\max}-I_{\min}}{I_{\max}+I_{\min}}
        = \frac{I_1}{I_0}= \frac{3 \F_z}{4\pi I_0}
        =-3D_{zz} \: \frac{1}{cn}\frac{\partial n}{\partial z} \,.
\ee

The experimental results for $\delta$ provide thus information on
$D$. For an estimate
we set $\partial n/\partial z\ap n/h$ where $h$ is the characteristic
scale on which $n$ changes. With $h\sim 500\:$pc and using
$\delta\sim 3\times 10^{-4}$, it follows
\be
 D \ap \frac{1}{3}\: \delta ch \ap 5\times 10^{27}\, {\rm cm}^2/{\rm s}\,.
\ee
Next we can estimate the mean free path of cosmic rays from $D=cl_0/3$
and obtain $l_0\sim 5\times 10^{17}$cm or 0.15~pc. Finally, we estimate
the residence time of cosmic rays in the disc as 
\be
\tau=\frac{h^2}{2D}\ap 8\times 10^6 {\rm yr}\,.
\ee
This means that the prediction for the residence time in our simple 
diffusion picture agrees in magnitude with observations. Since cosmic
rays are confined in the Galactic plane, and the Sun is located
somewhat above the plane, this diffusion flux should be directed
towards the northern Galactic hemisphere.

\paragraph{Complete cascade equation}

A rather general set of  equations describes the evolution of $N$ species
coupled by interactions $k\leftrightarrow i$ are the following transport or
cascade equations,
\ba   
 \frac{\partial n_i(E,\vx,t)}{\partial t} - \nabla (D \nabla n_i(E,\vx,t)) 
 &= & Q(E,\vx,t) 
\nonumber
\\ & - & 
\left(c\rho\lambda_{i,\rm inel}^{-1}(E) + \lambda_d^{-1} \right) n_i(E,\vx,t)
\nonumber\\ &-& \frac{\partial}{\partial E} \left( \beta_i n_i(E,\vx,t) \right) 
\nonumber\\ & + & 
\sum_k  \int_E^\infty dE' \;\frac{\d\sigma_{ki}(E',E)}{\d\!E} \:
n_k(E',\vx,t) \,.
\label{transport}
\ea
The first line describes diffusion, while the second line accounts for
the loss of particles in the energy interval $[E:E+\d\!E]$ because of
interactions or decays with $\lambda_d=\Gamma\tau_{1/2}$. 
The third line describes continuous energy
losses $\beta=\d E/\d t$ of a particle $i$: Two important examples are
synchrotron radiation and adiabatic redshift losses due to the
expansion of the Universe. Since $\beta n$ is the particle flux in energy space
(with ``velocity'' $\d E/\d t$), $\partial/\partial E (\beta_i n_i(E,\vx)$
is the divergence of this flux. Hence it is the analogue in energy
space to $\nabla_x \vj$ in coordinate space.

\paragraph{Leaky box model} 
This model assumes that cosmic rays inside a confinement volume
(e.g. the disc) have a constant escape probability per time,
$\tau_{\rm esc}\gg c/h$. Neglecting all other effects,
\be
 \frac{\partial n_i(E,\vx)}{\partial t} =
 - \frac{n_i(E,\vx)}{\tau_{\rm esc}} = D\Delta n_i(E,\vx) \,,
\ee
and the distribution of cosmic ray path lengths in the disc is
$n_i=n_0\exp(-t/\tau_{\rm esc}) =n_0\exp(-z/\lambda_{\rm esc})$. 
Hence one can replace
\be
 D\Delta n_i(E,\vx) \to  - \frac{n_i(E,\vx)}{\tau_{\rm esc}} 
\ee
in this model.
Physically, the diffusion coefficient is a function of the distance to
the disc, $D=D(z)$, and the escape probability increases
and the cosmic ray density decreases for increasing $|z|$, while in
the leaky box model both are constant inside the confinement volume. 
Therefore, the leaky box model can describe only average values of
observables that depend in general on $z$ and care has to be taken
comparing quantities calculated within the two different models.

If we consider now the steady-state solution, $\partial n_i/\partial t=0$, 
and replace the diffusion term by $n_i/\tau_{\rm esc}$, then
\be
 \frac{n_i(E)}{\tau_{\rm esc}} = 
  Q_i - \left(
   \frac{c\rho}{\lambda_i}+\frac{1}{\Gamma\tau_{1/2}}\right) n_i(E)
  +  \sum_k  \int_E^\infty\d\!E' \;\frac{\d\sigma_{ki}(E',E)}{\d\!E} \:
  n_k(E') \,.
\ee
For primary types like protons or iron, the decay term vanishes and
production via fragmentation can be neglected. Introducing
$\lambda_{\rm esc}=\beta c\rho \tau_{\rm esc}$ as the amount of matter
traversed by a particle with velocity $\beta c$ before escaping,
we obtain
\be
 n_i = \frac{Q_i\tau_{\rm esc}}{1+\lambda_{\rm esc}/\lambda_i}  \,.
\ee
The escape time $\tau_{\rm esc}$ in the leaky-box model should be,
similar to $D$ in the diffusion model, energy dependent. For the
simplest hypothesis that $\tau_{\rm esc}(E)$ of different elements
depends only on $Z$, one obtains from a fit to data
\be
  \lambda_{\rm esc}\ap 11 \frac{\rm g}{\rm cm^2} \left(\frac{4Z
      {\rm GeV}}{p}\right)^\delta 
\qquad{\rm for}\quad
 p\geq 4Z {\rm GeV}
\ee
with $\delta\ap 0.6$ and $\lambda_{\rm esc}=\const$ at lower energies.

For protons, $\lambda_p=55\,$g/cm$^2\gg\lambda_{\rm esc}$ for all
energies, and thus
\be
 n_p = Q_p\tau_{\rm esc}\propto Q_p E^{-\delta} \,.
\ee
Hence the generation spectrum of protons should be steeper than the one
observed, $Q_p\propto E^{-2.7+\delta} =E^{-2.1}$. 

For the other extreme case, iron, the interaction length is
$\lambda_{\rm Fe}=2.6\,$g/cm$^2$. Hence at low energies, iron nuclei are 
destroyed by
interactions before they escape, $\lambda_{\rm Fe}\ll\lambda_{\rm esc}$,
and therefore the iron spectrum reflects the generation spectrum, 
$n_{\rm Fe}\propto Q_{\rm Fe}$. Starting from the energy where
$\lambda_{\rm Fe}\sim\lambda_{\rm esc}$, the iron spectrum should become
steeper. The observed iron spectrum is indeed flatter at low energies and
steepens in the TeV range.

A main test for propagation models are radioactive isotopes with
life-time $\tau_{1/2}\sim\tau_{\rm esc}$. 
The abundance of such
isotopes, e.g. $^{10}$Be with $\tau_{1/2}=3.9\times 10^6\,$yr, can be
used to deduce separately $\tau_{\rm esc}$ and the density of gas.
One finds as mean density of the traversed gas $n\sim 0.3/$cm$^3$,
i.e.\ just one third of the value in the disc, supporting the idea
that cosmic rays are confined not only within the disc but in an
extended halo.

\paragraph{Knee} The knee marks a break in the cosmic ray spectrum at
$E\sim 3\times 10^{15}\,$eV. There are three classes of explanations: 
$(i)$ a break in the diffusion coefficient as function of energy, 
$(ii)$ a signature of the $E_{\max}$ distribution of sources, 
$(iii)$ a change in the interactions of cosmic rays in the atmosphere
at $\sqrt{s}\sim\:$few~TeV.  
The possibilities $(i)$ and $(ii)$ produce both a rigidity-dependent knee,
i.e.\ the position of the knee for different nuclei should be connected
by $E_Z=ZE_p$, while for $(iii)$ the position of the knee depends on $A$.

\begin{figure}
\epsfig{width=.65\textwidth,angle=0,file=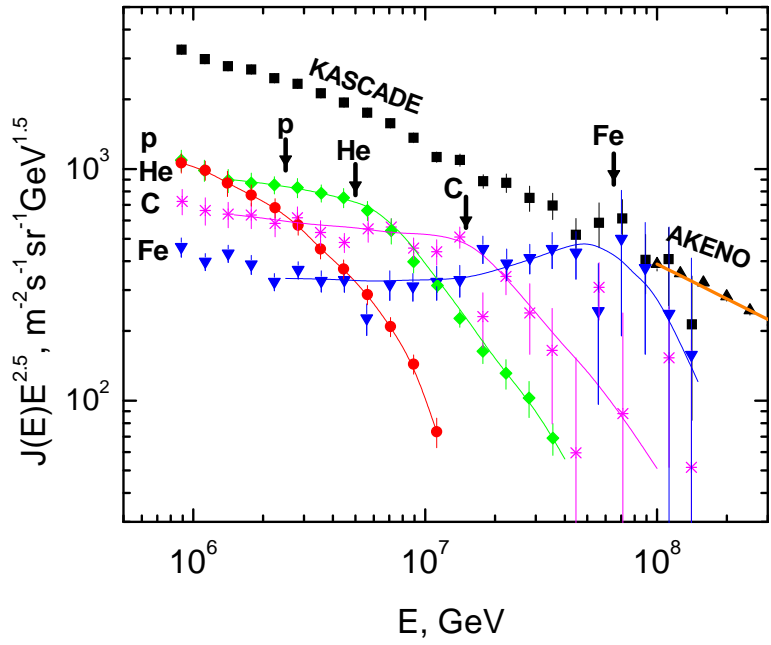}
\caption{\label{knee_comp}
The chemical composition measured by the KASCADE experiment
in the knee region. Additionally, the position of a rigidity-dependent
knee is for p, He, C and Fe indicated~\cite{Berezinsky:2004wx}.}
\end{figure}

Experimentally, a rigidity-dependent knee is favored as it is visible
from Fig.~\ref{knee_comp}. The energy of the knee agrees roughly with 
the critical energy, when the Larmor radius of a proton becomes equal
to the maximal length of magnetic field domains. Thus the knee might 
correspond to a transition from $D\propto E$ to $D\propto E^2$, 
or more generally to
an increased leakage of cosmic rays out of the galaxy.

\begin{figure}
\epsfig{width=.65\textwidth,angle=270,file=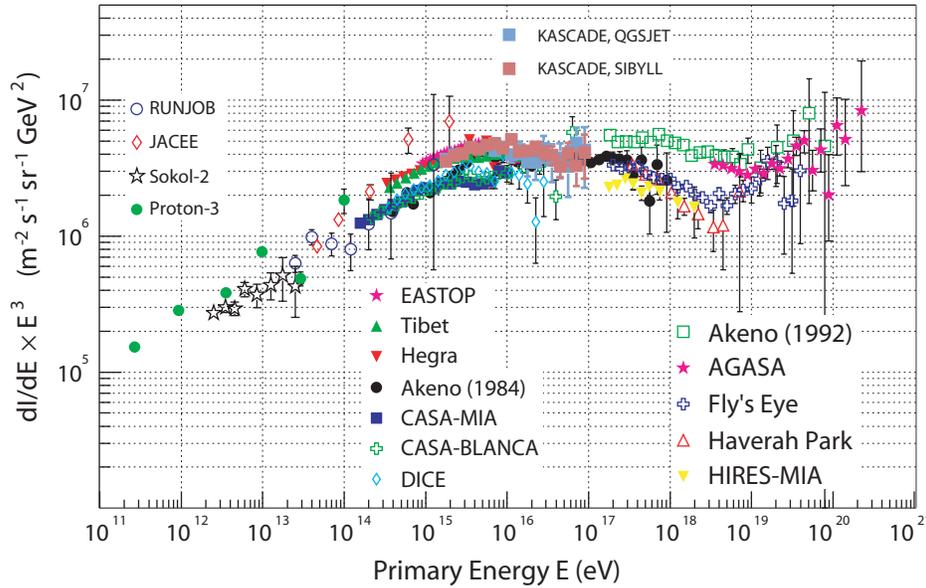}
\vskip-0.5cm
\caption{\label{knee}
The cosmic ray intensity in the knee region~\cite{Kampert:2006kg}.}
\end{figure}

\section{Exercises}
 
\begin{enumerate}
\item
Derive the solution of Eqs.~(\ref{spallation}) and that the observed
ratio 0.25 is reproduced for $X\ap 4.3\,$g/cm$^2$.
\item Show that the diffusion equation with a point source at $r=0$
and a space-independent diffusion coefficient has the solution 
$n(E,r)=1/(4\pi r)[Q(E)/D(E)]$.
\end{enumerate}

%% file: accel.tex
\chapter{Sources and acceleration of high energy cosmic rays}

\section{Sources of high energy cosmic rays}
The changes in the slope of the energy spectrum and in the composition
from protons to iron in the knee region suggest that there are at least two 
different kinds of cosmic ray sources. It is natural to associate the
bulk of cosmic rays with Galactic sources, while the highest energy
cosmic rays have an extragalactic origin, produced by more powerful
galaxies than ours. A further argument favoring an extragalactic
origin of cosmic rays with the highest energies is the observed
isotropy of their arrival directions on large scales that suggests a
cosmological distribution of their sources.

\subsection{General arguments}
\paragraph{Energy argument favoring supernova remnants (SNR)}
The luminosity $L_{\rm CR}$ of Galactic cosmic ray sources has to fit the
observed energy density $\rho_{\rm CR}\sim 1$~eV/cm$^3$ of cosmic rays, taking
into account their residence time $\tau_{\rm esc}\sim 6\times 10^6$~yr 
in the Galactic disk. With $V_D=\pi R^2 h\sim 
4\times 10^{66}$~cm$^3$ for $R=15\,$kpc and $h=200\,$pc
as volume of the Galactic disc, 
the required luminosity is $L_{\rm CR}= V_D\rho_{\rm CR}/\tau_{\rm esc}\sim
5\times 10^{40}$~erg/s.

In a successful core-collapse supernova (SN) around
$10\,M_\odot$ are ejected with $v\sim 5\times 10^8$~cm/s. Assuming
$1/(30\,$yr) as SN rate in the Milky Way, the average output in
kinetic energy of
Galactic SNe is $L_{\rm SN, kin} \sim 3\times 10^{42}$~erg/s.
Hence, if the remnants of SNe can accelerate particle with efficiency
$O(0.01)$, they could explain all galactic  cosmic rays as it was
suggested first by Ginzburg and Syrovatskii in the early 1960s.

\paragraph{Hillas argument}
\begin{figure}
\begin{center}
\epsfig{file=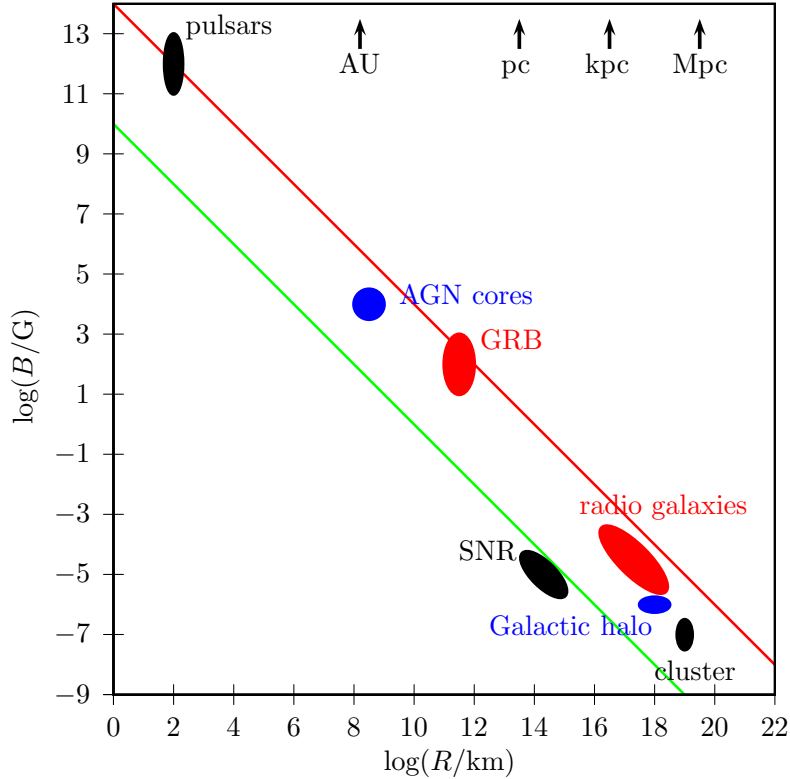,width=0.65\textwidth,angle=0}
\end{center}
\vskip-0.5cm
\caption{\label{Hillas}
Magnetic field strength versus size of various suggested  cosmic ray sources.}
\end{figure}
The Larmor orbit $R_L=E/(ZeB)$ of accelerated particles has to fit
inside the accelerator of size $R_s$, $R_L=E/(ZeB)\leq R_s$. 
For known magnetic fields and source sizes, one
can constrain thus the maximal achievable energy as $E_{\max}=\Gamma ZeBR_s$,
as it is done in Fig.~\ref{Hillas} for a compilation of potential
cosmic ray sources. The Lorentz factor $\Gamma$ introduced in $E_{\max}$ 
accounts for a possible relativistic
bulk motion of the source and is probably only for gamma ray bursts a
significant correction.  In Fig.~\ref{Hillas}, a so-called ``Hillas plot,''
sources able to accelerate protons to $E>10^{21}\,$eV should lie above the 
solid red line, while sources above the green line can accelerate iron up 
to $10^{20}\,$eV. Few candidate sources for acceleration to $E=10^{21}\,$eV
seem to be compatible with Hillas' argument.

On one hand, this constraint looks like a solid upper
limit, because energy losses are neglected, the maximal acceleration
time is finite and no  accelerator is 100\% efficient. On the other
hand, nonlinear processes may lead to an amplification of magnetic
fields inside the source.
In general, sources neither too small (minimizing energy losses) nor too big 
(avoiding too long acceleration times) are favored.

\paragraph{Blandford argument}

The acceleration of a proton to the energy $E=10^{20}\,$eV by regular
electromagnetic fields requires the potential difference $U=10^{20}\,$V.
What is the minimal power $P$ dissipated by such an accelerator?
In order to use the basic equation $P=UI=U^2/R$ known from high-school 
physics, we have to know the appropriate value of the resistance $R$. 
Since the acceleration region is
in most cases nearly empty, we use $R\sim 1000\:\Omega$ (lead by the 
``impedance of the vacuum'', $R=4\pi k_0/c=1/(\epsilon_0 c)\ap 377\,\Omega$).
Hence a source able to produce protons with $E=10^{20}\,$eV by
regular acceleration in
electromagnetic fields has the minimal luminosity~\cite{Blandford:1999hi} 
\be
 L = U^2/R\gsim 10^{37}\,{\rm W} =10^{44}\, {\rm erg/s} \,.
\ee
This can be transformed into an upper limit on the density $n_s$ of
ultrahigh energy cosmic rays (UHECR) sources, since the observed 
UHECR intensity
fixes the required emissivity ${\cal L}$, i.e.\ the energy input per
volume and time, as  ${\cal L}\sim 3\times 10^{46}$erg/(Mpc$^3$yr).
Hence, the density of UHECR sources able to accelerate protons to
$E=10^{20}\,$eV should be smaller than $n_s={\cal L}/L\sim
10^{-5}$/Mpc$^3$, if the acceleration is by regular
electromagnetic fields. For comparison, the density of normal galaxies
is $n_s\ap 10^{-2}$/Mpc$^3$, while the most common type of active
galactic nuclei in the nearby Universe, Seyfert galaxies, 
has the density $n_s\ap (1-5)\times
10^{-5}$/Mpc$^3$ within redshift $z\lsim 0.02$.

\subsection{Specific sources}

Most galactic astrophysical sources are connected with type II (or
core-collapse) supernovae (SN) and their remnants (SNR): 
Examples are the direct acceleration in the magnetosphere of young
pulsars and shock acceleration in SNRs. Above $E\gsim 10^{18}\:$eV,
the signature of galactic sources would be an enhanced flux towards
the galactic plane.  Practically all extragalactic sources except
gamma ray bursts are associated with active galactic nuclei (AGNs).

\paragraph{SNe II} 
Type II or core collapse supernovae occur at the end of the
fusion process in very massive stars, $M\gsim (5-8)M_\odot$. 
Theses stars develop an onion-like structure with a degenerate Fe
core. After the core is completely fused to iron, no further processes
releasing energy are possible. Instead, photo-disintegration  destroys
the heavy nuclei, e.g. via $\gamma+\Fe56\to\He4 + 4n$, and removes
the thermal energy necessary to provide pressure support. 
In the following collapse of the star, the density increases and the free
electrons are forced together with protons to form neutrons via
inverse beta decay, $e^-+ p\to n + \nu_e$: A proto-neutron star forms.
When the core density reaches nuclear density, the equation of state of
nuclear matter stiffens and infalling material is ``reflected,'' a
shock wave propagates outwards heated by neutrino emission from the
proto-neutron star. If the SN is successful, a neutron star is left over;
otherwise a black hole remains.

The released gravitational binding energy,
\be
\Delta E =
\left[-\frac{-GM^2}{R}\right]_{\rm star} -
\left[-\frac{-GM^2}{R}\right]_{\rm NS}
\sim 5\times 10^{53} {\rm erg
 \left( \frac{\rm 10km}{R} \right)
 \left( \frac{M_{NS}}{1.4M_\odot} \right)}
\ee
is emitted mainly via neutrinos (99\%). Only 1\% is transferred into kinetic
energy of the exploding star and only 0.01\% goes into photons.

\paragraph{Pulsars} are the left-over of the former iron core, now
  transformed into a neutron star with $R_{\rm NS}/R_\odot\sim
  10^{-5}$. They may be born fast rotating with a strong magnetic
  field, because of conservation of  
\begin{itemize}
\sitem
 angular momentum, $L=I\omega=\const$, with $I\sim MR^2$: 
A star like the sun, rotating once per month, $P\ap 10^6\,$s,  would
rotate with a ms period when contracted down to 10 km in size. 
\sitem
magnetic flux, $\phi_B=BA=\const$, for an ideal
conductor, where magnetic field lines are frozen in the plasma.
As the core collapses, the magnetic field lines are pulled more closely
together, intensifying the magnetic field by a factor 
$(R_\odot/R_{\rm NS})^2\sim 10^{10}$. Magnetic A stars have surface
fields up to $10^4\,$G, while the maximal observed field strengths of 
neutron stars are of the order $B \ap 10^{12}\,$G.
\end{itemize}

\paragraph{Rotating dipole model}

The energy of a rotating sphere is $E=I\omega^2/2$ and the energy
change is
\be
 \dot E=  I\omega \dot\omega \,.
\ee
The rotational kinetic energy of the Crab nebula, the remnant of a SN
observed by Chinese astronomers in 1054, is $E=\frac{1}{2}I\omega^2\ap 3\times
10^{49}\,$erg with $I=a MR^2$ and $a\ap 2/5$ for a homogeneous sphere;
its energy-loss per time is the time-derivative 
$\dot E=I\omega\dot\omega\ap 7\times 10^{38}\,$erg/s,
with $I=1.5\times 10^{45}\,$g\,cm$^2$ and  $\dot\omega=4\times
10^{-4}\,$yr$\,\omega$.

If the energy is lost as electromagnetic dipole radiation, then
\be
 \dot E = -\frac{B^2R^6\omega^4\sin^2\alpha}{6c^3}
\ee
where $\alpha$ denotes the angle between rotation and dipole axis, and
we obtain an estimate for Crab's magnetic field, $B\ap 7\times
10^{12}\,$G for $\sin\alpha=1$.  Thus pulsars have indeed extremely
strong magnetic fields that, if they are fast rotating, may accelerate
particles to high energies.

\paragraph{Direct acceleration of particles by pulsars}

The light-cylinder around a pulsar is the surface at
$R_c=c/\omega$. If the magnetic field lines would rotate rigidly with
the pulsar, then the linear velocity reaches the speed of light, $v=c$, 
at $R_c$.

If the magnetosphere is filled sufficiently with plasma, the electric
conductivity $\sigma\to\infty$, and Ohm's law $J=\sigma(E+vB/c)$ implies 
\be
 {\bf E} = -{\bf v}\times {\bf B}/c \,,
\ee
where $\vv=\omega\times \vr$. 

For a magnetic dipole field pointing along $\theta=0$,
\be
 \vB(r) = 
 \frac{B_0R^3}{r^3} \:\left( 2\e_r\cos\theta+e_\theta\sin\theta\right) \,,
\ee
the potential difference between the polar cap and infinity follows as
\be \label{pulsarmax}
 \Delta\phi = \int  {\bf E}\cdot  d{\bf s} 
            = -\frac{1}{c}\int {\bf v}\times {\bf B} \cdot  d{\bf s} 
            =\frac{B_0R^2\omega}{c}\: \sin^2\theta \,.
\ee
Open field lines that extend beyond the light cylinder $c/\omega$
start near the polar cap, within the cone 
$\theta_0\sim \sqrt{\omega R/c}$.  Inserting $\theta_0$ into
Eq.~(\ref{pulsarmax}) we obtain as maximal acceleration energy
achievable by a pulsar
\be
 E_{\rm max} \ap \frac{ZB_0R^2\omega\sin^2\theta_0}{c} 
             \ap  \frac{ZB_0R^3\omega^2}{c^2} 
\ap 8\times 10^{20}~{\rm eV} \,\frac{Z\,B}{10^{13}~{\rm G}}
                   \left( \frac{\Omega}{3000~{\rm s}^{-1}} \right)^2\,.
\ee                      
Thus a young, fast rotating pulsar appears to be a very good particle
accelerator. The main problems are that in realistic models the
potential difference $\Delta\phi$ that can be used for particle
acceleration is much smaller and the extreme energy losses due to,
e.g. curvature radiation. The magnetosphere of a pulsar may be also
filled by $\gamma\to e^+e^-$ with a plasma. 
Finally, pulsars as main source of UHECRs would predict a strong
anisotropy of the UHECR intensity, because neutron stars are concentrated in
the Galactic plane.

\paragraph{Shock acceleration in SN remnants} is the standard paradigm
for cosmic ray acceleration up to $E\sim 10^{15}-10^{17}$eV in our
galaxy and is discussed in the next section.

\paragraph{Active galaxies} is a common name for all galaxies with
unusual emission that is not associated with stars. In contrast to
normal galaxies whose total luminosity is the sum of the thermal emission
from each of the stars found in the galaxy, a large fraction of the
total luminosity of an active galaxy is non-thermal and is emitted by
the nuclei of the galaxy.  The modern view is that the common
mechanism behind the energy generation in AGNs is accretion on the
SMBH in their center.  

Main reason for this view is the fast variability of their spectra,
from time-scales of months for quasar spectra down to daily variations
for blazars. Causality limits then the size of the emission region as
$ct\lsim 200\,$AU. Secondly, the enormous energy output
of AGNs requires an extremely efficient energy generation mechanism.
For accretion on a BH, the maximal energy gain is $E_{\max}\sim GmM/R_S$,
where the Schwarzschild radius is $R_S=2GM/c^2$, and thus
$E_{\max}=mc^2/2$. A large part of this energy will be lost in the BH,
while the remainder heats up via friction an accretion disc
around the black hole. Modeling the accretion process gives an
efficiency of $\epsilon=10\%$--20\%. Thus the luminosity from accretion is 
\be
 L=  \frac{\epsilon c^2 \d m}{2\d t} \,.
\ee
For a rather modest mass consumption of the BH, $\d m/\d t=1M_\odot/$yr, 
one obtains $6\times 10^{45}$erg/s or $L\sim 10^{12}L_\odot$.

\begin{figure}
\begin{center}
\epsfig{width=.75\textwidth,angle=0,file=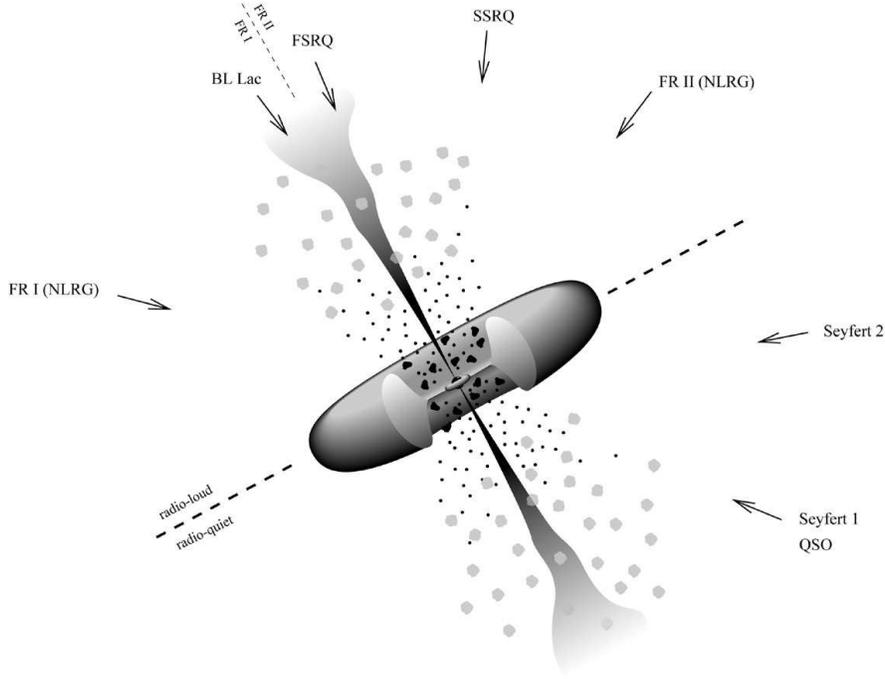}
\end{center}
\caption{\label{AGN}
The unified scheme of AGN.} 
\end{figure}

The unified picture of AGNs is illustrated in Fig.~\ref{AGN}. The
different AGN types are only facets of the same phenomenon--accretion
on a SMBH---viewed from different angles, at different stages of activity
(small or large $\d m/\d t$) and evolution in time (e.g.\ from a quasar
phase at redshift $z\sim 1$--4 towards a Seyfert galaxy at present).  
Blazars are AGN with a relativistic jet that is pointing in the
direction of the Earth and are therefore often ranked among the most 
promising sources of ultrahigh energy cosmic rays. 
According Fig.~\ref{AGN}, blazars are either 
Flat Spectrum Radio Quasars (FSRQ) or BL Lac objects.

\paragraph{Gamma Ray Bursts} come in two (main) sub-varieties,
depending on their duration that varies from fraction of a second to
many minutes. While short duration GRBs are most likely
the result of binary collisions between e.g.\ neutron stars, long
duration GRBs which make up about 2/3 of all GRBs are associated with
supernova events in extremely massive stars. GRBs are highly beamed
sources of gamma-rays and perhaps also of high energy neutrinos and
cosmic rays. A distinctive feature is the high Lorentz
factor of shocks in GRBs.

\section{Acceleration of cosmic rays}

\subsection{Second order Fermi acceleration}
\begin{figure}
\vspace*{3.2cm}
\begin{center}
\epsfig{file=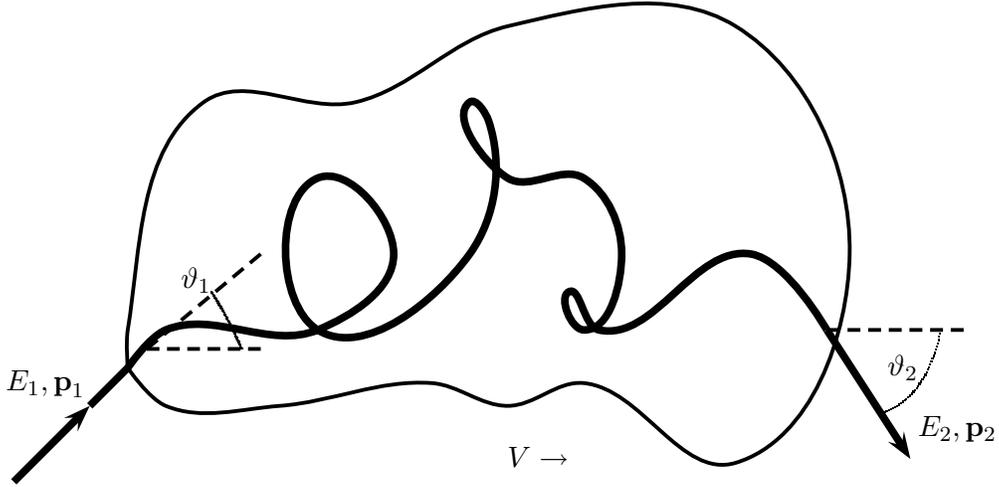,width=0.85\textwidth,angle=0}
\end{center}
\caption{\label{Fermi2}
A cosmic ray  ``scattering'' elastically at a magnetic cloud moving with
velocity $V$.}
\end{figure}

We consider a cosmic ray with initial energy $E_1$ ``scattering''
elastically on a magnetic cloud that moves with velocity $V\ll c$. 
We want to derive the
energy gain $\xi\equiv (E_2-E_1)/E_1$ per scattering.
The variables $(E_1,\theta_1)$ and  $(E_2,\theta_2)$ are shown in
Fig.~\ref{Fermi2}, which we label with a prime in the cloud system and 
without a prime in the lab system.

With a first Lorentz transformation between the lab and cloud system
we connect the variables characterizing the cosmic ray entering the cloud, 
\be
E_{1}^{\prime} = \gamma E_{1} (1 - \beta \cos \theta_{1})
\quad\mbox{where}\quad 
\beta = V/c \quad\mbox{and}\quad \gamma = 1/\sqrt{1-\beta^{2}} \,,
\ee
and with a second Lorentz transformation the exit variables,
\be
E_{2} = \gamma E_{2}^{\prime} (1 + \beta \cos \theta_{2}^{\prime}) \,.
\ee
Since the scattering off magnetic irregularities is collisionless and 
the cloud is very massive, energy is conserved, $E_{2}^{\prime} =
E_{1}^{\prime}$. Hence we can eliminate $E_2^\prime$ and obtain as
relative energy gain
\be \label{xi}
\xi = \frac{E_{2}-E_{1}}{E_1} = 
\frac{1 - \beta \cos \theta_{1} + \beta \cos \theta_{2}^{\prime}
- \beta^{2} \cos \theta_{1} \cos \theta_{2}^{\prime}}{
1 - \beta^{2}} -1 \,.
\ee

To proceed, we need average values of $\cos\theta_1$ and
$\cos\theta_2^{\prime}$. Since the cosmic ray scatters off magnetic
irregularities many times in the cloud, its exit direction  
is randomized, $\langle \cos \theta_{2}^{\prime} \rangle =0$.
The collision rate of the cosmic ray with the cloud is  proportional to their 
relative velocity $(v - V \cos \theta_{1})$. For ultrarelativistic
particles, $v\to c$, and the collision rate is
\be
 \frac{\d n}{\d\Omega_{1}} \propto (1 - \beta \cos \theta_{1}) \,.
\ee
We obtain $\langle \cos\theta_{1} \rangle$ averaging $\cos\theta_{1}$
weighted by $\d n/\d\Omega_{1}$ over all angles,  
\be
 \langle \cos \theta_{1} \rangle = 
 \int \cos \theta_{1} {\d n \over \d\Omega_{1}} \:\d\Omega_{1} /
 \int {\d n \over \d\Omega_{1}} \:\d\Omega_{1} = - {\beta \over 3} \,.
\ee
Plugging $\langle \cos \theta_{2}^{\prime} \rangle =0$
and $\langle \cos \theta_{1} \rangle = - \beta/3$
into Eq.~(\ref{xi}) and taking into account that $\beta \ll 1$ 
gives as average gain
\be
\langle\xi\rangle = {1 + \beta^{2}/3 \over
1 - \beta^{2}} -1 \simeq {4 \over 3} \beta^{2} \,.
\ee
Thus $\langle\xi\rangle\propto\beta^{2}>0$ and we have shown that on average a
cosmic ray gains energy scattering on ``magnetic clouds'' with an
ordered bulk velocity $V$. The energy gain per scattering is however
only of second order in the small parameter $\beta$, and acceleration
is therefore rather inefficient.
Moreover, one can show that the resulting energy spectrum would depend
strongly on the cloud parameters, making thereby the observed 
feature-less power spectrum of cosmic rays difficult to understand. 
Acceleration at shocks that we consider next avoids both disadvantages.

\subsection{Shock or first order Fermi acceleration}

\paragraph{Ideal fluids and shocks}
As first step in gaining some understanding of shocks we recall the
the ideal fluid equations. The three basic equations are the
conservation equation for mass, for momentum, and the Poisson
equation, 
\begin{subequations}
\be
\partial_t\rho+\nabla\cdot(\rho\vv) = 0 \,,
\ee
\be \label{euler}
 \rho\frac{\partial\vv}{\partial t} + \rho\vv\cdot(\nabla\vv) =\vF-\nabla P \,,
\ee
\be
 \Delta\phi = 4\pi G\rho \,.
\ee
\end{subequations}
The LHS of (\ref{euler}), the Euler equation, 
measures the velocity change $\d v/\d t$ of a fluid element,
summing up the change at a fixed coordinate, $\partial_t v$, and the  
change due to the movement of the element, while the RHS consists of
an external force $\vF$ and a force due to a pressure gradient $\nabla
P$. The Poisson equation connects the mass density $\rho$ with the
gravitational potential $\phi$. 

We consider now small perturbations $x_1$ around a static background, 
$\rho_0=\const$, $P_0=\const$ and $v_0=0$. Restricting us to small 
perturbations,
$x_1\ll x_0$, allows us to neglect all quantities quadratic in the 
perturbations. Sound waves propagate in most circumstances
adiabatically, i.e.\ without production of entropy, $\d S=0$. 
Thus changes in $P$ and $\rho$ are connected via 
\be
 P=P_0 + \left(\frac{\partial P}{\partial\rho} \right)_S \d\rho +
 \left(\frac{\partial P}{\partial S} \right)_P \d S= P_0 +  c_s^2 \d\rho \,.
\ee
Inserting $x=x_0+ x_1$ into the fluid equations and neglecting
quadratic term gives
\begin{subequations}
\be
\partial_t\rho_1+\rho_0 \nabla\cdot\vv_1 = 0 \,,
\label{Euler3a}
\ee
\be
\Delta\phi_1 = 4\pi G\rho_1 \,,
\label{Euler3b}
\ee
\be
\partial_t\vv_1 + 
\frac{c_s^2}{\rho_0}\nabla\rho_1 + \nabla \phi_1 = 0 \,,
\label{Euler3c}
\ee
\end{subequations}
where we used  also
$\partial_x P = \frac{\partial\rho_1}{\partial x}\,
\frac{\partial}{\partial\rho_1}\: P_1 =  
\frac{\partial\rho_1}{\partial x} c_s^2$.
These three equations can be combined into one second-order
differential equation for $\rho_1$. We multiply Eq.~(\ref{Euler3c}) by
$\rho_0$ and apply $\nabla$ on it,
\be
c_s^2 \Delta\rho_1 =- \rho_0 ( \partial_t\nabla\cdot\vv_1 + \Delta\phi_1) \,.
\ee
Then we insert (\ref{Euler3b}) for $\Delta\phi_1$ and 
(\ref{Euler3a}) for $\nabla\cdot\vv_1$, and obtain a linear, inhomogeneous wave
equation 
\be
\partial^2_t\rho_1 - \underbrace{c_s^2 \Delta\rho_1}_{\rm pressure}
= \underbrace{4\pi G\rho_1\rho_0}_{\rm grav. force} 
\ee
for the density perturbation $\rho_1$. The dispersion relation of
plane waves $\exp(-i(\omega t -kx))$,
\be
\omega^2=c_s^2k^2-4\pi G\rho_0 \,,
\ee
confirms that $c_s=(\partial P/\partial\rho_1)^{1/2}$
is the sound speed.

For a mono-atomic gas, the equation of state is $P=K\rho^\gamma$ with
$\gamma=5/3$. Thus, the sound speed is $c_s=(\gamma P/\rho)^{1/2}$ and
if an adiabatic compression with density $\rho_2=\eps\rho_1$
propagates, then $ c_s \propto \eps^{(\gamma-1)/2}$.
Hence the sound speed increases for a compression, the dense region
overruns uncompressed regions and becomes even denser: A discontinuity
(=''shock'') develops in some hydrodynamical variables like the density.

\begin{figure}
\vspace*{1.5cm}
\begin{center}
\epsfig{file=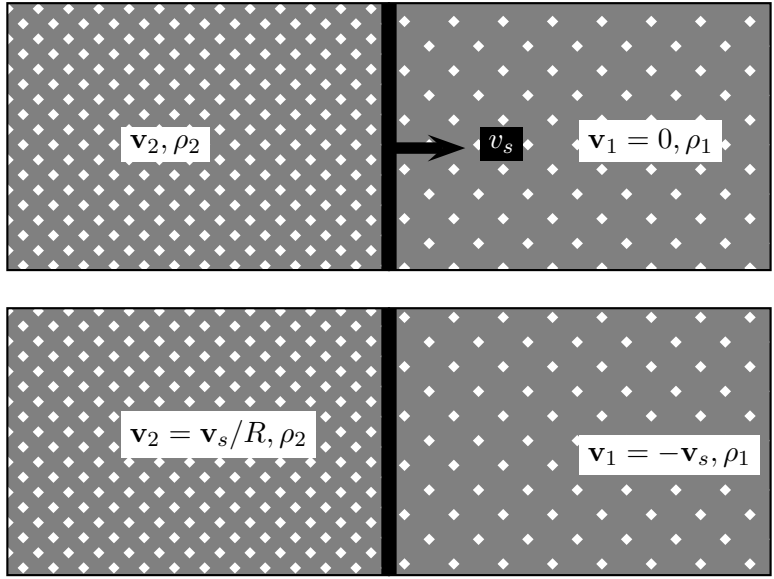,width=0.55\textwidth,angle=0}
\end{center}
\vspace*{1.cm}
\caption{\label{shockrest}
Conditions on the down-stream (left) and the up-stream (right) side of
a shock in the lab system (top) and in the shock rest frame with 
$\vv_1=-\vv_s$ (bottom).}
\end{figure}

\paragraph{Mach number of a (strong) shock}
Our main aim is to derive the Mach number ${\cal M}=v/c_s$  of a
shock that as we will see determines the slope of the energy spectrum
of accelerated particles. We will consider only the properties of an
one-dimensional, 
steady shock in its rest frame and assume that magnetic or gravitational
fields can be neglected. Then the continuity equation for mass, 
$\partial_t\rho+\nabla\cdot(\rho\vv)=0$, becomes simply 
\be  \label{mass1}
 \frac{\d}{\d x}  \left( \rho v \right) =0 \,.
\ee
The Euler equation simplifies using the same assumptions and taking
into account Eq.~(\ref{mass1}) to 
\be \label{mom1}
  \frac{\d}{\d x}  \left( P+\rho v^2\right) =0 \,.
\ee
Additionally to the conservation laws for mass (\ref{mass1}) and  momentum
(\ref{mom1}) we need the conservation law for energy,
\be
 \frac{\partial}{\partial t} \left( \frac{\rho v^2}{2}+\rho U +\rho\Phi
\right)
+\nabla\cdot\left[\rho \vv \left(  \frac{v^2}{2}+U +\frac{P}{\rho}+\Phi \right)\right] = 0 \,.
\ee
Here, the first bracket accounts for the change of kinetic, internal
and potential energy with time which has to be balanced by the energy flux
through the boundary of the considered volume. Specializing again to
the case of an one-dimensional, stationary flow with $\Phi=0$ gives
\be
 \frac{\d}{\d x}  \left( \frac{\rho v^3}{2} + (U+P)v\right) =0 \,.
\ee
Integrating these equations over the discontinuity of the shock results in
the ``Rankine-Hugoniot'' jump conditions,
\begin{subequations}
\ba
  \left[ \rho v \right]_1^2 =0 \,, \\
  \left[ P+\rho v^2\right]_1^2 =0\,,  \label{sh2b}\\
  \left[ \frac{\rho v^2}{2} + \frac{\gamma}{\gamma-1}Pv\right]_1^2 =0\,,
  \label{sh2c}
\ea
\end{subequations}
where we used also $U=P/(\gamma-1)$. Since we assumed a steady flow,
these jump conditions have to be evaluated in the shock rest frame,
cf.~Fig~\ref{shockrest}  -- otherwise time-derivatives should be included. 

Inserting first $\rho_2=(v_1/v_2)\rho_1$ into (\ref{sh2b}) gives 
$P_2=P_1+\rho_1v_1(v_1-v_2)$. Next we use these two expressions to
eliminate $\rho_2$ and $P_2$ from Eq.~(\ref{sh2c}). Reordering the
resulting equation according to powers of $v_2$,
\be
 \left(\frac{\gamma+1}{\gamma-1}\right) v_2^2 + \frac{2\gamma}{\gamma-1}
 \left(\frac{P_1+\rho_1v_1^2}{\rho_1v_1}\right) v_2 +
 v_1^2 \frac{2\gamma}{\gamma-1}\frac{P_1}{\rho_1} =0 \,,
\ee
replacing $P_1$ by the sound speed and dividing by
$v_1^2$, we obtain finally a quadratic equation for $t=v_2/v_1$,   
\be  \label{shockRM}
 \left(\frac{\gamma+1}{\gamma-1}\right) t^2 + \frac{2\gamma}{\gamma-1}
 \left(\frac{c_1^2}{v_1^2}+\gamma \right) t +
 \left( 1+\frac{2}{\gamma-1}\frac{c_1^2}{v_1^2}\right) =0 \,.
\ee
Now we recognize $v_1/c_1$ as the Mach number ${\cal M}$. Since we are
interested in fast flows, $v_1\gg c_1$, we can neglect the two  
$1/{\cal M}^2$ terms and obtain as approximate solutions
\begin{subequations}
\ba
  t=1 &\qquad{\rm or }\qquad& v_1=v_2 \,, \\
  t=\frac{\gamma-1}{\gamma+1}\equiv R &\qquad{\rm or }\qquad& Rv_2 =v_1 \,.
\ea
\end{subequations}
The first solution is obviously trivial, while the second one is the
strong shock solution. The compression ratio $R$ indicates how strong
the density and the velocity in the up- and down stream regions differ.
Since we are using as reference frame the shock frame, we have
$v_s=v_1$ and thus for $\gamma=5/3$ and $R=4$, 
\begin{subequations}
\ba
  v_2=v_s/R=v_s/4 \,, \\
  \rho_2=R\rho_1=4\rho_1 \,, \\
  P_2=3\rho_1 v_s^2/4 \,.
\ea
\end{subequations}
Hence no matter how strong a shock is, it can compress a mono-atomic gas
only by a factor four. The same factor relates the velocity of the
shock and of the matter after the passage of the shock. It is this
universal ratio of the up- and down-stream velocities for any strong
shock that results in the generic prediction of an $1/E^2$
spectrum produced by acceleration at shocks.

\paragraph{Acceleration at shocks}
We consider again a cosmic ray with initial energy $E_1$ ``scattering'' 
at magnetic irregularities that are now separated by a planar
shock moving with velocity $v_s$. The only change to the previous
discussion are the resulting different angular averages. The
(normalized) crossing
rate is given by the projection of an isotropic flux on the planar shock,
\be
 \frac{\d n}{\d\cos\theta_1} = \left\{ \begin{array}{ll}
                               2\cos\theta_1 & \cos\theta_1<0\\
                               0             & \cos\theta_1>0
                              \end{array} \right. \,.
\ee
while the crossing rate $\d n/\d\cos\theta_2$ is non-zero for
$\cos\theta_2>0$. 
Thus $\langle\cos\theta_1\rangle=-\frac{2}{3}$ and 
$\langle\cos\theta_2^\prime\rangle=\frac{2}{3}$ and hence
\be
\langle\xi\rangle \ap \frac{4}{3}\beta =\frac{4}{3} \frac{v_1-v_2}{c} \,.
\ee
The average gain $\langle\xi\rangle$ is now linear in $\beta$ and
justifies the name ``first-order'' Fermi (or shock) acceleration.

\paragraph{Energy spectrum produced by Fermi acceleration}
The energy $E_n$ of a cosmic ray after $n$ acceleration cycles is
\be
 E_n=E_0(1+\xi)^n
\ee
and the number of cycles needed to reach $E_n$ is thus
\be
 n=\ln\left(\frac{E_n}{E_0}\right) / \ln\left(1+\xi\right) \,.
\ee
If the escape probability $p_{\rm esc}$ per encounter is constant, then the
probability to stay in the acceleration region after $n$
encounters is $(1-p_{\rm esc})^n$. 
We obtain the fraction $f$ of particles with energy $>E_n$ as
\be
f(>E) = \sum_{m=n}^\infty (1-p_{\rm esc})^m 
      =\frac{(1-p_{\rm esc})^{\rnode{n2}{n}}}{p_{\rm esc}}
      \propto \frac{1}{p_{\rm esc}} \left(\frac{E}{E_0}\right)^\gamma \,,
\ee
where
\be
\gamma=\ln\left(\frac{1}{1-p_{\rm esc}}\right)/\ln(1+\xi)\ap p_{\rm
  esc}/\xi 
\ee
with $\xi\ll 1$ and $p_{\rm esc}\ll 1$. Hence, both first and second
order Fermi acceleration produces a power-law energy spectrum. 

The maximal energy achievable by a specific source is
determined by several factors. First, the finite life-time limits the
number of cycles $n$ and thus $E_n$. Second, the escape probability is
energy dependent and increases generally for increasing energy. Third,
energy losses like synchrotron radiation increase with energy and balance
at a certain point the energy gain.

\paragraph{Exponent $\gamma$ for shock acceleration}
Since we know already that
$\xi\ap\frac{4}{3}\beta=\frac{4}{3}\frac{v_1-v_2}{c}$, we have to
determine only the escape probability $p_{\rm esc}$ in order to
estimate the exponent $\gamma\ap p_{\rm esc}/\xi$ of the integral
spectrum produced by shock acceleration.
The particle flux $\F$ through an infinite, planar shock front is 
(cf.~Eq.(\ref{fluxplanar}))
\be
 \F(E)=\pi I(E) =\frac{cn(E)}{4} \,,
\ee
assuming $v_s\ll c$ and an efficient isotropization of cosmic rays
up-stream. 

In the shock rest frame, there is a particle flow $\F_{\rm esc}(E)=v_2n(E)$
downstream away from the shock front that will be lost for the
acceleration process. Thus the escape probability $p_{\rm esc}$ as the
ratio of the loss and the crossing flux is
\be
 p_{\rm esc} = \frac{\F_{\rm esc}}{\F} = \frac{v_2n}{cn/4}=\frac{4v_2}{c}
\ee
and the spectral index of the integral energy spectrum follows as
\be
 \gamma \ap p_{\rm esc}/\xi \ap \frac{3}{v_1/v_2-1} \,.
\ee
Typical values for the sound speed in the interstellar medium around a
SNR are $c_1\sim 10\,$km/s, while the shock velocities are $v_1\sim
10^4\,$km/s. Thus the Mach number ${\cal M}=v_1/c_1\gg 1$ and we can
use the result $v_1=4v_2$ in the strong shock limit. 
The exponent predicted by Fermi acceleration at non-relativistic
shock is therefore independent of the shock parameters and agrees
with the value needed to explain the spectrum of Galactic cosmic rays.
These are the two main reasons for the popularity of shock
acceleration. 

We can check that possible corrections are small by solving 
Eq.~(\ref{shockRM}) for arbitrary ${\cal M}$. Then we obtain
$R=4{\cal M}^2/(3+{\cal M}^2)$ and thus
\be
 \gamma \ap \frac{3+{\cal M}^2}{{\cal M}^2-1} 
 \ap  1+ \frac{1}{4{\cal M}^2} + {\cal O}({\cal M}^{-4}) \,.
\ee
Clearly, other effects like energy losses or an energy dependence of
$p_{\rm esc}$ will have a larger impact on the resulting spectrum
than the ${\cal O}({\cal M}^{-2})$ corrections.
Moreover, we worked within an extremely simplified picture and should
expect therefore deviations from a power-law with $\gamma\sim 1$. For
instance, we used the test particle approach that neglects the
influence of cosmic rays on the plasma. 

Finally we remark that the observed flux from extragalactic sources
might require a steeper injection spectrum than $\gamma\sim 1$, and 
that the acceleration 
mechanism responsible for cosmic rays beyond $\sim 10^{18}\,$eV is
still open. A discussion of more recent developments in the
understanding of cosmic ray acceleration can be found in
Ref.~\cite{blasi}.

%% file: gamma.tex
\chapter{Gamma-ray astronomy}

\section{Experiments and detection methods}

The Earth's atmosphere is opaque to photons with energy above 10\,eV, meaning
that to observe gamma rays directly requires placement of a detector
above the earth atmosphere. A major turning point in gamma-ray astronomy was
therefore the launch of the first satellite-borne telescope, SAS-2, in
1972.  

The heretofore most successful  gamma-ray satellite was the Compton
Gamma Ray Observatory, taking data from 1991--2000 with its 
four different experiments. Some of the main results were the
proof that a large fraction of gamma ray bursts are isotropically
distributed and thus extragalactic in origin, the detection of
discrete sources of extragalactic $\gamma$-ray emission, and an upper
limit on the diffuse extragalactic $\gamma$-ray background. 
A new satellite experiment, GLAST, will be launched 2008.

Main limitation of satellite experiments is their small collection
area,  which limits their use to energies $\lsim 100\,$GeV.
On the other hand, 100\,GeV is the limit
where the electromagnetic cascade in the earth's atmosphere from the
initial photon can be detected. While the cascade dies out high in
the atmosphere below $10\:$TeV, showers are still detectable via the 
Cherenkov emission of relativistic electrons and muons.
The main difficulty in ground-based gamma-ray astronomy
is the presence of an almost overwhelming background of charged
cosmic rays. 

\paragraph{Discrimination}

\begin{figure}
\epsfig{width=.95\textwidth,angle=0,file=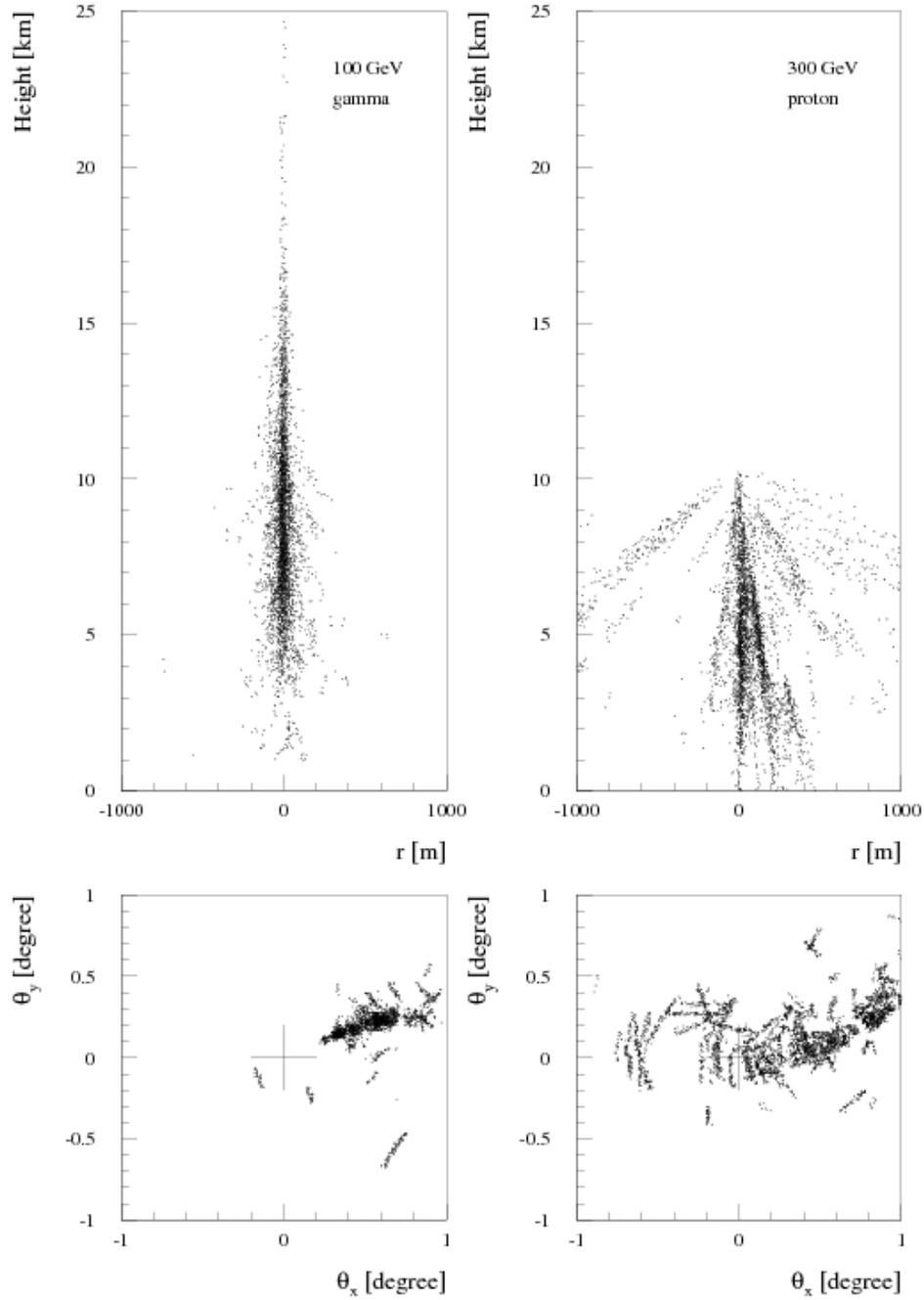}
\caption{\label{g_p_seperation}
The longitudinal development (top) and the projection (bottom) of a
shower initiated by a photon (left) and by a proton (right) \cite{hess}.}
\end{figure}

The image of an air shower in an atmospheric Cherenkov telescope can
be modeled as a two-dimensional ellipse. The two main
parameters that distinguish an air shower initiated by a photon and a proton
are the length and the width of this ellipse: While a photon
shower is narrow with a long elongation, a proton shower of comparable
energy is much wider and shorter, cf. Fig.~\ref{g_p_seperation}.
The resulting discrimination power would be not sufficient to separate
photons from the cosmic ray background, if photon sources
like e.g.\ blazars would not appear point-like. 

\section{Electromagnetic processes}

\subsection{Synchrotron radiation}

The power $P$ of emitted synchrotron radiation by a particle with mass
$m$, charge $e$ and momentum $p_\perp$ perpendicular to the magnetic
field is in the classical limit 
\be 
 P_{\rm cl} = \frac{2}{3} \: \alpha m^2 
              \left( \frac{p_\perp}{m}\frac{eB}{m^2} \right)^2 
            = \frac{2}{3} \: \alpha m^2 
              \left( \frac{p_\perp}{m}\frac{B}{B_{\rm cr}} \right)^2  \,.
\ee
The energy loss per time is $\beta=\d\!E/\d t=-P\propto m^{-4}$ and hence
is most severe for the lightest charged particle, the electron. 
The typical length-scale $l_{\rm syn}$ of synchrotron losses  
($B_{\rm cr}=m_e^2/e=4.41\times 10^{13}\,$G is the critical field
strength for an electron)
\be 
 l_{\rm syn} = \left( \frac{1}{E}\,\frac{\d\!E}{\d t}\right)^{-1}
 =\frac{3}{2\alpha m} \frac{1}{\gamma} \left( \frac{B_{\rm cr}}{B}\right)^2 \,.
\ee
For electrons with $E=10^{15}\,$eV in the Galactic disc, $B\sim
3\mu$G, we have $ l_{\rm syn}\sim 100\,$pc. Thus one of the reasons
why we have not discussed electrons as a major component of cosmic
rays is that their energy losses are 
too severe. On the other side, synchrotron radiation of relativistic
electrons is in the radio range and explains the strong radio emission
of many AGNs: Synchrotron radiation peaks at $0.29\omega_c$ with
\be
 \omega_c = \frac{3\gamma^2 B}{2B_{\rm cr}} \: m  \,.
\ee
For $B\sim 0.1\:$G close to a source and $\gamma=10^9$, the peak of
synchrotron radiation is at $10^{-10}m_e\sim 100\:$GHz. Synchrotron
radiation is the explanation of the first peak visible in
Fig.~\ref{BLLac}, showing the observed spectrum from a Blazar.

\subsection{(Inverse) Compton scattering}

The cross section for Compton scattering $e^-+\gamma\to e^-+\gamma$ is
with $\sigma_{\rm Th}=8\pi\alpha^2/(3m_e^2)$ 
\ba
 \sigma &=& \sigma_{\rm Th} \left( 1- s/m^2 +\ldots \right) 
 \quad{\rm for}\quad
 s/m^2\ll 1
\\
 \sigma &=& \frac{3m^2}{4s} \sigma_{\rm Th} \left( \ln s/m^2
 +\frac{1}{2} +\ldots \right) 
 \quad{\rm for}\quad
 s/m^2\gg 1
\ea
i.e.\ Compton scattering is suppressed for $s\gg m^2$. 

While in the classical Compton experiment an energetic photon hits an
electron at rest and transfers part of its energy, in astrophysics
``Inverse Compton scattering'' is more important: A fast electron
hitting a low-energy photon transfers a large fraction $y$ of its energy
to the photon,
\be
 y \ap 1/\ln(s/m_e^2) \,.
\ee
This is the explanation of the second peak visible in
Fig.~\ref{BLLac}, showing the observed spectrum from a blazar.

\section{Hadronic processes}

\paragraph{Neutral pion decay $\pi^0\to 2\gamma$}

In the rest system of the $\pi^0$, the two photons are emitted
back-to-back, ${\bf p}_1=-{\bf p}_2$ and $E_1=E_2=m_\pi/2$.
How are the photon energies distributed, if the pion was moving with
velocity $v$?

Let us use the Lorentz transformation between energies in the rest and the
lab system. In the lab system,  
\be \label{pion1}
E=\gamma(E'+\beta p'\cos\theta)
\ee
where $\theta$ is the angle between the velocity $\beta$ of the pion
and the emitted photon. The maximal/minimal values of $E$ follow
directly  as $E=\gamma(E'\pm\beta p')$ for $\cos\theta=\pm 1$,
i.e. if the photons are emitted parallel and anti-parallel to the
direction of flight of the pion. 

Inserting $E'=p'=m_\pi/2$ gives 
\be \label{pion2}
 E^{\max}_{\min} = \frac{1}{2}\: (\gamma m_\pi \pm \beta\gamma m_\pi) =
 \frac{1}{2}\:E_\pi (1\pm\beta)
\ee
or $ E_{\max}\to E_\pi$ and $E_{\min}\to 0$ in the ultra-relativistic
limit $\beta\to 1$.

What is the distribution of the emitted photons? Since the pion is a
scalar particle, the photon distribution is isotropic in the lab system,
\be
 \d N=\frac{1}{4\pi} \d\Omega=\frac{1}{2} \d|\cos\theta| \,.
\ee
Using (\ref{pion1}), one can express $\d E$ as $\gamma\beta p'\d|\cos\theta|$
and thus
\be \label{pionspec}
 \d N=\frac{1}{4\pi} \d\Omega = \frac{\d E}{2\gamma p'} \,.
\ee
Hence $\d N/\d E=\const$, and the photon spectrum produced by a pion
beam with uniform velocity is a box between $E_{\min}$ and $E_{\max}$
from Eq.~(\ref{pion2}). If we use $\log(E)$ as $x$ coordinate, the
boxes are symmetric with respect to $m_\pi/2$.

\paragraph{Photon spectrum from `` many pion decays''}
Consider now the photon spectrum not from a single pion, but from a
beam of pions with arbitrary energy spectrum. Each photon
spectrum from a single decay is as we saw a box centered at
$m_\pi/2$. 
Hence also the total photon spectrum from pions with an arbitrary energy 
distribution is symmetric with respect to $m_\pi/2$.

\section{Source models}

\begin{figure}
\begin{center}
\epsfig{width=.75\textwidth,angle=0,file=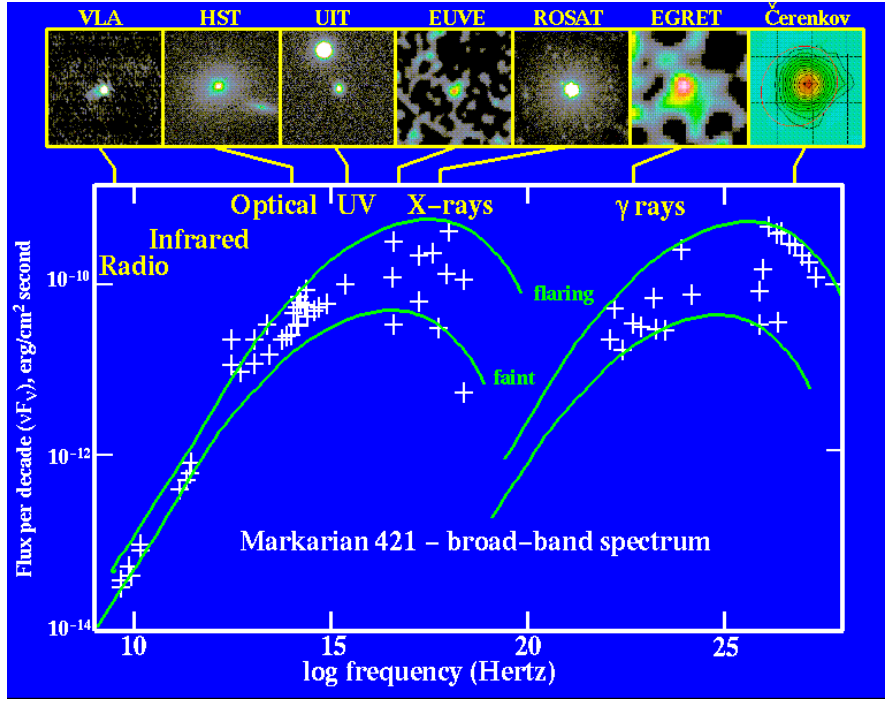}
\end{center}
\caption{\label{BLLac}
Spectrum from the blazar Markarian 421 in a flaring and a quiet 
state~\cite{keel}.} 
\end{figure}

The energy spectrum observed from blazars, i.e.\ AGN which jets are
pointing towards us, covers all the
electromagnetic spectrum from radio frequencies up to TeV
energies. It is characterized by two bumps, cf.\ Fig.~\ref{BLLac},
one peaking between the IR and the X-ray band, another one 
at gamma-ray energies. The first peak can be explained as synchrotron
radiation from relativistic electron, while the second one is
presumably due to inverse Compton scattering -- between the
same electrons and either soft photons or synchrotron photons.

Although sources like blazars should accelerate not only electrons but
also protons, synchrotron radiation and inverse Compton scattering of
electrons are sufficient to explain in most cases the observed
spectra. There are only few exceptions where an additional photon component
from pion decay at the highest energies might be needed.

The proof that a source accelerates indeed protons might be easier for
Galactic sources, where the source region can be resolved by modern
atmospheric Cherenkov telescopes. An example is
the recent observation of the SNR RX~J1713.7-3946 by the HESS array 
that has been interpreted as the
first direct  evidence that SNRs accelerate cosmic rays up to 
TeVs~\cite{hessSNR}. The image of this source shows an increase of TeV
photons in the directions
of known molecular clouds. This is in line with the expectation that, if
photons are produced as secondaries in pp collisions, their flux is
correlated with the matter density in the SNR.  

\section{Electromagnetic cascades}
\label{elamgcascades}

A high energy particle interacts at the top of the atmosphere and
initiates a cascade. In the 1930's, it was recognized that the
observed cosmic rays on ground are only the secondaries produced in
this cascade.
\label{Simplestmodel}
Heitler introduced the following simple toy model of an
electromagnetic cascade: 
\begin{itemize}
\item
In each interaction characterized by the constant interaction length
$\lambda$, a parent particle splits into two daughter particles. Each
particle takes half of the energy.
\sitem
After $n=X/\lambda$ generations, the cascade consists of
$N(X)=2^{X/\lambda}$ particles.
\sitem
The energy per particle is $E(X)=E_0/N$, where $E_0$ is the energy of
the primary.
\sitem 
Particle production stops, when $E(X)<E_{\rm cr}\sim 0.1\,$GeV. Below
this energy, photons loose energy mainly by ionization
losses. Unstable particles decay and the number of particles in the
shower decreases. 
\end{itemize}

The number of particles at the shower maximum $ X_{\max}$ is
\be
 N(X_{\max})=\frac{E_0}{E_{\rm cr}} \qquad {\rm with}\quad
 X_{\max}=\frac{\lambda}{\ln 2} \ln\frac{E_0}{E_{\rm cr}} \,.
\ee
Thus the energy $E_0$ of the primary particle is connected via
\be
 N(X_{\max}) \propto E_0 \qquad {\rm and}\quad
 X_{\max}\propto  \ln E_0 
\ee
to measurable quantities. Note that $E_{\rm cr}$---the quantity that mainly
controls $N(X_{\max})$---is determined by well-known low energy
processes.

\paragraph{Cascade equations and scaling}
The cascade equation for electron ($=e^+,e^-$) and photons
are
\be
  \frac{\partial N_i(E,r)}{\partial r} = 
  -n\sigma_{i,\rm tot}(E) N_i(E,r) +  
  n\int_E^\infty\d\!E' \;\frac{\d\sigma_{ki}(E',E)}{\d\!E} \: N_k(E',r)
\ee
where $i=e,\gamma$. Two possible boundary conditions are
\begin{subequations}
\ba
 N_i(E,0) & = & \delta(E-E_0) \,, \\
 N_i(E,0) & = & KE^{-\alpha}  \,,\label{boundary}
\ea
\end{subequations}
i.e.\ the study of a single shower or the general properties of
electromagnetic cascades. We consider next only one species (i.e.\ 
we are not interested which particles are produced in the interactions) and
the simpler second case for the boundary condition. Under which
conditions may we hope for an analytical solution? The energy and
depth dependence should factorize, $N(E,r)=g(E)f(r)$, i.e.\ the
shower should have at each depth the same energy dependence. This
happens, if $(i)$ the cross sections are not energy dependent, and 
$(ii)$  the differential cross sections scale,  
\be 
 f(z) = 
 \frac{1}{\sigma} \,\frac{\d\sigma}{\d z} = 
 \frac{E'}{\sigma}\,\frac{\d\sigma(E',E)}{\d\!E}
\ee
with $z=E/E'$. Then, we can rewrite
$N(E',r)=KE^{\prime-\alpha}=Kz^{\alpha}E^{-\alpha}$ for boundary
condition (b) and it follows 
\be \label{csc}
 N(E,r) = 
 N(E,0) \exp[ -n\sigma r (1- Z(\alpha) ] 
\ee
with the spectrally averaged energy loss fraction
\be \label{Z_spec}
 Z(\alpha) = \int_0^1 \d z \: z^{\alpha-1} f(z) \,.
\ee
Equation (\ref{csc}) is straightforward to interpret: For $z\to 0$,
one obtains the limiting case of full absorption, while interactions
without energy transfer do not change $N(E,r)$ at all. In the
intermediate case, Eq.~(\ref{Z_spec}) weights the energy loss $z$
properly not only with $f(z)$ but also with the spectral shape
$N(E)\propto E^{-\alpha}$.  

For a steeply falling spectra, $\alpha>1$, the region $z\to 1$
gives the dominating contribution to the integral defining $Z(\alpha)$. 
This illustrates that for the modeling of air shower
the knowledge of cross sections in the ``forward'' region $z\to 1$ is
crucial.

\subsection{Cascades in the atmosphere}

\paragraph{Probabilities}

The differential probability $P(z)$ for an electron with energy $E$
scattering on an atomic nucleus to emit a photon with energy $zE$
traversing the depth $\d t=\d X/X_0$ is in the high energy limit 
\be \label{peg}
 P_{e\gamma}(z)=z+\frac{1-z}{z}\left(\frac{4}{3}+2b\right) \,,
\ee
where $b=b(Z)\ap 0.012$ depends on the average charge $Ze$ of air. 
This process is infrared-divergent, i.e.\ 
the  probability to emit a soft photon diverges, $P_{e\gamma}(z)\to\infty$ for
$z\to 0$. Only if we consider the energy loss rate,
\be
 \frac{\d\!E}{\d X} = -\frac{1}{X_0} \int_0^1\d z \:zE \,P_{e\gamma}(z) =
-\frac{E(1+b)}{X_0} \ap -\frac{E}{X_0} \,,
\ee
we obtain a finite result. By contrast, the probability for the pair
production of a massive $e^+e^-$ pair in the Coulomb field of a
nucleus is infrared-finite,
\be \label{pge}
 P_{\gamma e}(z)= \frac{2}{3}-\frac{1}{2}b+\left(\frac{4}{3}+2b\right)
\left(z-\frac{1}{2}\right)^2 \,,
\ee
and can be integrated to
\be \label{pair}
 \frac{1}{\Lambda_{\rm pair}} = \int_0^1 \d z \: P_{\gamma e}(z) = 
 \frac{7}{9}-\frac{b}{3}\ap\frac{7}{9} \,.
\ee

Both electrons and photons lose their energy exponentially,
$E(X)=E_0\exp(-X/X_0)$, with the radiation length $X_0\ap
37.2$g/cm$^2$ (in air) and $7X_0/9$ as scales, respectively.
Below the critical energy $E_{\rm cr}\ap 84.2$~MeV, ionization losses
become more important than pair production and the growth of the
cascade stops.

\paragraph{Shower profile in the atmosphere}
The solution of the coupled cascade equations for electrons and
photons in the high-energy limit, i.e.\ using Eq.~(\ref{peg}) and
~(\ref{pge}) as splitting probabilities, and for a power-law boundary
condition (\ref{boundary}) at the top of the atmosphere is lengthy but
straightforward. Here we mention only the two essential steps, for
details see the classic paper of Rossi and Greisen~\cite{rg}: 
First, one obtains an infrared-finite expression
writing the divergent total bremsstrahlung probability as an integral
and combining it with the gain term. Second, in the scaling limit the
functions $Z$ are for a fixed $\alpha$ simple numbers and one has to
solve therefore only two coupled first-order differential
equations. Hence, the solutions are of the type 
$A\exp(\lambda_1(X))+B\exp(\lambda_2(X))$.

\begin{figure}
\begin{center}
\epsfig{width=.75\textwidth,angle=0,file= 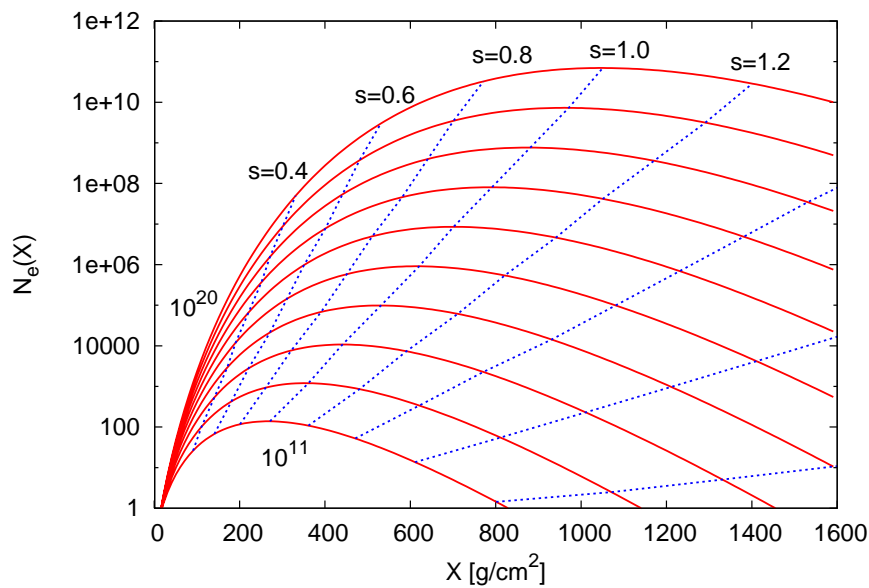}
\end{center}
\vskip-0.3cm
\caption{\label{elmag_shower}
Number of electrons $N_e(X)$ in an electromagnetic shower initiated by a
photon as function of the depth $X$ for primary energies
$10^{11},10^{12},\ldots, 10^{20}\,$eV; the shower age $s$ is also
indicated.} 
\end{figure}

The number of electrons $N_e(X)$ in an electromagnetic shower initiated by a
photon as function of the shower age 
\be
 s=\frac{3t}{t+2\beta} 
\ee
is approximately
\be  \label{profile}
 N_e = \frac{0.31}{\sqrt{\beta}} \: 
       \exp\left[ t\left(1-\frac{3}{2}\ln s\right)\right] \,,
\ee
where  $t=X/X_0$ and $\beta=\ln(E/E_{\rm cr})$. This approximation takes into 
account also low-energy processes; $N_e(X)$ is shown for several shower 
energies  in Fig.~\ref{elmag_shower}. A shower initiated by a primary with 
energy $E=10^{20}\,$eV requires about one atmospheric depth to develop fully 
and contains around $10^{11}$ particles at its maximum.

\subsection{Cascades on diffuse photon backgrounds}

\begin{figure}
\begin{center}
\epsfig{file=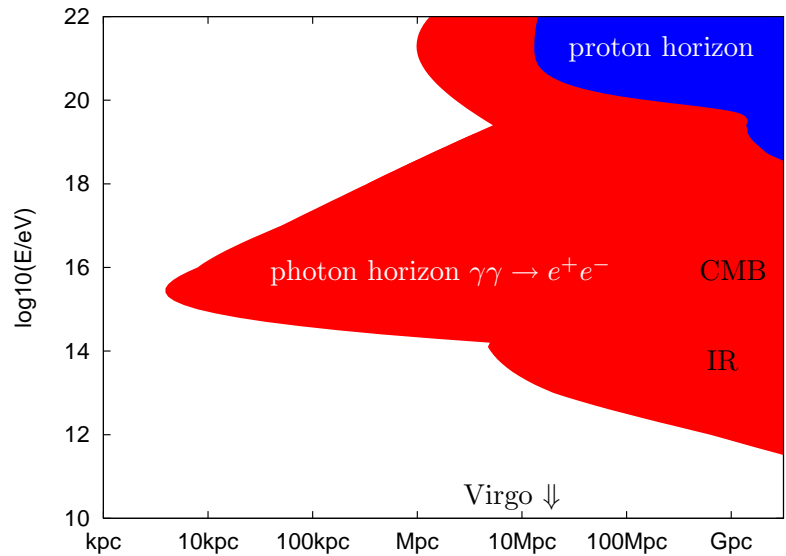,width=11.cm}
\end{center}
\caption{\label{gamma_horizon}
The gamma and the cosmic ray horizon, i.e.\ the free mean path of
photons and cosmic rays as function of energy: The (red/blue) shaded
area is inaccessible for photons/cosmic rays. The distance to the
nearest galaxy cluster is also indicated.} 
\end{figure}

A high energy photon scattering on a low energy background photon
(e.g. from the cosmic microwave or infra-red background) can start an
electromagnetic cascade, via the two main processes
\be
 \gamma+\gamma_b\to e^++e^-
\ee
\be
 e+\gamma_b\to e^+\gamma \,.
\ee
Pair production stops for $s=4E_\gamma\eps_\gamma<4m_e^2$, where
$\eps_\gamma$ is the typical energy of the background photons. 
Electrons continue to scatter on cosmic microwave 
photons in the Thompson regime, producing photons with
average energy
\be  \label{thompson}
 E_\gamma = \frac{4}{3}\frac{\eps_\gamma E_e^2}{m_e^2} 
 \ap 100\:{\rm MeV} \: \left( \frac{E_e}{1{\rm TeV}} \right)^2 \,.
\ee 
Hence the Universe acts as a calorimeter for electromagnetic
radiation, accumulating it in the MeV-GeV range.  
Moreover, it means that the Universe is opaque for high energy photons.
Figure~\ref{gamma_horizon} show the absorption length of photons on
various sources of background photons. Starting from energies in the TeV
range, high-energy photon astronomy can study only the local (and
therefore also recent) Universe. This is one of the main motivations
to explore also high energy neutrinos, despite of the experimental
challenges in neutrino detection.
%

\section{Exercises}
\begin{enumerate}
\item
Show that $P_{\rm cl}$ for synchrotron radiation has the correct
Lorentz transformation properties by expressing it as function of
$F_{\mu\nu}$ and $p_\mu$.
\item
Rewrite the energy losses due to synchrotron radiation and Compton
scattering as function of the energy density of the magnetic field and
real photons and show that the two expressions agree in the classical
limit.
\item
Find the number of particles in the shower maximum, the corresponding
shower age $s$ and depth $X_{\max}$ from Eq.~(\ref{profile}). 
\end{enumerate}

%% file: exgal.tex
\chapter{Extragalactic cosmic rays}

\section{Propagation of extragalactic particles}

\subsection{Energy losses of protons and nuclei}

There are three main energy loss processes for protons propagating
over cosmological distances: Adiabatic energy losses due to the
expansion of the universe, $-(\d\!E/\d t)/E=H_0$, $e^+e^-$ pair-, and
pion-production on  photons of the cosmic microwave background (CMB).

The relative energy loss per time of a particle (due to interactions 
with the CMB) can be estimated as
\be
 \frac{1}{E}\:\frac{\d\!E}{\d t} = \langle y\sigma n_\gamma\rangle \,
\ee
where $y=(E-E')/E$ is the energy fraction lost per interaction, 
$n_\gamma\ap 410/$cm$^3$ is the density of CMB photons with
temperature $T\ap 2.7\,$K and the brackets
$\langle\ldots\rangle$ remind us that we should perform an average of
the differential cross section with the momentum distribution $n(\vp)$
of photons. In our estimates, we avoid this complication considering
only reactions well above the threshold energy when essentially all
photons participate in the reaction.

\paragraph{Electron-positron pair-production} Since the produced
$e^+e^-$ pair in the process $p+\gamma\to p+e^++e^-$ is light,
this energy loss process has a low threshold energy but leads in turn
only to a small energy loss\footnote{Close to the threshold, the energy
  fraction lost can be obtained considering the ``decay'' of the intermediate
state at rest.} per interaction, $y= 2m_e/m_p\ap 10^{-3}$.
The threshold energy on CMB photons follows from
$(k_p+k_\gamma)^2\geq(m_p+2m_e)^2$ as 
$E_p\geq m_em_p/E_\gamma\sim 2\times10^{18}$~eV.
The cross section of the reaction is $\sigma=\alpha/32 \sigma_{\rm Th}
f(s)$, i.e.\ is a factor $\sim 2\times 10^{-4}$ smaller than Compton
scattering.

\paragraph{Pion-production and the GZK effect}
Greisen and Zatsepin and Kuzmin noticed that pion-production should
introduce a ``cutoff'' in the cosmic ray spectrum. Since $y\ap m_\pi/m_p\ap
0.2$ close to threshold, but the cross-section is of similar size as
the one of pair-production, the energy losses of protons suddenly
increase around $5\times 10^{19}$~eV. 

\begin{figure}
\vspace*{-0.4cm}
\epsfig{width=.45\textwidth,angle=0,file=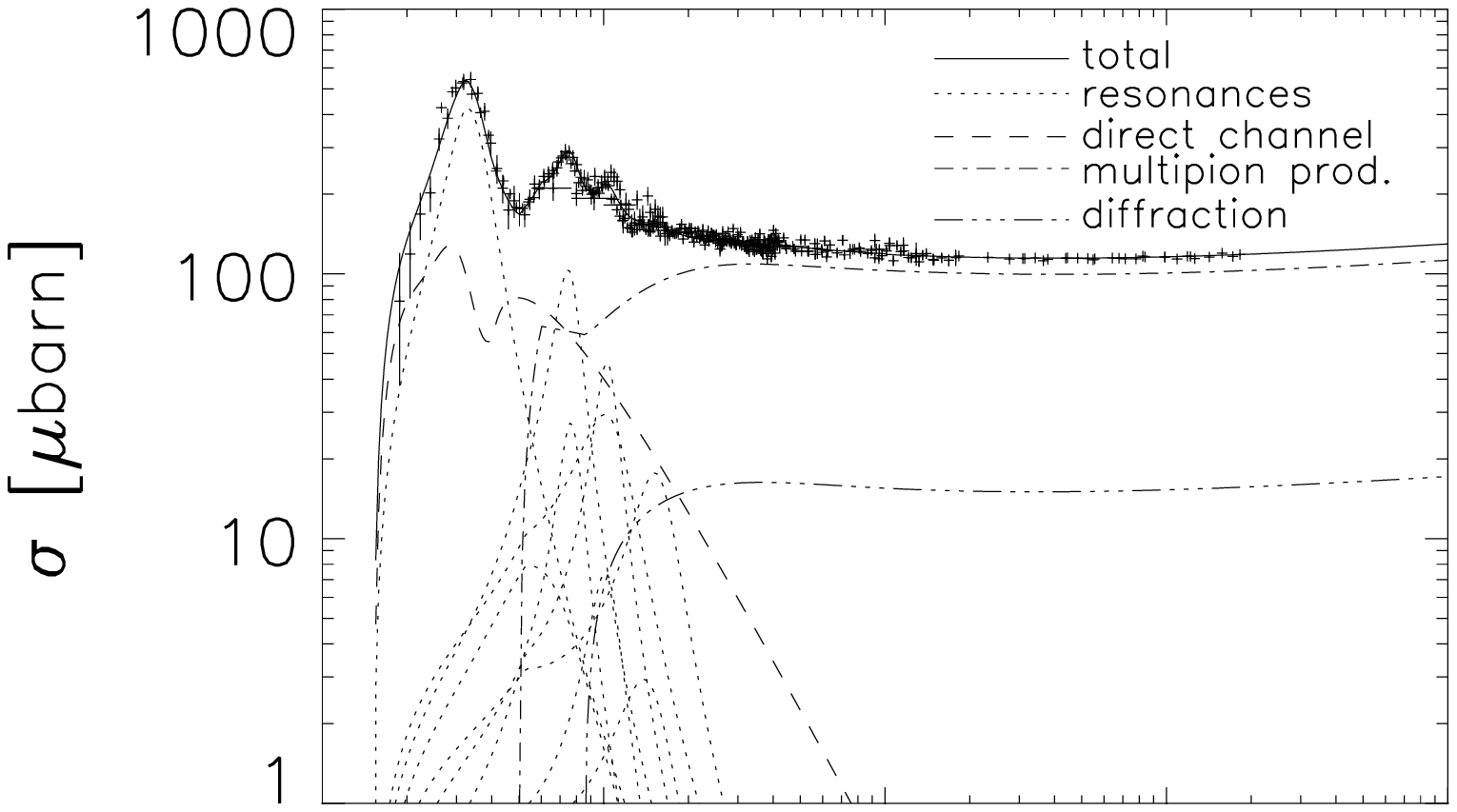}
\vspace*{0.4cm}
\epsfig{width=.45\textwidth,file=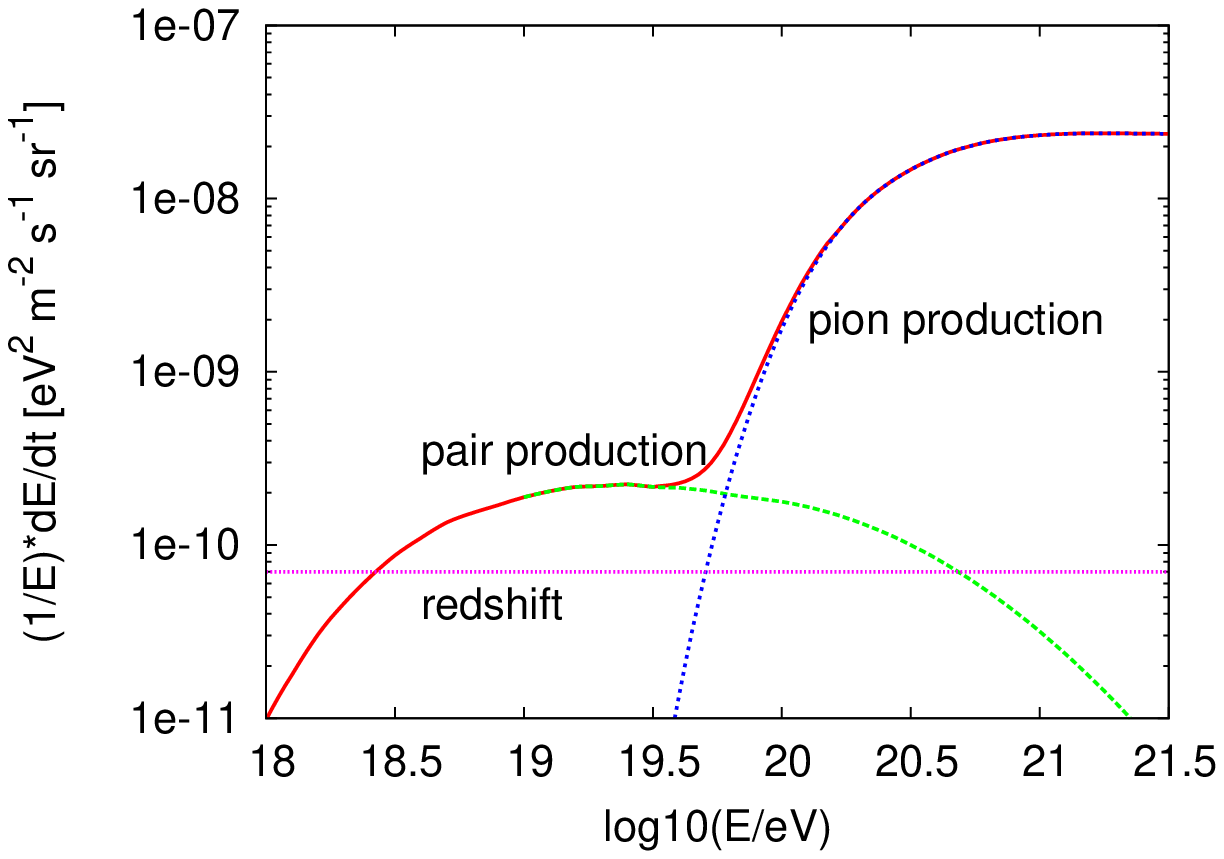} 
\caption{\label{Delta_res}
Left: Experimental data for the photo-nucleon cross-section together with
a theoretical prediction for the contributions of various
subprocesses, from~Ref.~\cite{sophia}.
Right: Comparison of the different contributions to the energy losses
of a proton.} 
\end{figure}

For an estimate of the cross section for pion production we use the
Breit-Wigner formula with the lowest lying nucleon resonance $\Delta^+$ as 
intermediate state, 
\begin{equation}  \label{breitwigner}
 \sigma_{\rm BW}(E) = \frac{(2J+1)}{(2s_1+1)(2s_2+1)} \, 
 \frac{\pi}{p_{\rm cms}^2} \, 
 \frac{b_{\rm in}b_{\rm out}\Gamma^2}{(E-M_R)^2 + \Gamma^2/4}\ ,
\end{equation}
where $p_{\rm cms}^2=(s-m_N^2)^2/(4s)$, $M_R=m_\Delta=1.230\,$GeV,
$J=3/2$, $\Gamma_{\rm tot}=0.118\,$GeV, $b_{\rm in}=0.55\%$. At resonance, we
obtain $\sigma_{\rm BW}\sim 0.4\:$mbarn in good agreement with
experimental data, cf. Fig.~\ref{Delta_res}. We estimate the 
energy loss length well above threshold $E_{\rm th}\sim 4\times
10^{19}\:$eV with $\sigma\sim 0.1$mbarn and $y=0.5$ as 
\be
 l_{\rm GZK}^{-1}=\frac{1}{E}\:\frac{\d\!E}{\d t} \ap
 0.5\times 400/{\rm cm}^3 \times 10^{-28}\:{\rm cm}^2 \ap  
 2\times 10^{-26}\:{\rm cm}^{-1} 
\ee
or $ l_{\rm GZK}\sim 17\:$Mpc. Thus the energy loss length of a proton
with $E\gsim 10^{20}\:$eV is comparable to the distance of the closest
galaxy clusters and we should see only local sources at these
energies. Note that it is this smallness of the  ``horizon scale'' that 
makes the identification of UHECR sources feasible, despite the rather poor 
angular resolution of UHECR experiments, $\delta\theta\sim 1^\circ$, 
and deflections of charged particles in magnetic fields.

The right panel of Fig.~\ref{Delta_res} compares the relative energy
losses of a proton due to redshift,  $e^+e^-$ pair- and
pion-production on CMB photons. One clearly recognizes the increase of
the energy losses by two orders of magnitude around $5\times
10^{19}\:$eV and expects therefore a corresponding suppression (not a
cutoff) of the
cosmic ray flux above this energy. (Remember that each shell of thickness
$\Delta r$ contributes the same fraction to the total intensity, if
energy losses and general relativistic effects can be neglected.)   
Thus the name GZK cutoff is somewhat misleading, and the exact
strength of the suppression depends on various details as the number
density of the sources.

The AGASA experiment detecteded in the 90'ies an excess of events 
above $10^{20}$~eV compared to predictions. There had been extensive 
discussions, if this result requires some kind of ``new physics'', but 
meanwhile new data of the Hires and Pierre Auger Observatory (PAO) 
confirmed the UHECR flux expected in the presence of the 
GZK effect.

\paragraph{Attenuation of nuclei}
The dominant loss process for nuclei of energy $E\gsim 10^{19}\,$eV
is photodisintegration $A+\gamma\to (A-1)+N$ in the CMB and the infrared
background due to the giant dipole resonance. The threshold for this
reaction follows from the binding energy per nucleon, $\sim 10\:$MeV.
Photodisintegration leads to a suppression of the flux of nuclei
above an energy that varies between $3\times 10^{19}\:$eV for He and
$8\times 10^{19}\:$eV for Fe.

\subsection{Galactic and extragalactic magnetic fields}

Magnetic fields affect the propagation of charged particles deflecting
them, leading in turn to synchrotron radiation. While even for ultrahigh energy
protons synchrotron losses are negligible except in the relatively 
strong magnetic fields close to sources, the magnitude of deflections  
limits the possibility to perform charged particle astronomy.
The deflection angle in a regular magnetic field after the distance $d$ 
is
\be
  \theta\simeq\frac{d}{R_L}\simeq 0.52^\circ Z
  \left(\frac{p_\perp}{10^{20}\,{\rm eV}}\right)^{-1}
  \left(\frac{d}{1\,{\rm kpc}}\right)
  \left(\frac{B}{\mu{\rm G}}\right) 
\ee
for a particle with momentum $p_\perp$ perpendicular to $B$.

\begin{figure}
\hskip1.5cm
\epsfig{file=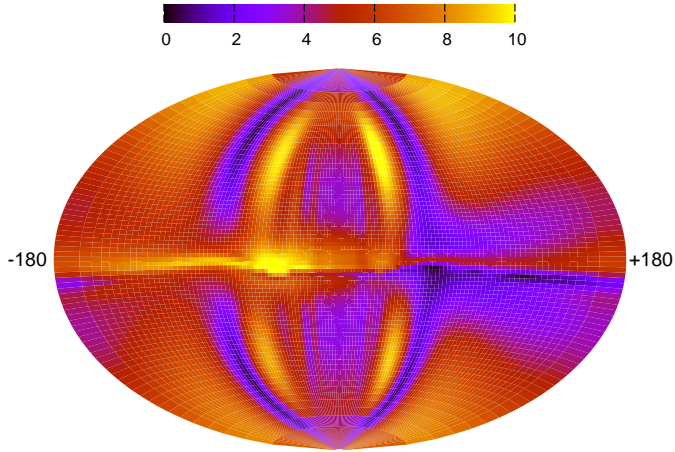,width=0.75\columnwidth} 
\vskip-1.4cm
\caption{Deflection map  of the GMF for a rigidity
  of 4$\times$ 10$^{19}$~V from Ref.~\cite{Kachelriess:2005qm}; 
  The deflection scale at the top is in degree. 
\label{fig:defl}}
\end{figure}

Figure~\ref{fig:defl} shows a deflection map for a specific model for
the Galactic magnetic field  and a rigidity of $4\times 10^{19}\:$V;
for details of the model see Ref.~\cite{Kachelriess:2005qm}.   
The map  uses a
Hammer-Aitoff projection of galactic coordinates with the Galactic center
in the middle. The expected deflections depend strongly on the direction, but
typically trajectories passing the Galactic center or plane suffer
larger deflections. 

Magnetic fields beyond the Galactic disk are poorly known and include  
possibly an extended regular and a turbulent field in the halo of our 
Galaxy and a large scale extragalactic magnetic field (EGMF). 
In the latter two cases, the magnetic field is simplest characterized 
by an r.m.s. strength $B$ and a
correlation length $l_c$, i.e. it is assumed that the field is smooth 
on scales below $l_c$  in real space. If we neglect energy loss processes, 
then the r.m.s. deflection angle $\delta_{\rm rms}=\langle\delta^2\rangle^{1/2}$ 
over the distance $d\gg l_c$ is
\be   \label{deflec}
  \delta_{\rm rms}\simeq\frac{(2dl_c/9)^{1/2}}{R_L}\simeq0.8^\circ\,
  Z\left(\frac{E}{10^{20}\,{\rm eV}}\right)^{-1}
  \left(\frac{d}{10\,{\rm Mpc}}\right)^{1/2}
  \left(\frac{l_c}{1\,{\rm Mpc}}\right)^{1/2}
  \left(\frac{B}{10^{-9}\,{\rm G}}\right) \,.
\ee

For a calculation of $\delta_{\rm rms}$ one needs either
observational data or reliable theoretical predictions both for the magnitude
and the structure of EGMFs. At present, no single theory for the
generation of magnetic field has become widely accepted. 
Observational evidence for EGMFs has been
found only in a few galaxy clusters observing 
their synchrotron radiation halos or performing Faraday rotation
measurements. The two methods give somewhat different results for the
field strength in clusters, with $B\sim 0.1$--1$\:\mu$G and $B\sim
1$--10$\:\mu$G, respectively. Outside of clusters only upper limits
exist for the EGMF. The combination of poor observational data and of
a missing consistent theoretical picture prevents at present a reliable
estimate of the influence of EGMFs on the propagation of UHECRs. 
It is likely that future UHECR data will provide the first
information about EGMF outside the core of galaxy clusters.

Deflections also imply an average time delay of
\begin{equation}
  \tau\simeq \delta_{\rm rms}^2 d/4\simeq1.5\times10^3\,Z^2
  \left(\frac{E}{10^{20}\,{\rm eV}}\right)^{-2}
  \left(\frac{d}{10\,{\rm Mpc}}\right)^{2}
  \left(\frac{l_c}{1\,{\rm Mpc}}\right)
  \left(\frac{B}{10^{-9}\,{\rm G}}\right)^2\,{\rm yr}
  \label{delay}
\end{equation}
relative to the rectilinear propagation of cosmic rays with the speed of 
light. As a result, even a bursting extragalactic source like a gamma ray
burst is seen over time-scales that exceed vastly possible observation
times.

\section{The dip and the Galactic--extragalactic transition}

\paragraph{Transition energy} 

\begin{figure}
\epsfig{file=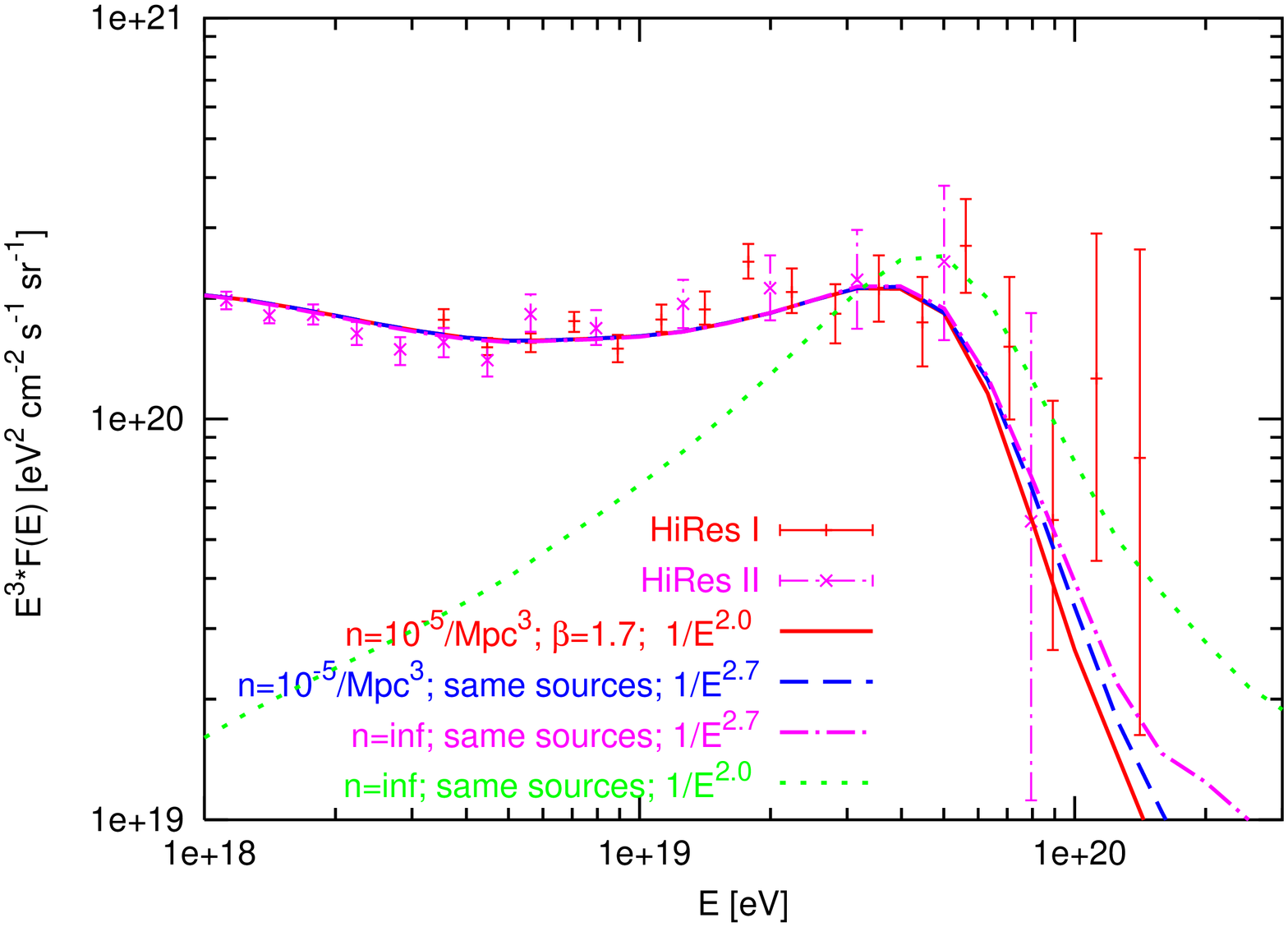,height=0.45\textwidth,width=6cm,angle=270} 
\vskip-6cm\hskip7.5cm
\epsfig{file=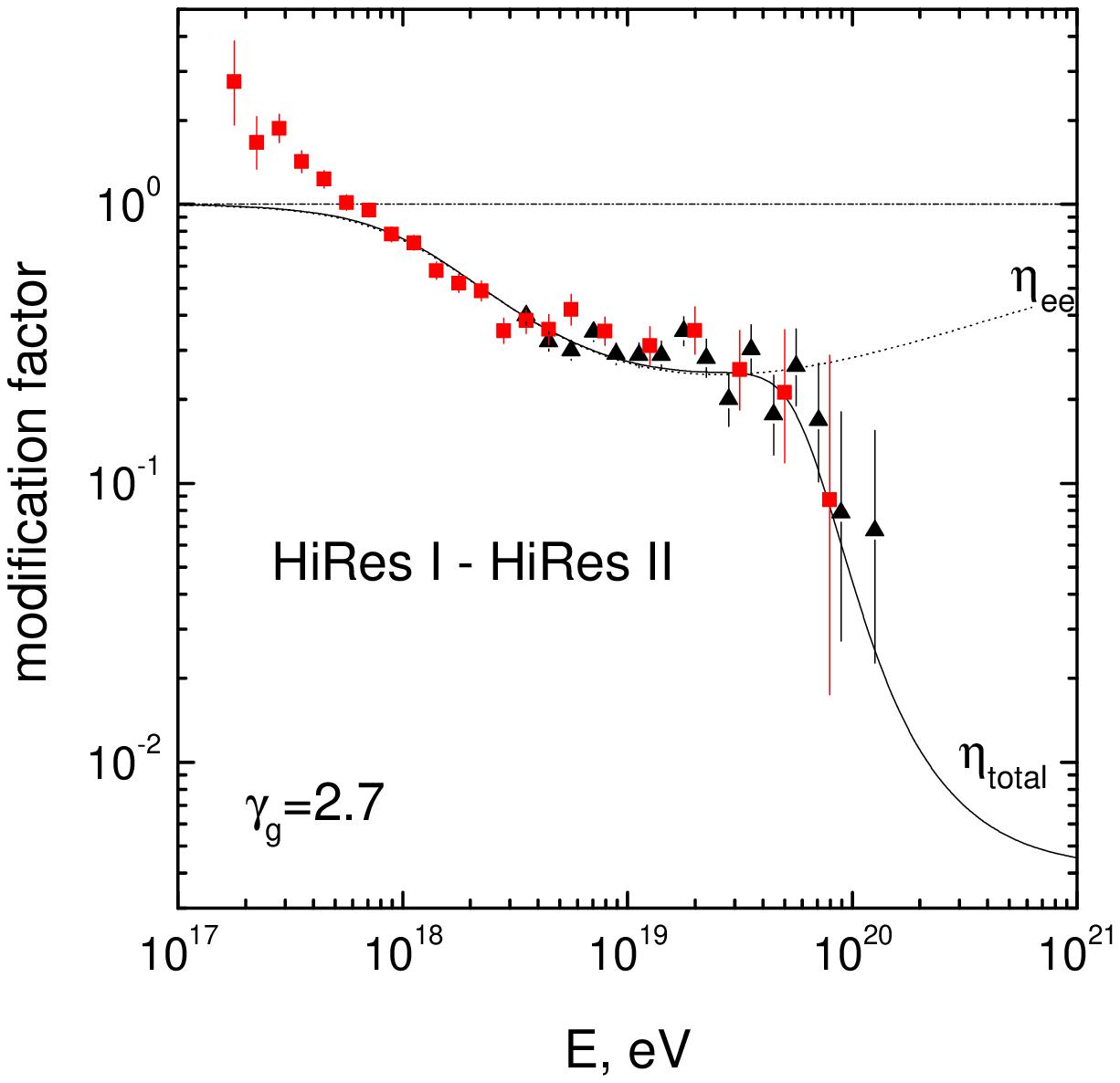,width=0.45\textwidth}
\caption{\label{modif}
Left: Proton spectrum for different generation spectra~\cite{emax}. Right: 
Modification
factor for the proton spectrum  from uniformly distributed sources~\cite{dip}.}
\end{figure}

The traditional point of view of the transition between Galactic and
extragalactic sources assumes that the generation spectrum of both Galactic 
and extragalactic cosmic rays
is $\d\!N/\d\!E_g\propto 1/E^\alpha$ with $\alpha=2.0$--2.2 as
predicted by Fermi acceleration. In this picture, extragalactic
sources dominate only above $E\sim 3\times 10^{19}\,$eV the cosmic ray flux,
cf. the green line in the left panel of Fig.~\ref{modif}. This
scenario requires therefore the existence of new, isotropically
distributed Galactic sources able to accelerate to $E\sim 10^{19}\,$eV.
In a more recent view, one uses as generation spectrum of extragalactic 
cosmic rays $\d\!N/\d\!E_g\propto 1/E^\alpha$ with
$\alpha=2.6$--2.7 and can then explain the shape of the cosmic ray spectrum
down to a few$\times 10^{17}\,$eV with extragalactic sources~\cite{dip}. 
How well the shape is described becomes clearer plotting the 
``modification factor $\eta$'', i.e.\ the ratio of the diffuse fluxes 
calculated taking into account all losses and including only redshift 
losses. In the right panel of Fig.~\ref{modif}, experimental data from the 
HiRes experiment are compared to the theoretically
predicted modulation factor 
assuming a simple power-law as generation spectrum. 
Galactic sources dominate the spectrum, when $\eta>1$: In this case, 
the observed cosmic ray intensity is larger than expected from
extragalactic sources alone. Hence Fig.~\ref{modif} motivates the view
that Galactic sources dominate the spectrum only below a (few$\times
10^{17}\,$--$10^{18})\,$eV.  

There exist at least two different explanations  for
the steep generation spectrum with $\alpha=2.6$--2.7:
Either it indicates that the acceleration to energies 
above $E\gsim 10^{17}\,$eV is not mainly based on Fermi shock
acceleration, or that the diffuse and the source spectra differ not
only by propagation effects. In the latter case, it is sufficient that
there is a distribution of maximal source energies as 
$\d n/\d\!E_{\max}\propto 1/E^{1.6}$. This assumption is plausible,
since the number of sources able to accelerate to higher and higher maximal 
energies $E_{\max}$ should decrease with $E_{\max}$.

\paragraph{Magnetic Horizon} The transition between Galactic and extragalactic
dominance in the cosmic ray spectrum may be caused by a
magnetic horizon hiding extragalactic sources below $E\lsim
10^{18}\,$eV. The maximal distance a cosmic ray can travel is in the
diffusion picture given by
\be
 r_{\rm hor}^2 = \int_0^{t_0} \d t \:D(E(t)) 
               = \int_{E_0}^{E} \frac{\d\!E'}{\beta}\: D(E'(t)) \,,
\ee
where $t_0$ is the age of the  Universe. If we consider cosmic rays
with energy below
$E_0\lsim 10^{18}$eV, the energy losses are mainly due to the expansion
of the Universe, $\beta=\d\!E/\d t=-H$. We use a ``quasi-static'' Universe,
$H(t)=H_0$ and $t_0H_0=1$. Then $E(t)=E_0\exp(-H_0t)$ and
\be
 r_{\rm hor}^2 = \frac{cl_c}{6H_0} \: \left(\frac{E}{\Ecr}\right)^2 
                (\exp(2)-1) \,,
\ee
using $D(E)=cl_c/3 \:(E/\Ecr)^2$ valid for $R_L\gsim l_c$. (Remember
that $\Ecr$ was defined by $R_L(\Ecr)=l_c$.) 

If we assume that a magnetic field with coherence length $l_c\sim\,$Mpc
and strength $B\sim 
0.1\,$nG exists in a significant fraction of the Universe, then the size
of the magnetic horizon at $E=10^{18}\:$eV is $r_{\rm hor}\sim 100\,$Mpc.
Hence, similar to the GZK suppression above $6\times 10^{19}\,$eV, we
see a smaller and smaller fraction of the Universe for lower and lower 
energies.
As a consequence, the spectrum of extragalactic cosmic rays visible to us
steepens below $10^{18}\,$eV and the extragalactic component becomes
sub-dominant. 

Energetic reasons suggest that not only the observed diffuse spectrum 
but also the intrinsic spectrum of extragalactic 
UHECR steepens not far below $10^{17}\,$eV: The total luminosity of a 
source depends on its minimal energy as 
$L\propto E_{\min}^{2-\alpha}$. The combination of large 
$\alpha$ and small $E_{\min}$ cuts across the energy budget of even as 
powerful and numerous sources as AGNs. This problem is softened if
there is a transition to an $1/E^2$ generation spectrum at low energies.

\section{Extensive air showers initiated by hadrons}

A proton or a nuclei interacting at the top of the atmosphere 
produces an air shower consisting of a core of high energy hadrons 
that transfer continuously part of their energy to new electromagnetic
sub-cascades, mainly by the process $\pi^0\to 2\gamma$. 
A smaller fraction of the
energy is transferred by decays of charged pions to muons and
neutrinos. We will not discuss these processes in detail, but refer
instead only to the results of the simple model discussed in 
Sec.~\ref{Simplestmodel}:
The primary energy $E_0$, the depth of the shower maximum $ X_{\max}$ and
the number $N(X_{\max})$ of particles at the shower maximum are
connected by 
\be
 N(X_{\max}) \propto E_0 \qquad{\rm and}\quad
 X_{\max}\propto  \ln E_0 \,.
\ee

\paragraph{Differentiating between primaries}

\begin{figure}
\begin{center}
\epsfig{file=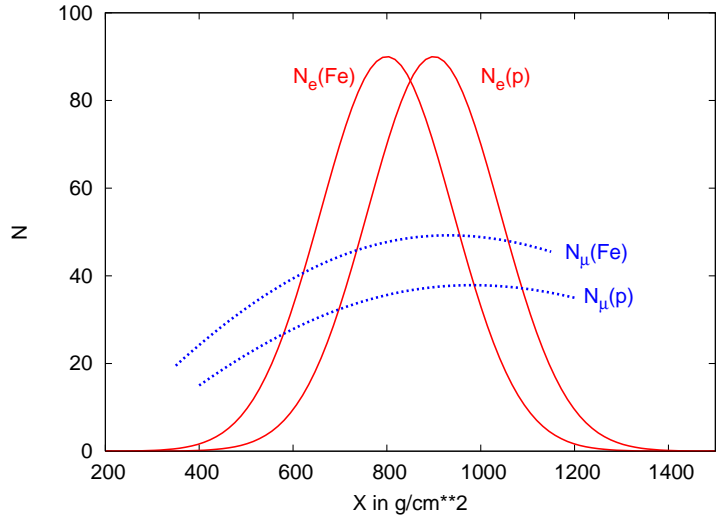, height=7cm,width=10cm,angle=0} 
\end{center}
\caption{\label{Ne_Nmu}
Electron number $N_e$ and muon number $N_\mu$ (in arbitrary units) for
two proton and iron showers as function of the depth $X$.}
\end{figure}

According to the superposition model, a nuclei of mass number $A$ and
with energy $E_0$ interacts as $A$ independent nucleons with energy $E_0/A$. 
Thus the number of particles at the shower maximum is unchanged but
its position $X_{\max}\propto  \ln E_0/A$ depends on $A$. As it is
schematically shown in Fig.~\ref{Ne_Nmu}, an iron shower develops faster
and reaches is maximum earlier than a proton shower. 
Moreover, the shower-to-shower fluctuations in the case of heavy nuclei 
as primaries are smaller compared to the ones for primary protons. 
This is very intuitive in the simple superposition picture where a shower 
initiated by a nuclei is described as a superposition of $A$ independent 
showers. However, the simple superposition picture fails to describe
quantitatively EAS fluctuations, because of the large fluctuations of 
elementary nucleus-nucleus interactions.

Figure~\ref{Ne_Nmu} shows also the muon number $N_\mu$ for a proton
and iron shower. First, because muons do not multiply through electromagnetic
processes (for example, the cross section for the process 
$\gamma+ A\to A+\mu^++\mu^-$ is strongly suppressed by the muon mass), 
their number is much smaller than the one of electrons and photons. 
As a result of the slower increase
of $N_\mu$ and the large attenuation length, the muon number has a much
weaker dependence on the depth. This reduces fluctuations and
makes, at least in principle, the muon number a better estimator for
the primary energy. However, this advantage is counterbalanced by the larger
theoretical uncertainty of $N_\mu$ and the smaller number of muons.
Figure~\ref{Ne_Nmu} illustrates also that
observing both $N_e$ and $N_\mu$ at a fixed depth $X$ provides
information about the chemical composition of cosmic rays.

\paragraph{Experimental techniques}
Present experiments use mainly two different techniques to detect
ultrahigh energy cosmic rays. The traditional one uses an array of detectors
on the ground to sample secondaries. Two commonly used detector types are 
scintillation detectors mainly sensitive to electrons and water tanks 
where the Cerenkov light emitted manly by muons is recorded by 
photomultipliers. The array of detectors is built as sparse as 
possible, to achieve a large aperture within fixed time and money 
constraints. Therefore only a tiny fraction of all particles is measured. 
Timing
information allows one to reconstruct the arrival direction and the
core of the shower. Then a theoretical formula for the number density
$\rho(r)$ of
particles as function of the distance to the shower core and of the
primary energy is fitted to the measurements. Empirically, it is found
that $\rho(r)$ at rather large distance from the core is least
sensitive to the primary type (i.e.\ the mass number $A$ of nuclei) and
has the smallest fluctuation from shower to shower. Depending on the
detector type and the spacing of the array, this optimal distance varies
and amounts e.g.\ to 1000\,m  in the case of the PAO. The conversion
factor between, say $\rho(1000)$, and the primary energy has to be
found either from Monte Carlo simulations or using different 
experimental techniques.

An example for a method that is calorimetric and therefore, at least 
in principle, allows one a model-independent determination of the primary 
energy is the fluorescence light method.
As the shower passes through the air, electrons excite nitrogen molecules 
that emit isotropically fluorescence light. Optical cameras on the ground
can trace therefore the shower evolution. If the excitation probability
of nitrogen by a beam of electrons is known, the emitted light
integrated over the shower trajectory can be used as a measure for the energy of
the primary particle. Disadvantages of this method are the restriction to 
moonless nights, the need for a careful control of the atmospheric 
conditions, and the more difficult calculation of the aperture of the 
experiment.

\section{Astronomy with charged particles}

Possible anisotropies expected for extragalactic cosmic rays can be
classified into four subclasses: 
\begin{itemize}
\item 
At such high energies that deflections in the Galactic and
extragalactic magnetic fields are sufficiently small,  
point sources may reveal
themselves as small-scale clusters of UHECR arrival directions. This
requires additionally a rather low density of UHECR sources so that
the probability to observe several events of at least a subset of
especially bright sources is large enough.\\
Even if one observes from most sources only one event, a
correlation analysis with specific sources types may reveal the UHECR
sources. This requires however to choose a very special class of
astronomical sources and/or a high threshold energy, in order to keep
the number of potential sources within the horizon sufficiently small.
\sitem 
Moving to lower energies, the energy-loss horizon of UHECRs and
thereby the number of sources visible increases. Moreover, deflections
in magnetic fields become more important. As a result, the
identification of single sources is not possible anymore. Instead,
anisotropies on medium scales should reflect the inhomogeneous
distribution of UHECR sources that is induced by the observed 
large-scale structure (LSS) of matter, cf.~Fig.~\ref{pscz}. 
Since different source types
have different clustering properties, this offers an alternative way
to identify the sources of UHECRs.
\begin{figure}
\begin{center}
\epsfig{file=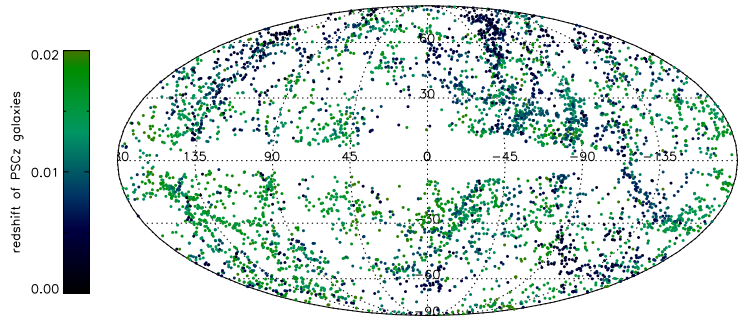,width=0.74\textwidth,angle=00}
\end{center}
\vspace*{-0.6cm}
\caption{
The distribution of normal galaxies color coded according their redshift,
in galactic coordinates. The empty band along the Galactic plane is caused
by extinction and does not reflect an intrinsic lack of 
objects~\cite{Cuoco:2007id}.
\label{pscz}}
\end{figure}
\sitem 
At even lower energies, also the LSS of sources disappears, both
because the inhomogeneities in the source 
distribution will be averaged out due to the increased  energy-loss
horizon of UHECRs  and because of deflections in the EGMF. Thus the
cosmic ray sky would 
appear isotropic, if the Earth would be at rest with respect to the
cosmological rest frame. As the observation of the CMB dipole shows,
this is not case, and a dipole anisotropy is expected if the
cosmic ray flux is dominated by sources at cosmological distance. This
is the ``cosmological variant'' of the Compton-Getting effect discussed
earlier. 

Since the Sun moves with $u=368\pm 2$~km/s towards the great attractor,
the predicted anisotropy is $\delta_{\rm CG}=(2+2.7)\,u\simeq 0.6\%$ taking 
into account the observed spectrum $I(E)\propto E^{-2.7}$ of
cosmic rays above the ankle.  
Deflections of UHECRs in the Galactic magnetic field should only displace 
the dipole axis, but not affect the magnitude of this effect.
For instance, at energies (2--3)$\times
10^{19}$~eV and for proton primaries, the dipole position should be 
aligned to the one observed in the CMB within about 10$^\circ$. 
Observing the Compton-Getting effect at only one energy provides combined
information on the intervening Galactic magnetic field and the charge
of the cosmic ray primaries. However, observations at two or more
energies break this degeneracy. For example, the determination of the
average primary charge is straightforward as long as cosmic rays propagate
in the quasi-ballistic regime and given by the relative shift of the
cosmic ray and CMB dipole axis at two different energies.
\sitem 
Finally, analogous to the latitude effect in the geomagnetic field,
the Galactic magnetic field (GMF)  can induce anisotropies in the
observed flux of extragalactic UHECRs, for rigidities low enough that
blind regions exist. Anisotropies of this kind
should be expected in models that invoke a dominating extragalactic
proton component already at $E\simeq 4\times 10^{17}$~eV  or
extragalactic iron nuclei at $E\lsim 10^{19}$~eV~\cite{Kachelriess:2005qm}.
\end{itemize}

It is not guaranteed that all these four anisotropies can be observed. 
If EGMFs are large, UHECR primaries are nuclei and/or the source density
is large, the integrated flux above the energy where point sources
become visible may be for the present generation of UHECR experiments 
too small. Similarly, a transition from galactic to extragalactic sources 
at a relatively high energy reduces the chances to observe the cosmological 
Compton-Getting effect.

\subsection{Clustering on medium and small scales}
\label{msc}

The cosmological Compton-Getting effect requires that inhomogeneities in
the source distribution of cosmic rays are averaged out. As the free mean
path of cosmic rays decreases for increasing energy, anisotropies connected to
the large-scale structure (LSS) of cosmic ray sources become more prominent
and replace the cosmological Compton-Getting effect. 
The exact value of this transition energy $E_\ast$ depends
both on the amount of clustering in the source distribution and the
free mean path  $\lambda_{\rm CR}$ of cosmic rays, i.e. also the primary type. 
For the specific case of proton primaries and a source distribution
proportional to the density of baryons, 
Ref.~\cite{napoli} found $E_\ast\ap 5\times 10^{19}$~eV and
a minimal number of order 100 events for a detection.

\begin{figure}
\begin{center}
\epsfig{file=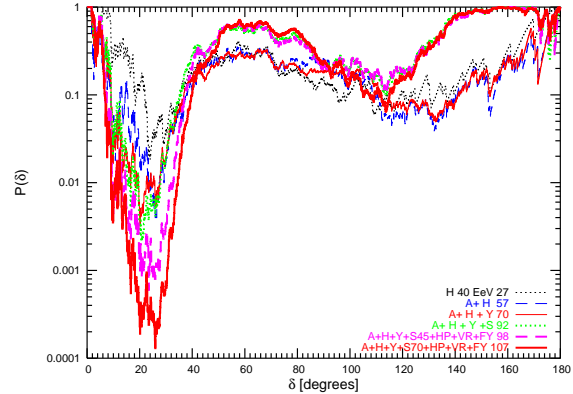,width=0.34\textwidth,angle=270}
\end{center}
\vspace*{-0.6cm}
\caption{
Chance probability $P(\delta)$ to observe a larger value of the autocorrelation
function as function of the angular scale  $\delta$ for different
combinations of all publicly available experimental data; 
from Ref.~\cite{Kachelriess:2005uf}.
\label{p_ch_ex}}
\end{figure}

The first strong evidence for medium-scale clustering was found
combining all available data sets of UHECR events with
energy $E\geq 5\times 10^{19}\,$eV and published arrival directions 
until 2005 in~\cite{Kachelriess:2005uf}. The most convenient tool to 
search for deviations from anisotropy is in this case the cumulative 
auto-correlation function,
\be
  W(\theta)=\frac{N(<\theta)}{N'(<\theta)}-1 \,,
\ee
where $N$ and $N'$ denote the number of pairs with separation angle
$<\theta$ in the observed data and in a randomly generated data
set, respectively. Figure~\ref{p_ch_ex} shows 
the chance probability $P_{\rm ch}$ to
observe a larger number of pairs the random data than in the observed
data. 
Clearly, the observed arrival directions show a surplus of clustering
that is most significant on a typical scale of $\sim 25^\circ$.
Similar findings have emerged recently from an analysis of
the preliminary data from the Pierre Auger Observatory (PAO)~\cite{ICRC}. For
64 events with $E>4\times 10^{19}\:$eV the data presented in
Ref.~\cite{ICRC} show a surplus of clustering in the broad range
from 7 to 30 degrees. Thus the LSS of matter, probably modified by
extragalactic magnetic field, is reflected in the arrival directions
of UHECRs. The obvious question is how one can extract the information
contained in a plot like Figure~\ref{p_ch_ex}.

\begin{figure}
\begin{center}
\epsfig{file=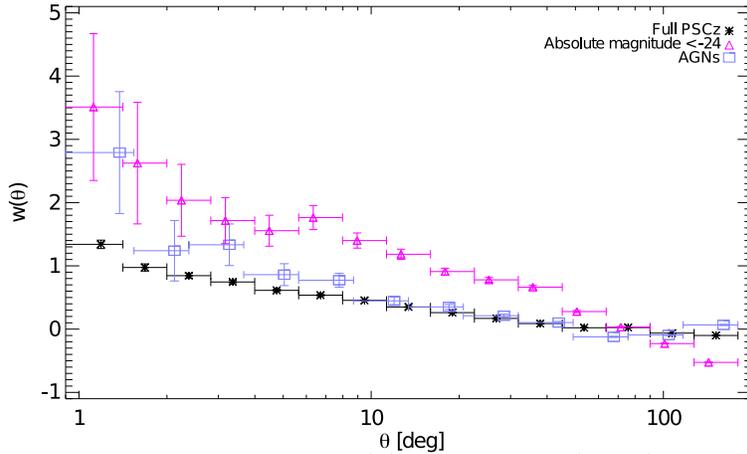,width=0.64\textwidth,angle=0}
\end{center}
\vskip-1.3cm
\caption{
The auto-correlation function $w(\theta)$ for galaxies (black),
galaxies with $M<24$ (magenta), and AGNs (blue) as function of $\theta$
with $1\,\sigma$ error bars, from Ref.~\cite{Cuoco:2007id}.
\label{auto}}
\end{figure}

In order to address this issue, we compare the non-cumulative
auto-correlation $w(\theta)$, i.e.\ considering the number of pairs
$N$ and $N'$ per bin
$[\theta-\Delta\theta:\theta+\Delta\theta]$, of galaxies and AGNs
in Fig.~\ref{auto}. 
By construction, $w(\theta)>0$ indicates overdense regions and one
recognizes that the clustering at smaller angular scales increases for
normal galaxies much slower than for AGNs. Tightening the cut in
absolute magnitude to $M<24$ makes the clustering of normal galaxies on
the smallest scales comparable in strength to the one of AGNs. 
Both effects are easily understandable: More massive objects are
stronger clustered, and the overlap of the brightest (optical) galaxies
and AGN is large so that these two sets should have quite similar
properties. A comparison of the auto-correlation function of UHECR
arrival directions with the one from proposed source catalogues  
seems to be most promising on intermediate angular scales: First, the
statistical errors both in cases of sources and cosmic rays become
sufficiently small above $\sim 10^\circ$. Second, the effect of
systematic errors like experimental errors and deflections in magnetic
fields decreases on larger scales. Turning the second argument around,
one can use the scale below which deviations between the measured and
the expected autocorrelation functions appear as measure for the
combined effect of deflections in extragalactic and Galactic magnetic
fields.  

A study of the auto-correlation function of UHECR arrival directions
should allow one with a rather small event sample to distinguish
between GRBs and AGNs as sources. Mainly long duration GRBs 
are discussed as sites for the UHECR acceleration. They occur in star 
forming galaxies that are less clustered as average galaxies.
Thus GRBs should be distinguishable from AGNs by their clustering 
properties.

The strong clustering of AGNs as the paradigm source of UHECRs
requires a re-interpretation of the traditional view of small-scale
clustering.  Often, one considered for simplicity uniformly
distributed sources and distinguished between chance coincidences 
and true clusters, i.e.\ events from the same source. 
For clustered sources, the probability to observe multiplets from
several sources in an overdense region is much increased and this
might be the main source of small-scale clusters.

\subsection{Correlations with astrophysical sources}
\label{corr}

The angular resolution of modern cosmic ray experiments reaches 
$\sim 1\,$degree,
and thus even for the UHECRs with the highest energies observed, 
$E\sim 10^{20}\,$eV, deflections in magnetic fields are likely to be more 
important.
In order to avoid a too large number of potential sources per angular
search bin, one has thus to choose either a very specific test
sample, e.g.\ a small subset of all AGNs,
or a very high energy cut such that the cosmic ray horizon scale is
sufficiently small, say of the order 100~Mpc. Thus we see that the GZK
effect is a crucial ingredient for a successful correlation analysis.

\begin{figure}
\begin{center}
\epsfig{file=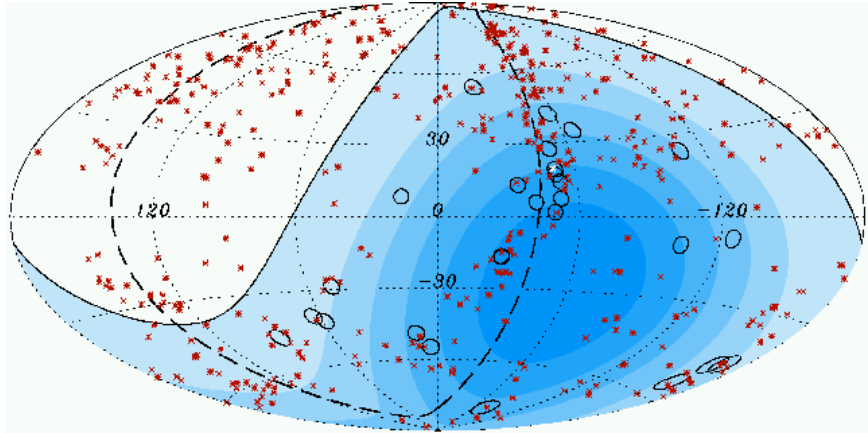,width=0.74\textwidth,angle=0}
\end{center}
\caption{
Skymap in galactic coordinates showing the arrival directions of the 
27 highest energy cosmic rays with $E>5.7\times  10^{19}\,$eV 
(black circles) detected by Auger and 472 AGN within 75\,Mpc 
(red stars). The blue region indicates the field
of view of Auger, from Ref.~\cite{pao_sc}.
\label{agn}}
\end{figure}

Until now, all claims for correlations of UHECR arrival directions
with specific source classes have remained controversial. It was
anticipated that the increased aperture of the PAO should improve this
situation. The recent analysis of the Auger experiment~\cite{pao_sc} 
claims indeed correlations between the arrival directions of the
highest energy cosmic rays with $E>5.7\times  10^{19}\,$eV and nearby AGNs.
The red stars in Fig.~\ref{agn} mark the positions of AGNs with distances 
$\leq 75\,$Mpc, while the open circles show the search bin with opening
angle $3.1^\circ$ around the CR arrival directions used in this
analysis. 
An overdense
region of AGNs along the supergalactic plane (dashed line) is clearly
visible, where the angular distance between AGNs is considerably
smaller than the size of the search bin used. The observation of a
larger number of correlations between UHECR arrival directions and
AGNs than expected from a random distribution, as in Ref.~\cite{pao_sc},  should be therefore 
interpreted only as evidence that the UHECR sources have a similar
distribution as AGNs. The differentiation among different sources
(e.g.\ AGNs vs. GRBs or among different subtypes of AGNs) would require
a comparison of the correlation (and/or auto-correlation) signal
for different source types.

%% file: he_nu.tex
\chapter{High energy neutrino astrophysics}

\section{Connecting neutrino, $\gamma$-ray and cosmic ray fluxes}

Neutrinos are copiously produced by nuclear reactions in stars and
SNe, and as secondaries in cosmic ray interactions. Because of the small 
distance to the production site, solar and atmospheric
neutrinos are experimentally easiest accessible and the only ones
regularly detected at present. The smallness of their interactions that makes
neutrino detection experimentally so challenging turns neutrinos 
into ideal messengers to study the interiors of stars, the surroundings of 
supermassive black holes in AGNs or the edge of the universe.

As we noted in the introduction, any process involving hadronization, i.e.\
the formation of colorless hadrons out of quarks and gluons, 
leads mainly to the production of pions. Isospin symmetry
fixes then the ratio of charged and neutral pions. For instance, close
to the threshold, the two possible channels of photo-meson production,
$p+\gamma\to p+\pi^0$ and $p+\gamma\to n+\pi^+$ have a ratio that is
close to 2:1. 
While the $\pi^0$ decay produces two photons, the final state
of the second channel contains after the decays $n\to p+e^-+\nu_e$ and 
$\pi^+\to \mu^+ +\nu_\mu\to e^+ +\nu_e +\bar\nu_\mu+\nu_e$
two electrons and four neutrinos. Clearly, the production of neutrinos
is thus intimately tied to the one of photons and electrons, and both
depend in turn on the flux of primary cosmic rays.

While photons and electrons start electromagnetic cascades, either on 
photons in the source or from the CMB, neutrinos reach us 
suffering only from redshift losses. As a result, the measurement of
X- and $\gamma$-rays fluxes  can be used to limit possible high energy 
neutrino fluxes.
The connection between the observed cosmic rays and neutrinos is more
model-dependent. A popular set of assumptions is the following one:
Protons are accelerated in a source region, confined by magnetic
fields. They interact on background photons (say in the UV range) in the
source, neutrons from $p+\gamma_{\rm UV}\to n+\pi^+$ escape and provide
the observed cosmic rays, while the $\pi^+$ decays produce neutrinos.
This argument does not exclude that (a subset of) sources exists with 
sufficiently large interaction depth $\tau$ such that also neutrons 
interact frequently inside the source. The cosmic ray flux from these
sources would be strongly suppressed and therefore this type of sources
has been coined ``hidden sources.''
Before we discuss the connection of neutrinos, cosmic rays, and $\gamma$-ray 
fluxes, we consider the yield of neutrinos and photons for a single 
source.

\paragraph{Photon and neutrino yields at the source} 
The yield $Y_i(E)$ of secondaries is defined as the dimensionless ratio of 
the secondary flux $\phi_i(E)$, $i=\gamma,\nu$, and the product of the  
initial proton flux $\phi_p(E)$  and the interaction depth $\tau$,
\be
 Y_i(E) =  \frac{\phi_i(E)}{\tau\phi_p(E)} \,. 
\ee
We consider only in detail the simpler case of photons as secondaries. 
Generalizing Eq.~(\ref{pionspec}), the total photon flux from neutral 
pion decays that are in turn produced as secondaries in $pp$ or $p\gamma$ 
scatterings is 
\be 
 \phi_{\gamma}(E_\gamma) = 
 2 \int_{E_\gamma}^\infty\frac{\d E_\pi}{E_\pi}\:
   \int_{E_\pi}^\infty\d E_p \:N_{\pi^0}(E_p)\: 
   \frac{\d\sigma(E_p,E_\pi)}{\d E_\pi}
   \:\tau(E_p)\phi_{p}(E_p) \,,
\ee
where $N_{\pi^0}$ denotes the multiplicity of neutral pions 
and $\tau$ the interaction depth.
For the same assumptions as used in the derivation of Eq.~(\ref{Z_spec}), 
we obtain as approximation for the photon yields
\be
 Y_\gamma(E) =  \frac{\phi_\gamma(E)}{\tau\phi_p(E)} =
 \frac{2}{\alpha}\: Z_{p\pi^0}(\alpha) \,,
\ee
where now in the definition of $Z_{p\pi^0}(\alpha)$ the neutral pion 
multiplicity is included. 
The calculation of the neutrino yields proceeds along the same lines. 
However, the resulting expressions are rather lengthy, because
the decays are non-isotropic, the decay chains are more complex and 
the final-state particles massive. The relevant expressions can be found in 
Refs.~[E1,E2].

\begin{figure}[h]
\begin{center}
\includegraphics*[width=0.45\textwidth,clip]{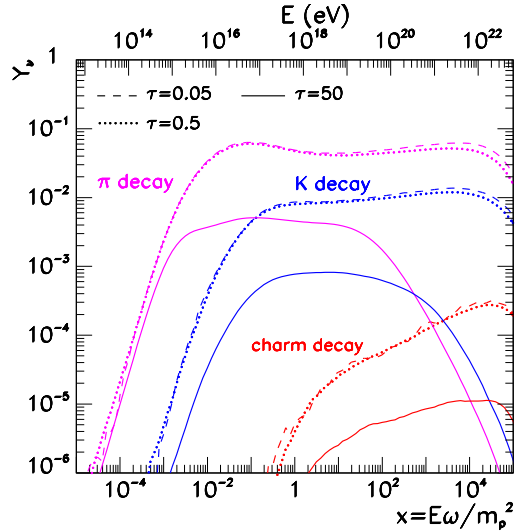}
\caption{\label{Y1} 
Neutrino yield $Y$ as function of $x=E\omega/m_p^2$ for $T=10^4$~K from pion, 
charged and neutral kaon, and charm decays for $\tau=0.05, 0.5$ and 50; 
from Ref.~\cite{Y}.}
\end{center}
\end{figure}

In Fig.~\ref{Y1}, the neutrino yields $Y_\nu(E)$ from various subprocesses
(pion, kaon and charm decays) are shown for a thermal distribution of photons
with temperature $T=10^4\,$K and three different interaction depths. 
As long as $\tau\lsim 1$, i.e.\ multiple scatterings can be 
neglected, the yields are by construction independent from $\tau$. 
The yields are also independent from $T$, if one uses 
instead of the energy $E$ the parameter $x=E\omega/m_p^2$, where
$\omega=1.6T$ is the maximum of the Planck distribution, 
Therefore they are a convenient tool to estimate photon
or neutrino yields as long as $\tau\lsim 1$. 
Tabulated values of $Y$ for different $\alpha$ can be found in 
Ref.~[E1,E2].

\paragraph{Cascade limit}
The general concept of electromagnetic cascades discussed in 
Sec.~\ref{elamgcascades} has several important applications. Firstly, we 
can use it to derive a simple bound on the UHE neutrino flux.
Using the branching ratio 2:1 for $p+\gamma\to p+\pi^0$ and 
$p+\gamma\to n+\pi^+$, the approximation that the four leptons produced 
in muon decay share equally the energy as well as the fact that in 
neutron decay almost all energy 
is taken by the proton, the ratio of energy transferred to neutrinos and the
electromagnetic channel follows as 1:3, $\omega_\nu=3\omega_\gamma$.
  
If we approximate the neutrino intensity in the range $[E,E_{\max}]$ by a 
power-law, $I_\nu(E)=AE^{-\alpha}$, then the integral neutrino intensity is
connected to $\omega_\nu$ via
\be
 I_\nu(>E) = \frac{\alpha}{\alpha+1}
            \left[1-\left(\frac{E_{\max}}{E}\right)^{-\alpha}\right]^{-1}
            \frac{c}{4\pi}\,\frac{\omega_\nu}{E} \,,
\ee
where $\omega_\nu$ is the energy density in neutrinos in the same energy 
range $[E,E_{\max}]$. The first two terms are bounded by one, provided that the 
energy range considered is large enough, $\alpha\ln(E/E_{\max})>1$.
With $\omega_\nu\leq 3\omega_\gamma$, we obtain thus as bound for the
integrated neutrino flux
\be \label{casc1}
 I_\nu(>E) \leq \frac{c}{12\pi}\,\frac{\omega_\gamma}{E} \,.
\ee

In order to apply this bound, we have to distinguish between different 
possible photon backgrounds on which cascades can develop. The only universal
one is the CMB leading to a guaranteed flux%
\footnote{If UHECRs are dominantly nuclei, a neutrino flux at lower 
energies from neutron decays (via $A+\gamma_{3K}\to (A-1)+n$) results.} 
of UHE neutrinos via the GZK reaction of UHECRs as well as to a contribution to 
the diffuse photon flux in the 100\,MeV--10\,GeV range. 
This energy range has been observed by the EGRET experiment and thus a bound 
on the diffuse extragalactic gamma-ray flux can be derived. Integrating
this maximal flux, the limit on $\omega_\gamma$ follows as
\be \label{casc2}
\omega_{\gamma} \leq 2\times 10^{-6}\, {\rm eV/cm}^3 \,.
\ee
If we assume that $I_\nu(E_\nu)=AE_\nu^{-2}$, we can combine Eqs.~(\ref{casc1})
and (\ref{casc2}) into a limit on the differential neutrino intensity,
\be \label{casc3}
  E_\nu^2 \,I_\nu(E_\nu) \leq
 2 \times 10^{-6} \,{\rm GeV \; cm^{-2}\; s^{-1} \; sr^{-1}} \,.
\ee

Electromagnetic cascades may happen also inside many sources.
Since the Thomson cross section is a factor 3000 larger than the
cross section for photo-meson production, 
$\sigma_{\rm Th}/\sigma_{p\gamma}\sim 3000$, 
sources where photo-meson reactions are relevant are likely optically thick.
To be specific, let us assume that the source of opacity are UV photons, 
$\eps_\gamma\sim 10\:$eV, as it is the case for many AGNs. 
Hence the critical energy for pair production is reduced to $\Ecr\sim 25\,$GeV 
compared to $\Ecr\sim 400\,$TeV for cascades on CMB photons.
Thus the picture of the escaping photon radiation is as follows:

\begin{itemize}
\sitem
Photons with $E\gg\Ecr$ escape from the source only from an outer layer of 
such a thickness $l$ that the interaction depth 
$\tau_{\gamma\gamma}=l\sigma_{\gamma\gamma}n_\gamma$  for pair-production is small,
$\tau_{\gamma\gamma}\sim 1$. 
\sitem
The source is bright in the $\sim 10-100\,$GeV range, where the last generation
of photons generated in the cascade can escape. 
\sitem
Electrons with energy below $E_{\rm cr}$ continue to scatter in the source 
in the Thomson regime, producing photons in the X-ray range with average 
energy
\be
 E_\gamma = \frac{4}{3}\frac{\eps_\gamma E_e^2}{m_e^2} 
 \sim 100\,{\rm eV} - 0.1\,{\rm MeV}
\ee 
for $E_e=1\,$GeV and 25\,GeV, respectively.
\end{itemize}

\paragraph{Cosmic ray limits} The total energy density in neutrinos can be
directly bounded by the energy density injected in cosmic rays by
extragalactic sources, if neutrinos are produced only in
transparent sources. If one assumes further that only neutrons can
leave the sources, this bound becomes a prediction. To proceed, one
has to specify the generation spectrum of extragalactic cosmic ray,
i.e.\ one has to choose essentially between the traditional point of
view (injection spectrum $\d E/\d N\propto E^{-2}$ and transition
energy around $3\times 10^{19}\,$eV) or the dip scenario
[(effective) injection spectrum $\d E/\d N\propto E^{-2.6}$ and transition
energy around $10^{18}\,$eV). Using the first choice, several authors
advocated bounds or expectations for the neutrino flux of order 
\be \label{wb}
  E_\nu^2 \,I_\nu(E_\nu) \sim 
 {\rm few}\times 10^{-8} {\rm GeV \; cm^{-2}\; s^{-1} \; sr^{-1} } \,.
\ee
Thus the cosmic ray limit is about a factor 100 stronger than the cascade 
limit. 
However, one should be aware that even within the framework of transparent
sources cosmic ray bounds obtained for a fixed set of parameters 
can only be indicative, because the sensitivity of the cosmic ray and the 
neutrino fluxes on several free parameters like the redshift evolution of 
their sources differs strongly.
\begin{figure}[h]
\begin{center}
\includegraphics*[width=0.65\textwidth,clip]{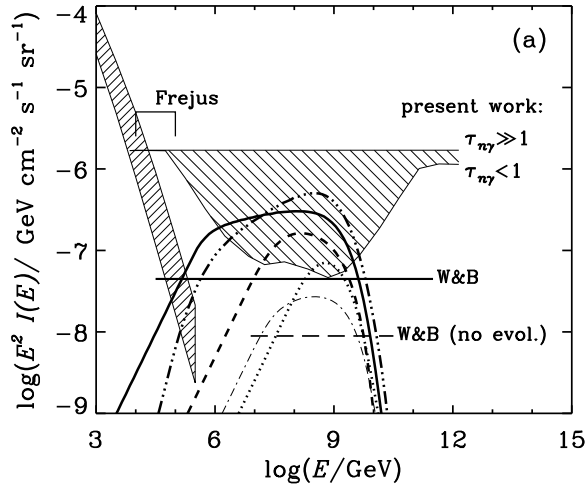}
\caption{\label{wbmpr} 
A collection of various bounds on neutrino fluxes: The solid line at the
top corresponds to the cascade limit~(\ref{casc3}), the lower dashed
horizontal line to the cosmic ray limit~(\ref{wb}), while the lines 
in-between correspond to varying assumptions used for the cosmic ray limit, 
from Ref.~\cite{MPR}.}
\end{center}
\end{figure}

\section{Neutrino interactions, masses and mixing}
One of the most surprising properties of the weak interactions is that
only left-chiral fermions $(P_-\psi)$ and right-chiral anti-fermions
$(\overline{\psi}P_+)$ participate in charged-current (CC) weak
interactions,
\be   \label{CC-SM}
  {\cal L}_{CC} = -
 \frac{g}{\sqrt{2}}\sum_\alpha \bar l_{L,\alpha} \gamma^\mu \nu_{L,\alpha} 
                          W^-_\mu +{\rm h.c.} \,.
\ee
Thus weak interaction violate parity P
leading to a doublet/singlet structure of fermions,
\be
\pmatrix{\nu_e \cr e^-}_L,
\pmatrix{\nu_\mu \cr \mu^-}_L, 
\pmatrix{\nu_\tau \cr \tau^-}_L 
\qquad e_R, \mu_R, \tau_R \,.
\ee
The current (\ref{CC-SM}) defines what we mean with electron neutrino:
It is the neutrino created together with an electron in a CC interaction.

Since ordinary Dirac mass terms
\be 
m_f\overline{\psi_f}\psi_f = 
m_f(\overline{\psi_f^{\mathrm{L}}}\psi_f^{\mathrm{R}}+
        \overline{\psi_f^{\mathrm{R}}}\psi_f^{\mathrm{L}})
\ee
are not gauge invariant, one introduces fermion masses in the standard model 
by gauge-invariant Yukawa interactions. These interactions need
not to know about weak interactions and are thus not diagonal in
flavor space, 
\be
 {\cal L}_{\rm mass} = - \sum_{\alpha,\beta} \bar \nu_{L,\alpha}
                       M_{\alpha\beta}^{(\nu)} \nu_{R,\beta}
                       - \sum_{\alpha,\beta} \bar l_{L,\alpha}
                       M_{\alpha\beta}^{({\rm ch})} l_{R,\beta} + {\rm h.c.},
\ee
i.e.\ the mass matrices $M_{\alpha\beta}$ 
are not hermitian or even diagonal in the basis of the weak
eigenstates. (We denote the weak eigenstates
$\nu_\alpha=\{\nu_e,\nu_\mu,\nu_\tau\}$  and $l_\alpha=\{e,\mu,\tau\}$ 
by Greek indices $\alpha,\beta\ldots$, 
and the mass eigenstates by Latin indices $i,j,\ldots$ Furthermore,
we have assumed that the neutrinos have only Dirac mass terms to simplify
the formulas). Since the mass matrices are not hermitian, they cannot be
diagonalized by a simple unitary transformation. However, arbitrary
mass matrices can be diagonalized by a biunitary transformation,
\be
 M_{\rm diag}^{(\nu)}=U^{(\nu)\dagger} M^{(\nu)} T^{(\nu)} \,,
\ee
where  $U^{(\nu)} U^{(\nu)\dagger} = T^{(\nu)} T^{(\nu)\dagger} ={\bf 1}$.
Then, the connection between weak and mass eigenstates is given by
\be  \label{U}
 \nu_{R,\alpha} = \sum_i T^{(\nu)}_{\alpha i} \nu_{R,i} \:, \qquad\qquad
 \nu_{L,\alpha} = \sum_i U^{(\nu)}_{\alpha i} \nu_{L,i} \:, 
\ee
and similar equations hold for the charged leptons.
 
Inserting the transformations of Eq.~(\ref{U}) into the charged
current Lagrangian (\ref{CC-SM}) of the SM,
results in
\be   \label{Hcc}
 {\cal L}_{CC} = -
 \frac{g}{\sqrt{2}}\sum_{i,j} \bar l_{L,i} \gamma^\mu V_{ij} \nu_{L,j} 
                              W^-_\mu +{\rm h.c.} , 
\ee
where $V=U^{({\rm ch})\dagger}U^{(\nu)}$ is the analogue of the CKM
matrix in the lepton sector. Since the charged current
interaction involves only left-chiral fields of both charged leptons and 
neutrinos, the product of the two mixing matrices of the right-handed
leptons, $T^{({\rm ch})\dagger}T^{(\nu)}$, is unobservable.
The mixing $V=U^{({\rm ch})\dagger}U^{(\nu)}$ between charged leptons and
neutrinos has two different sources: It could be ascribed either
completely to mixing in the neutrino ($V=U^{(\nu)}$) or in the charged
lepton sector ($V=U^{({\rm ch})\dagger}$), or most probably to some
superposition of both. Only the combination $V$ is observable and, 
by convention, we set $U^{(l)}={\bf 1}$.

In the case of massless neutrinos we can choose the neutrino mass
eigenstates arbitrarily. In particular, we can set
$U^{(\nu)}=U^{({\rm ch})}$ for any given $U^{({\rm ch})}$, hereby rotating
away the mixing. This shows that neutrino masses are a necessary
condition for non-trivial consequences of mixing in the lepton sector.

\subsection*{Neutrino oscillations in vacuum}
\begin{figure}
\begin{picture}(0,0)%
\epsfig{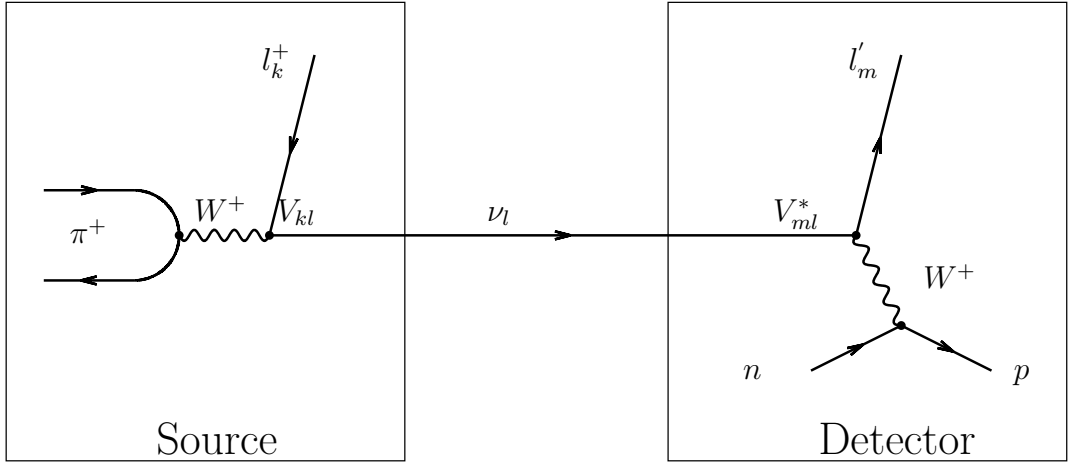}%
\end{picture}%
\setlength{\unitlength}{4144sp}%
\begingroup\makeatletter\ifx\SetFigFont\undefined%
\gdef\SetFigFont#1#2#3#4#5{%
  \reset@font\fontsize{#1}{#2pt}%
  \fontfamily{#3}\fontseries{#4}\fontshape{#5}%
  \selectfont}%
\fi\endgroup%
\begin{picture}(6369,2769)(304,-3223)
\put(811,-1906){\makebox(0,0)[b]{\smash{\SetFigFont{12}{14.4}{\rmdefault}{\mddefault}{\updefault}
$\pi^{+}$
}}}
\put(1216,-3166){\makebox(0,0)[lb]{\smash{\SetFigFont{17}{20.4}{\rmdefault}{\mddefault}{\updefault}
Source
}}}
\put(5176,-3166){\makebox(0,0)[lb]{\smash{\SetFigFont{17}{20.4}{\rmdefault}{\mddefault}{\updefault}
Detector
}}}
\put(1441,-1771){\makebox(0,0)[lb]{\smash{\SetFigFont{12}{14.4}{\rmdefault}{\mddefault}{\updefault}
$W^+$
}}}
\put(1936,-1771){\makebox(0,0)[lb]{\smash{\SetFigFont{12}{14.4}{\rmdefault}{\mddefault}{\updefault}
$V_{kl}$
}}}
\put(3196,-1771){\makebox(0,0)[lb]{\smash{\SetFigFont{12}{14.4}{\rmdefault}{\mddefault}{\updefault}
$\nu_l$
}}}
\put(1846,-871){\makebox(0,0)[lb]{\smash{\SetFigFont{12}{14.4}{\rmdefault}{\mddefault}{\updefault}
$l^{+}_{k}$
}}}
\put(5806,-2176){\makebox(0,0)[lb]{\smash{\SetFigFont{12}{14.4}{\rmdefault}{\mddefault}{\updefault}
$W^+$
}}}
\put(4726,-2716){\makebox(0,0)[lb]{\smash{\SetFigFont{12}{14.4}{\rmdefault}{\mddefault}{\updefault}
$n$
}}}
\put(6346,-2716){\makebox(0,0)[lb]{\smash{\SetFigFont{12}{14.4}{\rmdefault}{\mddefault}{\updefault}
$p$
}}}
\put(5356,-871){\makebox(0,0)[lb]{\smash{\SetFigFont{12}{14.4}{\rmdefault}{\mddefault}{\updefault}
$l^{'}_{m}$
}}}
\put(4906,-1771){\makebox(0,0)[lb]{\smash{\SetFigFont{12}{14.4}{\rmdefault}{\mddefault}{\updefault}
$V^{\ast}_{ml}$
}}}
\end{picture}
\caption{\label{pidecay}
Production of a superposition of neutrino mass eigenstates $\nu_l$ in
pion decay and subsequent detection of the neutrino flavour via the
secondary lepton $l^\prime_m$.}
\end{figure}
Let us consider e.g.\ neutrinos produced in charged pion decay.
The ratio $R$ of $\pi\to e\nu_e$ and $\pi\to\mu\nu_\mu$ decay rates is
\be
 R = 
 \frac{\Gamma(\pi\to e\nu_e)}{\Gamma(\pi\to\mu\nu_\mu)} =
 \frac{m_e^2}{m_\mu^2}\:\frac{(m_\pi^2-m_e^2)^2}{(m_\pi^2-m_\mu^2)^2} 
 \ap 1.28\times 10^{-4} \,,
\ee
since angular momentum conservation  in the pion rest frame requires a
helicity flip of the lepton. Similar, in neutron decay and in fusion
reactions in stars only $\nu_e$'s are emitted, because of energetic
reasons.  Hence, in many occasions we start with a (nearly) pure flavor
state. 

The time-evolution between creation of an arbitrary state at $t=0$ and 
detection at $t$ becomes simplest, if we decompose the weak
interaction eigenstate $\nu_\alpha$ into mass eigenstates $\nu_i$,
\be
|\nu(t)\rangle=\sum_i U_{\alpha i}^{(\nu)}| \nu_i\rangle \e^{-iEt} \,.
\ee
Neutrinos are in all applications ultra-relativistic,
\be
E_i =(p^2+m_i^2)^{1/2}\ap p + m_i^2/(2p) \,,
\ee
where we have assigned also a definite momentum to the states
$|\nu_i\rangle$. 
\be
|\nu(t)\rangle= \e^{-i pt} \sum_i U_{\alpha i}^{(\nu)}|
\nu_i\rangle \e^{-im_i^2/(2p)t} \,.
\ee
The probability for a transition from the flavor $\nu_\alpha$ to
$\nu_\beta$ after the distance $L=ct$ is  
\be
P_{\alpha\to\beta}(t) = \left| \sum_{k=1}^n
  U_{\beta k}^\ast  \exp(-i Et) U_{\alpha k} \right|^2 \,,
\ee
where we introduced also $\Delta m^2_{ij}=|m_i^2-m_j^2|$.

To be specific, we consider the simple case of two-flavor neutrino 
oscillations. Then the survival probability is 
\be \label{2nu_surv}
P_{\nu_e\to\nu_e} = 
1-\sin^2\left( \pi\,\frac{L}{l_{\rm osc}} \right) \:\sin^2 2\theta
\ee
with the vacuum oscillation length 
\be
l_{\rm osc} =  4\pi\frac{E}{\Delta m^2}\approx 2.5\;{\rm m}
                \left(\frac{E}{\rm MeV}\right)
                \left(\frac{{\rm eV}^2}{\Delta m^2}\right)
\ee
Thus the survival probability depends only on $L/E$, the mass squared
difference of the two neutrinos, and one mixing matrix element.
Most strikingly, the smallness of the neutrino masses makes it 
possible to observe the interference of quantum states on macroscopic 
length scales.

\section{Atmospheric neutrino oscillations}

The principle production modes for neutrinos produced by comic rays in the
atmosphere are pion, kaon and muon decays. With
$\pi^\pm\to\mu^\pm+\nu_\mu(\bar\nu_\mu)$ and 
$\mu^\pm\to e^\pm+\nu_\e(\bar\nu_\e)+\bar\nu_\mu(\nu_\mu)$,
one expects the ratio
\be
 \frac{\nu_\e+\bar\nu_\e}{\bar\nu_\mu+\nu_\mu}
 = \frac{N_\e}{N_\mu}\sim \frac{1}{2}\,,
\ee
if all particles can decay. Muon decays become suppressed for $E\gsim
2.5$~GeV ($\lambda_d>\lambda_{atm}$) and thus the ratio should
decrease for higher energies. Since most neutrinos are produced by
low energy cosmic rays, geomagnetic effect must be included in the flux
calculations. On the experimental side, the connection of the visible
energy and direction (of the secondary electron or muon) to the one of
the neutrino can be made only on a statistical basis and requires good
knowledge of detector response and acceptance. 

Part of these problems can be avoided by looking at the double ratio
\be
R = \frac{(N_\e/N_\mu)_{\rm data}}{(N_\e/N_\mu)_{\rm MC}} \,.
\ee
If in the Monte Carlo no oscillations are included, one expect $R\sim 1$ if
neutrinos are massless. The first conclusive deviation from $R=1$ was 
found by the Superkamiokande experiment in 1998 with $R=0.69\pm 0.06$.

\paragraph{Ratio of up- and down-going muons} A decisive test for the
oscillation hypothesis is the ratio of up- and down-going muons:
The oscillation length varies between 20km and 12800km, and as
function of the zenith angle the ratio should show the form predicted
by the neutrino oscillation formula. By fitting the experimental data
to the free parameters $\Delta m^2$ and  $\sin^2\theta$ of
the neutrino oscillation formula, one derived
$\Delta m^2\sim 2.2\times 10^{-3}$~eV$^2$ and $\sin^2\theta\sim 1$.

\begin{figure}
\epsfig{figure=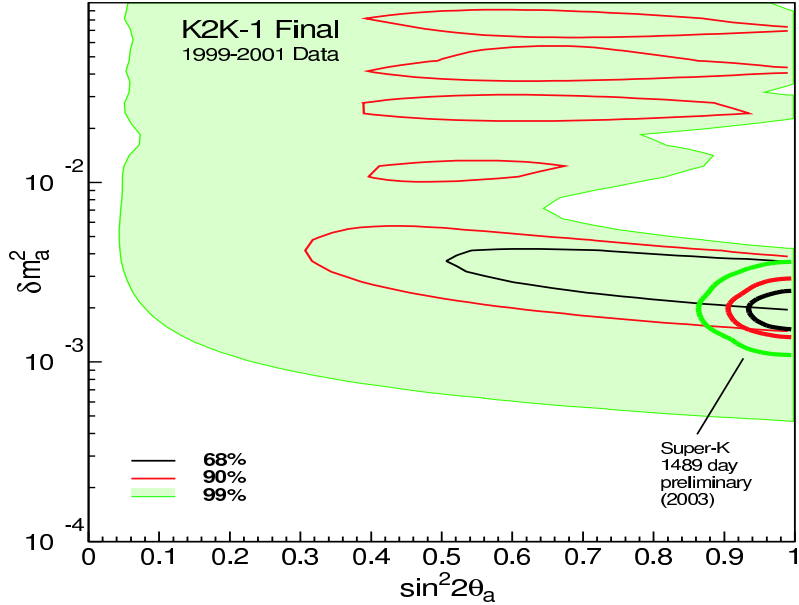,width=10.5cm,height=8cm} 
\caption{\label{K2K}
Overlay of the Superkamiokande and the K2K result.} 
\end{figure}

The atmospheric anomaly was later confirmed by sending a $\nu_\mu$
beam from KEK to Kamioka. The distance KEK-Kamioka is $L=250$~km, 
while the mean neutrino energy was $E\sim 1.3$~GeV. The combined
result of the two experiments is shown in Fig.~\ref{K2K}, yielding
the surprising result that the atmospheric neutrino mixing angle
is -- in contrast to the quark sector -- close to maximal. 

\section{High energy neutrino scattering}

\paragraph{Parton model and scaling violation}
Neutrino-nucleon scattering can be described in the parton model. This
model assumes that a nucleon probed at sufficiently high 
$Q^2\gsim Q_{\rm min}^2\sim 1$~GeV$^2$ behaves 
as a collection of independently interacting partons, i.e.\ quarks and
gluons. If the function $f_i(x)$ denotes the probability to find a
parton of type $i$ with momentum fraction $p_i=xp_N$ and
$\sigma(\hat{s})$
the corresponding parton cross section, then the neutrino-nucleon cross
section follows as
\be
 \sigma(s) = \sum_i \int_{x_{\rm min}}^1 \d x \,f_i(x) \hat{\sigma}(\hat{s})\,.
\ee
In its original formulation, the parton model assumed that the parton
distribution functions $f_i(x)$ depend only on $x$. We can convince
ourselves that $f_i(x)$ should depend also on $Q^2$ by looking at 
a Bremsstrahlung process.
Consider a quark as an external line of a Feynman diagram with momentum 
$p$ and arbitrary mass $m$ emitting a gluon of momentum $q$,
\be
 \frac{1}{(p+q)^2-m^2} = \frac{\pm 1}{2pq} =
 \frac{\pm 1}{2\omega E(1-\cos\theta)}  \,,
\ee
where $\omega$ is the energy of the gluon, $E$ and $v$ the velocity of
the parton and $\theta$ the emission angle. There is a soft
divergence for all $m$, if the energy of the emitted gluon goes to
zero, $\omega\to 0$. Additionally, there is a collinear singularity for
light partons, $v\to 1$, when $\theta\to 0$.

These singularities generate logarithmic divergences integrating over
the final phase space,
\be
 \d\sigma_{n+1} = 
 \frac{\alpha_s(Q^2)}{2\pi}\ln^2(Q^2/Q^2_{\min}) \d\sigma_{n} \,,
\ee
that can compensate the strong coupling although $\alpha_s$ is small for 
large $Q^2$.  As a result, parton cascades or jets develop in fragmentation 
processes. Similarly, the parton number inside a nucleon increases with 
increasing $Q^2$. 
Although the neutrino-quark cross section is constant above $s\gg m_W^2$, 
the neutrino-nucleon cross section continues to grow as $\sigma\propto E^{0.4}$,
 because the number of accessible partons with 
$x\geq x_{\min}\sim Q_{\rm min}^2/s$ increases as $xq(x,Q^2)\sim x^{-0.33}$.

\begin{figure}
\epsfig{figure=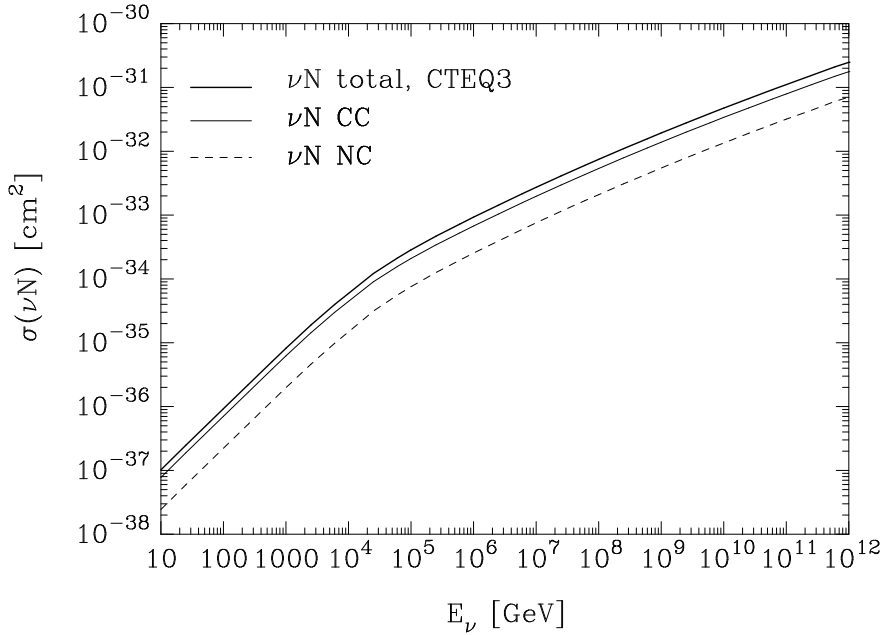,width=.75\textwidth} 
\caption{\label{fig:GQRS08} 
Cross sections for $\nu N$ interactions at high energies:
dotted line, $\sigma(\nu N \rightarrow \nu+\hbox{anything})$; thin
line,  $\sigma(\nu N \rightarrow \mu^{-}+\hbox{anything})$; thick
line, total (charged-current plus neutral-current) cross
section, from Ref.~\cite{Gandhi}.}
\end{figure}

\paragraph{DGLAP equations}

The $Q^2$-dependence of the parton distribution functions is described in 
perturbative QCD by 
the Dokshitzer-Gribov-Lipatov-Altarelli-Parisi (DGLAP) equations
\ba
\frac{\d q_i(x,Q^2)}{\d t} &=& 
\frac{\alpha_s(Q^2)}{2\pi} \int^1_x \frac{\d y}{y} \bigg[
  q_i(y,Q^2)  P_{qq} \left( \frac{x}{y} \right) + g(y,Q^2) P_{qg}
  \bigg( \frac{x}{y} \bigg) \bigg] 
\\
\frac{\d g_i(x,Q^2)}{\d t} &=&
\frac{\alpha_s(Q^2)}{2\pi} \int^1_x \frac{\d y}{y} \bigg[
  q_s(y,Q^2)] P_{gq}\bigg(
  \frac{x}{y} \bigg) + g(y,Q^2) P_{gg} \bigg( \frac{x}{y} \bigg) \bigg].  
\ea
with $t=\ln(Q^2/Q_0^2)$ and $q_s=\sum_{j=1}^{n_f}[q_j + \bar{q}_j]$.
The splitting functions $P_{ij} \big( \frac{x}{y} \big)$, with $i,j =
q,g$ give the probability that parton $j$ with momentum $y$ radiates a
quark or gluon and becomes a parton of type $i$ with fraction $\big(
\frac{x}{y} \big)$ of the momentum of parton $j$.  

If we compare this equation to the production term of the transport 
equation for cosmic rays, Eq.~(\ref{transport}), we see that the structure 
of the two equations is the same. The reason for this formal agreement 
(or the simplicity of the DGLAP equations) is that interference terms can be 
neglected in the QCD cascade, since it corresponds to a semi-classical 
evolution.

\section{Astrophysical sources of high energy neutrinos}

\paragraph{Experiments and techniques}
Main problem of high energy neutrinos physics is the atmospheric
neutrino background and the expected low fluxes from astrophysical 
neutrino sources. The search for neutrino sources is most promising 
at high energies, because the atmospheric neutrino flux is very steep:
At high energies, pions scatter and muons do not decay.
Main detection methods are at present Cherenkov light (in ice or water) emitted
e.g.\ by a relativistic muon produced in the reaction $\nu_\mu+N\to \mu+X$. In
the future, the radio or acoustic signal produced by a horizontal shower 
initiated by a neutrino reaction may allow one the supervision of even
larger detector volumes.

The event rate of muons with energy $E>E_{\min}$ in a detector of area
$A$ is 
\be \label{nurate}
 {\rm Rate} = A \, \int\d\!E \d\Omega \;
 P_\mu(E;E_\mu^{\min}) S(E) \, I_\nu(E_\nu) \,.
\ee
Here, the shadow factor $S$ takes the attenuation in the Earth into 
account, and $P_\mu(E;E_\mu^{\mathrm{min}})$ is the probability that a muon 
is created and reaches the detector with $>E_{\min}$.

\begin{figure}
\centering
\includegraphics[width=0.65\textwidth]{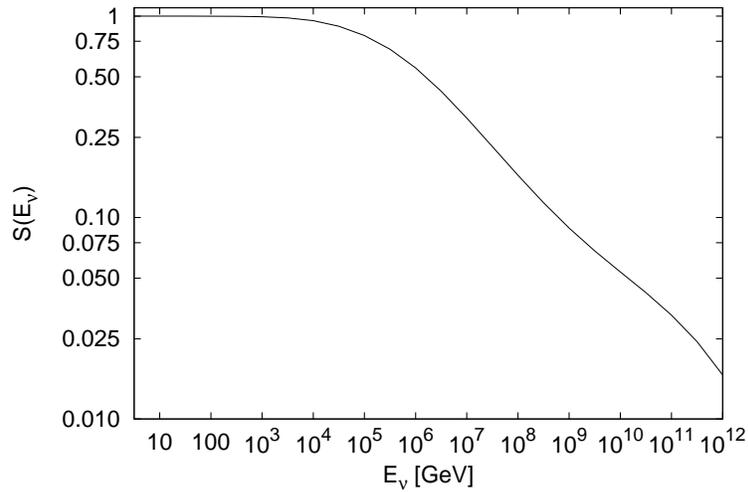}
\caption{Averaged shadow factor $S(E)$ as function of the neutrino
energy.}
\label{fig:shadow-nu}
\end{figure}

\paragraph{Attenuation and Shadow factor}
The growth of the neutrino-nucleon cross section with energy means that
above $E_\nu\sim 40\,$TeV the Earth becomes opaque even for neutrinos.
If we neglect any regeneration effects and assume that the flux is
isotropic, this attenuation can be represented by a shadow factor. The
shadow factor averaged over the upward-going hemisphere is
\be
 S(E) = \frac{1}{2\pi} \int_{-1}^{0} \d\cos\theta \int\d\phi \, 
 \exp [-X(\theta)/\lambda_{\mathrm{int}}(E)] 
\ee
and is shown in Fig.~\ref{fig:shadow-nu}.

\begin{figure}
\centering
\includegraphics[width=0.65\textwidth]{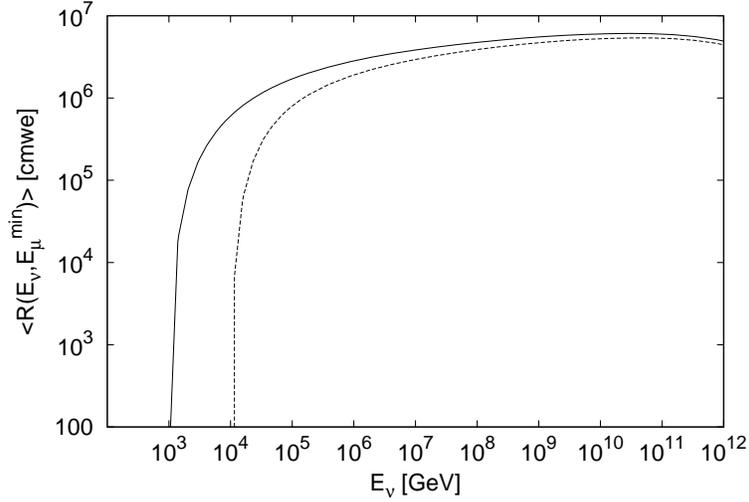}
\caption{Average ranges (in rock) for muons produced in
  charged-current interactions of neutrinos with energy $E_\nu$, at
  threshold energies $E_\mu^{\mathrm{min}} = 1 \,\textrm{and}\, 10
  \,\mathrm{TeV}$.} 
\label{fig:range-nu}
\end{figure}

\paragraph{Average range of muons}
The second ingredient in the rate calculation is the probability 
that the neutrino creates a muon that is energetic enough to arrive at the
detector with an energy $E_\mu$ larger than the detector's threshold
energy $E_\mu^{\mathrm{min}}$.   The probability that
a muon can be recorded in a detector depends on the average range
$\langle R \rangle$ of a muon in rock
\begin{equation}
\langle R(E_\nu;E_\mu^{\mathrm{min}}) \rangle =
\frac{1}{\sigma_{CC}(E_\nu)} \,
\int_{0}^{1-E_\mu^{\mathrm{min}}/E_\nu} \displaylimits {\, \mathrm{d}
  y \; R(E_\mu,E_\mu^{\mathrm{min}}) \frac{\mathrm{d}
    \sigma_{CC}(E_\nu,y)}{\mathrm{d} y}},   
\end{equation}
where the energy  of muons produced in a charged-current interaction of 
neutrinos with matter is $E_\mu = (1-y)E_\nu$. After a high energy muon is 
produced, it undergoes a continuous energy loss as it propagates.  
The range $R$ of an energetic muon follows from the energy-loss
relation
\be
-\, \mathrm{d} E_\mu/ \mathrm{d} X = \alpha(E_\mu) + \, \beta(E_\mu)E_\mu,
\label{eq:energy-loss}
\ee
where $X$ is the thickness of matter traversed by the muon in
$\mathrm{g}/\mathrm{cm}^2$. The first term represents ionization
losses, while the second term represents 
bremsstrahlung, $e^+e^-$ pair production and nuclear interactions. If
the coefficients $\alpha$ and $\beta$ are independent of energy, we
can approximate their values to be $\alpha = 2.0 \times \, 10^{-3} \,
\textrm{GeV} \, \textrm{cmwe}^{-1}\,$ ($\mathrm{cmwe} =
\mathrm{g/cm}^3$) and $ \, \beta = 3.9 \times \, 10^{-6}  \,
\textrm{cmwe}^{-1}$. Integrating Eq.~(\ref{eq:energy-loss}), the
muon range is 
\be
R(E_\mu,E_\mu^{\mathrm{min}}) \equiv X(E_\mu^{\mathrm{min}}) -
X(E_\mu) = \frac{1}{b} \ln{\frac{a + \,bE_\mu}{a +
    \,bE_\mu^{\mathrm{min}}}} \,. 
\ee
The average range of muons from charged current neutrino interactions
is shown in Fig.~\ref{fig:range-nu} for threshold energies 1 TeV and
10 TeV. A muon produced with $E_\mu = 10$ TeV will travel on average a few
kilometers until its energy is degraded to 1 TeV.

\begin{figure}
\centering
\includegraphics[width=0.65\textwidth]{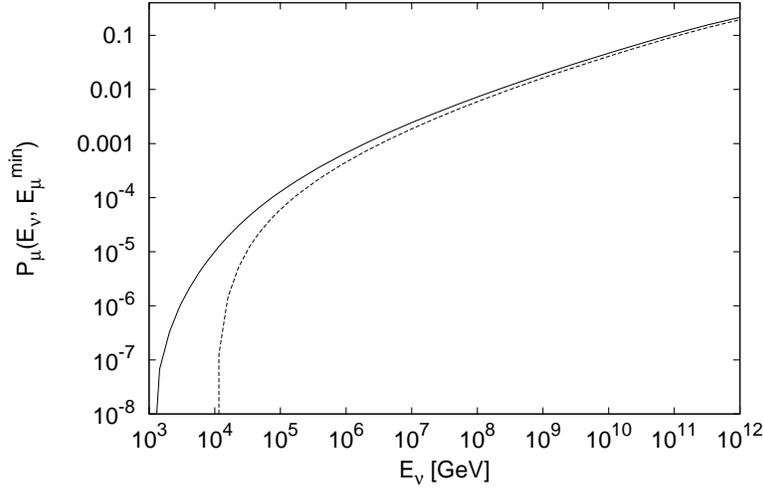}
\caption{Probability that a neutrino of energy $E_\nu$ produces an
  observable muon with energy exceeding $E_\mu^{\mathrm{min}} =
  1\,\mathrm{TeV}$ and $10\, \mathrm{TeV}$.} 
\label{fig:pro-nu}
\end{figure}

The probability that a muon neutrino of energy $E$ produces an observable
muon is then 
\begin{equation}
P_\mu(E,E_\mu^{\mathrm{min}}) = N_A \, \sigma (E)\langle
R(E;E_\mu^{\mathrm{min}} \rangle 
\end{equation}
and is shown in Fig.~\ref{fig:pro-nu}. A useful approximation  for
energies $E\gsim \,$TeV is $P_\mu(E,E_\mu^{\mathrm{min}})\sim 2\times
10^{-6} (E_\nu/{\rm TeV})^{0.8}$. 

\paragraph{Sensitivity and event rates} 
We derive now a rough estimate of the event rate
expected in a neutrino telescope. We assume $E_{\min}=10^6\,$GeV as minimal 
energy in order to avoid the background of atmospheric neutrinos and 
a neutrino intensity equal to the cascade limit (\ref{casc3}). 
Then we approximate the remaining ingredients of the integrand in 
Eq.~(\ref{nurate}) as
$P_\mu\sim 10^{-3}(E/E_{\min})^{0.8}$ and $S\sim (E/E_{\min})^{-0.2}$. Choosing
as area $A=(0.1\,{\rm km})^2$, i.e. the area covered by the current 
ICECUBE prototype AMANDA, we obtain of order 1\,event/year. Thus the 
sensitivity of this experiment corresponds roughly to the maximal
flux allowed by the cascade limit, while an 1\,km$^3$ neutrino telescope
like ICECUBE will probe fluxes down to the ones compatible with the 
cosmic ray limit.

\section{Exercises}
\begin{enumerate}
\item Derive from Fig.~6.3 the energy density of (extragalactic)
  cosmic rays above $10^{18}\,$eV for $\alpha=2$ and 2.7. 
 Estimate from Fig.~6.1 the ``life-time''
 of a cosmic ray  and thus the required emissivity ${\cal L}$.
\item
Derive for two neutrino the oscillation probability,
$P_{\nu_e\to\nu_\mu}$, starting from the mixing matrix
\be
U=\left(
\begin{array}{cc}
 \cos\theta&\sin\theta\e^{\i\phi} \\
 -\sin\theta&\cos\theta\e^{\i\phi}
\end{array}
\right) \,.
\ee
Show that that the Majorana phase $\phi$ does not enter the
oscillation probability. Convince yourself that probability is conserved.
\end{enumerate}

%% file: newphys.tex
\chapter{Cosmic rays as tool for particle physics}

\section{Air shower and new particle physics}

At present accelerator data from the Tevatron constrain new physics
directly only up $1\,$TeV, while in the near future LHC extends this 
range up to $14\,$TeV. Therefore one might wonder if cosmic ray
data can be used to constrain new particles physics at
energy scales that are not accessible to human accelerators. However 
we know from the discussion in Sec.~2.2 that a significant change of
hadronic cross sections requires the modification of scattering with
small four-momentum transfer $t$ -- while new physics is most often 
connected to the regime of large masses and therefore large $t$. 
More promising is neutrino scattering where at high energies the whole 
range $0\leq |t| \lsim m_W^2$  contributes uniformly to the 
neutrino-nucleon cross section on the parton level.

\paragraph{New resonances} In secs.~2.2 and 7.5 we considered mainly the 
$t$-channel contribution to cross sections. To complement this discussion, 
we assume now that a new particle with mass $m$ and
decay width  $\Gamma$  acts as a $s$-channel resonance in $\nu N$ 
scattering. (Concrete examples are models containing lepto-quarks or 
a sneutrino in $R$-parity violating supersymmetry.) In the
narrow-width approximation, $\Gamma \ll m$,  for the Breit-Wigner 
denominator 
\be
 \frac{1}{(\hat{s} - m^2)^2 + m^2\Gamma^2}  \to
 \frac{\pi}{m\Gamma} \delta(\hat{s} - m^2) 
\ee
in the partonic cross section, the integration over $x$ just picks
out partons with $\hat s=xs=m^2$. The total cross section is
\be
 \sigma_{\nu N}(s) \sim \sum_i \frac{g^2}{m^2} \, xq_i(x=m^2/s,Q^2=s)\,,
\ee
and hence is suppressed for large $m$. Experimental limits require either 
a small coupling $g$, a large mass $m$, $m^2\gsim M_W^2$ or a suitable
combination. This simple example illustrates that it is difficult to
increase the interactions of neutrinos by a large factor. An exception
may be models with large extra dimensions. 
An important ingredient for the large increase of the $\nu N$ cross 
section in these models is 
that the neutrino couples not only to quarks but also to 
gluons. 

Neutrinos as the particles with
the smallest cross section are the most sensitive tool to search with
cosmic ray data for deviations from the standard model predictions
caused by unknown physics. They produce mostly showers close above or
below the horizon. Looking at the zenith angle dependence 
of the event rate, one can disentangle the cross section and flux,
and thus test for deviations from the standard model prediction.

\paragraph{Understanding strong interactions}

We have seen that several aspects of strong interaction cannot be
calculated within QCD but are described by phenomenological models. 
Moreover, even usual methods and
results from  perturbative QCD may become unreliable at ultrahigh
energies. A simple example are the DGLAP equation and the parton distribution 
functions: In the standard picture, 
only splittings $X\to Y Z$ are taken into account, and for decreasing
$x$ the number of partons in nucleons increases monotonically.
As a result, the neutrino nucleon cross section increase
like a power, while for $s\to\infty$ only a logarithmic growth is
allowed. Physically, the density of partons cannot become infinite and
at some point recombinations, $Y Z\to X $, will balance splittings.

\begin{figure}
\hspace*{3cm}\epsfig{file=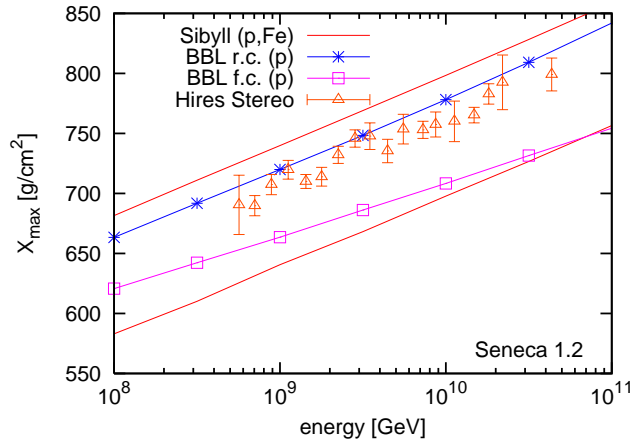,width=0.55\textwidth,angle=0}
\caption{\label{Xmax}
Experimental data (red, with errorbars) and model predictions for
$X_{\max}$ distribution as function of energy, from 
Ref.~\cite{Drescher:2004sd}.}
\end{figure}

Although the physical picture seems to be rather simple, its
formalization is highly  non-trivial and has led e.g.\ to ideas like 
the ``color glass condensate.'' A test of one of these proposals
against experimental cosmic ray data is shown in Fig.~\ref{Xmax}.
The two red curves are the prediction of an often used model for hadronic
interaction in cosmic ray physics for protons (upper) and iron (lower
line). These two curves enclose the experimental data (red, with
errorbars) and thus this model predicts that the cosmic ray flux is a
mixture of protons and heavier nuclei. The other two lines are models
where the growth of the parton density is slower than in perturbative
QCD. Both lines are for protons and the corresponding lines for iron
would be even lower. Hence the model producing the magenta line is
already excluded by cosmic ray data, while the other one predicts a pure
proton flux.

In summary, cosmic ray data can be used at present only to exclude 
extreme model predictions, since the chemical composition of cosmic 
rays is not independently fixed. The discovery of cosmic ray sources and 
the subsequent study of cosmic ray deflections in magnetic fields offers
an alternative way to measure the chemical composition, providing in turn
the possibility to test QCD models more precisely at ultrahigh energies.

\section{Relics from the early universe as cosmic rays sources}

\def\xf{x_{\rm f}}

\begin{figure}
\begin{center}
\includegraphics[width=0.55\textwidth]{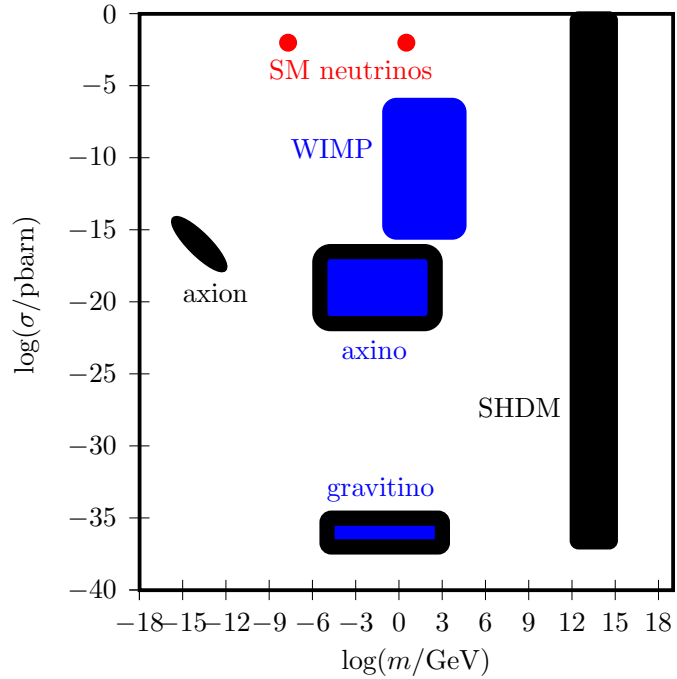}
\end{center}
\vskip-0.3cm
\caption{\label{fig:DM}
A selection of dark matter candidates in the plane cross section versus 
mass; blue and black corresponds to thermal and non-thermal
as main production channel.} 
\end{figure}

Evidence for dark matter comes from flat rotation curves of galaxies,
the virial mass of galaxy clusters, big-bang nucleosynthesis, structure
formation and the CMB: Combining all these pieces of evidence, one
concludes that there exists around six times as much cold dark
matter (CDM) than baryons. Many well-motivated dark matter
candidates that can have the correct abundance have been proposed, 
with a rather big variety in their
properties as shown in Fig.~\ref{fig:DM}. 
Therefore most of them have to be searched for by some specialized
methods. We shall discuss only two candidates in more detail that can 
serve also a source of cosmic rays. One of them are ``WIMPs,'' i.e.\ 
particles with masses and interactions similar to the weak scale. The 
second one are superheavy DM particles (SHDM) with masses 
around $10^{13}\,$GeV. Before, we comment briefly om the other candidates
shown in  Fig.~\ref{fig:DM}: Neutrinos with interactions as predicted
by the standard model would have either with $\sum_{i=1}^3m_i\ap 10\,$eV
or few GeV the correct abundances as thermal relics. Both possibilities
are excluded by various reasons. Axions were proposed as solution to
the strong CP problem. They are pseudo Goldstone bosons that mix with
pions, and therefore their masses and decay constants are connected by
$m_a=m_\pi f_\pi/f_a$. Gravitinos and axinos are supersymmetric partners
of the axion and the graviton, respectively. They can be the lightest 
supersymmetric particle that is protected in many models by a symmetry 
and thus stable.
Last but not least, the lightest neutralino, a mixture of the
supersymmetric partners of higgses, $Z$ and photon, is a typical example 
for a WIMP.

\subsection{WIMPs as thermal relics}

When the number $N=nV$ of a particle species is not changed by
interactions, then the expansion of the Universe dilutes their number
density as $n\propto R^{-3}$. The corresponding change in time is
connected with the expansion rate of the universe, the Hubble parameter
$H=\dot R/R$, as 
\be
 \frac{\d n}{\d t}=\frac{\d n}{\d R}\frac{\d R}{\d t}=-3 n \frac{\dot R}{R}
 =-3Hn \,. 
\ee 
Additionally, there might be production and annihilation
processes. While the annihilation rate 
$\beta n^2=\langle\sigma_{\rm ann}v \rangle\:n^2$ has to be proportional to
$n^2$, we allow for an arbitrary function as production rate $\psi$,
\be
 \frac{\d n}{\d t}=-3Hn - \beta n^2 + \psi \,.
\ee
In a static Universe, $\d n/\d t=0$ defines equilibrium distributions
$n_{\rm eq}$. Moreover, detailed balance requires $\beta n_{\rm eq}^2 = \psi$
and thus we can eliminate the unknown function $\psi$, 
\be
 \frac{\d n}{\d t}=-3Hn - \langle\sigma_{\rm ann}v\rangle ( n^2 -n_{\rm eq}^2) \,.
\ee
This equation together with the initial condition $n\ap n_{\rm eq}$ for
$T\to\infty$ determines $n(t)$ for a given annihilation cross section
$\sigma_{\rm ann}$. 
\begin{figure}
\begin{center}
\epsfig{figure=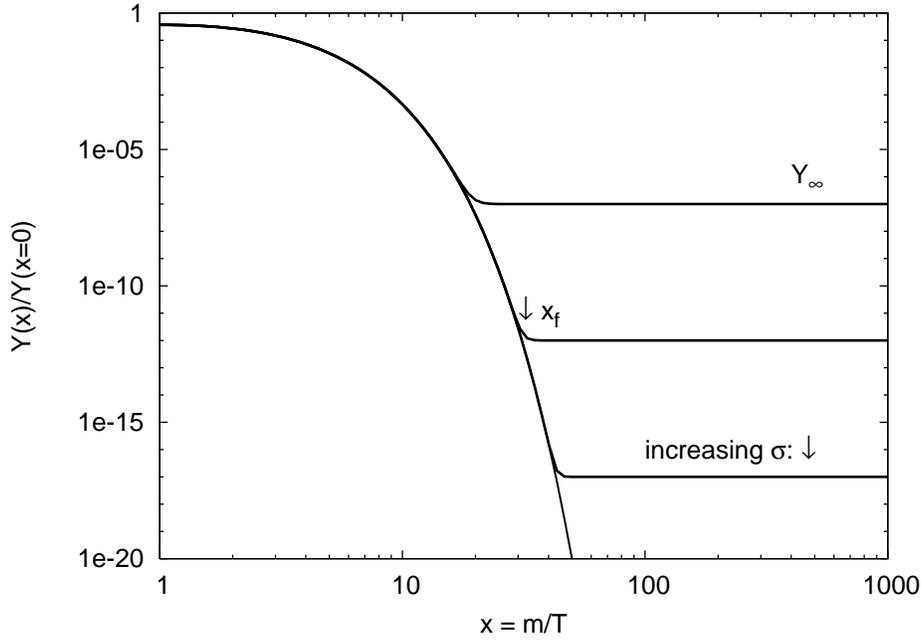,,width=0.8\textwidth}
\end{center}
\caption{Illustration of the freeze-out process. The quantity $Y=n_X/s$
is $n_X$ divided by the entropy density $s\propto R^{-3}$ to scale out
the trivial effect of expansion.
\label{freeze}}
\end{figure}
The evolution of $n_X$ (divided by the entropy density  $s\propto 1/R^3$ 
to scale out the trivial effect of expansion) is shown schematically in
Fig.~\ref{freeze}: As the universe expands and cools down, $n_X$ decreases 
at least
as $R^{-3}$. Therefore, the annihilation rate $\propto n^2$ quenches
and the abundance ``freezes-out:'' The reaction rates are not longer
sufficient to keep the particle in equilibrium and the ratio $n_X/s$ stays
constant.   

Numerically, one obtains for the relative abundance
$\Omega_X=\rho_X/\rho_{\rm cr}$ of CDM  
\be 
 \Omega_X h^2 = \frac{m_Xn_X}{\rho_{\rm cr}} \ap \frac{2\times 10^{-28}{\rm
 cm^3/s}}{\langle\sigma_{\rm ann}v \rangle}\,\xf \,,
\ee
where $\xf$ is the ratio of the mass and the 
freeze-out temperature, $\xf=M/T_{\rm f}$. 
Thus the abundance $\Omega_X$ is inverse proportionally to the thermally 
averaged annihilation cross section $\langle \sigma v_{\rm ann}\rangle$, 
since a more strongly
interacting particle stays longer in equilibrium. The abundance
depends only logarithmically on the mass $m$ via the freeze-out
temperature $\xf$.  Typical values of $\xf$ found 
numerically are $\xf\sim 20$. From the observed value $ \Omega_{\rm
  CDM}\ap 0.20 $ one obtains the annihilation cross section in the
early universe---that controls also the annihilation rate important for
indirect searches today. The size of the cross section corresponds
roughly to weak interactions and has led to the paradigm of ``WIMPs'',
weakly interacting massive dark matter.

Unitarity bounds each partial-wave $l$ of the thermally averaged 
annihilation cross section as  
$\langle\sigma_{\rm ann}v\rangle^{(l)}\leq {\rm const.}/(vm^2)$. 
Requiring $\Omega< 0.3$ leads to
$m<\;$(20--50)\,TeV. This bounds the mass of any stable particle that was
once in thermal equilibrium.

\subsection{SHDM as inflationary relic}

The energy density of non-relativistic particles decreases only as 
$\rho_X\propto 1/R^3$, while the one of radiation decreases as
$\rho\propto 1/R^4$. Therefore particles that never came in chemical
equilibrium with radiation should be created only in tiny amounts in 
the early universe. Already the production by gravitational
interactions at the end of inflation is sufficient to ensure an
abundance  $\Omega_X \sim 1$ for stable particles with masses
$m_X\sim 10^{13}$GeV. 

As an illustration, we consider the Klein-Gordon equation for the field
modes $\phi_k$ of a scalar field in the Friedman-Robertson-Walker
metric,
\be
 \ddot{\phi}_k(\eta) + m_{\rm eff}^2(\eta) \phi_k(\eta) = 0 \,,
\ee
where the effective mass depends on an unknown parameter $\xi$ and is
\be
 m_{\rm eff}^2 (\eta) = k^2 + M_X^2 R^2 + (6\xi - 1)\frac{\ddot{R}}{R}\,.
\ee
Since $m_{\rm eff}$ is time dependent, vacuum fluctuations will be
transformed into real particles. Thus, the expansion of the Universe
leads to particle production.  

The predicted abundance of $X$ particles today from inflationary cosmology is
\be
 \Omega_X h^2 \sim \bigg( \frac{m_X}{10^{11}\, \textrm{GeV}} \bigg)^2 
 \frac{T_{RH}}{10^9 \, \textrm{GeV}},
\ee
where $T_{RH}$ is the temperature at reheating. Since the production is 
gravitational, the result is model-dependent only through the
dependence on cosmological parameters like the value of
Hubble parameter at the end of inflation and the reheating 
temperature $T_{RH}$.

\section{Indirect detection of dark matter}

\subsection{Dark matter in the Galactic halo}

The average density of DM in the Galaxy is strongly increased compared to
the extragalactic space, $n_{\rm MW}/n_{\rm ex}\sim 10^5$. 
Therefore the annihilation rate of
DM can become again appreciable inside the Milky Way, and in particular, in
objects where DM is strongly accumulated.

\paragraph{DM distribution and clumps}
Typical results from $N$-body simulations are for the smooth DM mass
density profile $\rho_{\rm sm}$ 
\begin{equation} 
\rho_{\rm sm}(r)=\rho_\odot\left(\frac{r_\odot}{r}\right)
\left(\frac{r_\odot+a}{r+a}\right)^2,
\end{equation}
with $\rho_\odot =0.3\,$GeV/cm$^3$ as the dark matter density at the
solar distance from the Galactic center, and $a\sim 25\,$kpc as the 
characteristic scale where the slope changes from $r^{-3}$ to $r^{-1}$
(``Navarro-Frenk-White profile''). At small radii $r\lsim 1$~kpc, the
missing resolution of $N$-body simulations, the influence of baryonic
matter and of the galactic SMBH make a reliable estimate of the DM
density difficult.

According to the model of hierarchical structure formation, the first
objects to form were the smallest structures. For the case of WIMPs, 
$10^{15}$ Earth-mass DM halos about as large as the Solar System might 
be in the Milky Way. An important contribution to the total
annihilation signal can be given by small
clumps that have a denser core and may be at small distance. 

\paragraph{Photons, neutrinos and antiprotons from DM annihilation in
  the halo}
The secondaries of DM annihilations will be the stable particle of
the standard model, i.e.\ photons, neutrinos, electrons and protons. For the
latter two, only the anti-particles may provide some useful
information. 

The differential flux of the final state $i$ at the Earth from DM
annihilations is 
\begin{equation}
 I_{\rm sm}(E,\psi) = \frac{\d N_i}{\d E}\,
                      \frac{\langle\sigma v\rangle}{2\,m_X^2}\,
 \int_{\rm l.o.s.} \d s\,\frac{\rho_{\rm sm}^2[r(s,\psi)]}{4\pi}, 
\label{Ism}
\end{equation}
where $r(s,\psi)=(r_\odot^2+s^2-2\,r_\odot\,s\cos\psi)^{1/2}$,
$\psi$ is the angle between the direction in the sky and the
galactic center (GC), $r_\odot\approx 8.0\,$kpc is the solar
distance from the GC, and $s$ the distance from the Sun along the
line-of-sight (l.o.s.). In terms of galactic latitude $b$ and
longitude $l$, one has $\cos\psi=\cos b\cos l$.
The energy spectrum $\d N_i/\d E$ can be calculated only within a specific 
model for the DM particle.

\paragraph{SHDM as UHECR source}

The original motivation to introduce SHDM was the ``AGASA
excess'', i.e.\ a surprisingly large number of cosmic ray events with
energy $\gsim 10^{20}\,$eV and thus above the GZK cutoff. Since SHDM behaves
by definition as CDM, its abundance in the galactic halo is strongly
enhanced. As a result the cosmic flux is dominated by the halo
component and the GZK cutoff is absent in this model.

The hadronization spectra of superheavy particles can be reliably 
calculated using standard QCD methods. The predicted spectrum in the 
SHDM model, $\d N/\d E\propto E^{-1.9}$, cannot 
fit the observed UHECR spectrum at energies $E\leq (6$--$8)\times
10^{19}$~eV. Thus mainly events at $E\geq (6$--$8)\times 10^{19}$~eV, and most 
notably any excess at energies beyond the GZK cutoff, 
could be produced by SHDM decays. As for all hadronization processes,
the main component of the UHE flux are neutrinos and photons from pion
decay. As additional signature, one expects a  Galactic anisotropy, 
because the Sun is not in the center of the
Galaxy. The degree of this anisotropy depends on how strong the CDM is
concentrated near the Galactic center -- a question under debate.
First results from the PAO find neither the predicted Galactic
anisotropy nor the dominance of photon primaries at the highest
energies. If confirmed, SHDM can play only a sub-dominant role as
source of UHECRs.

\subsection{Neutrinos from dark matter annihilations in the Sun and the Earth}

Dark matter particles $X$ scatter on matter in the Sun or Earth, lose energy, 
and may become gravitationally bound. They continue to scatter, gaining or 
losing energy. If energy looses dominate, they sink down to the center where 
they eventually annihilate. The directed signal of high energy neutrino 
would provide a rather clear evidence of DM. In the following, we illustrate
the basic steps in the calculation of the resulting neutrino flux,
following rather closely Ref.~\cite{GS283}.

The total number $N_X$ of dark matter particles in a celestial body is
determined by three processes: The capture rate $R_C$, the evaporation 
rate $R_E$,  and the annihilation rate $R_A$,
\be  \label{balance}
 \dot N_X = R_C - R_E N_X - R_A N_X^2 
             = R_C - \Gamma_E - \Gamma_A \,.
\ee
If the DM particles are self-conjugated, $X=\bar X$, each annihilation  
reduces their number by two and thus one should replace $R_A N_X^2$ by
$2R_A N_X^2$.
The evaporation rate takes into account that an already captured dark 
matter particle might be ``kicked out'' 
in a scattering process and is only important for particles not much
heavier than nucleons, say $m_X \lsim {\rm few}\,$GeV. Therefore we will set
$R_E= 0$ and obtain as solution of Eq.~(\ref{balance})
\be  
  N_X(t) = N_{\rm eq} \tanh(t/\tau) 
\ee
with $\tau=(R_C R_A)^{-1/2}$ as time-scale to reach the
equilibrium number $N_{\rm eq}=(R_C /R_A)^{1/2}$.

We estimate first the capture rate $R_C$ of WIMPs, using where needed 
the numerical values appropriate for the Sun: The total
number of interactions is proportional to the number $N_i$ of scatters of
type $i$ in the celestial body, to the elastic cross section $\sigma_{iX}$, 
the relative velocity $v_{\rm rel}$ and the density of WIMPS $n_X$:
Hence $R_C$ should be of the form  
$R_C\propto n_X \sum_i N_i\sigma_{X i}v_{\rm rel}$. 
 Additionally, a factor 
$f=f(v_{\rm esc},v_{\rm rel})$ should count if the scattering was
successful, i.e. if the WIMP velocity after the scattering is
smaller than the escape velocity $v_{\rm esc}$ from the Sun. 
The function $f$ should be dimensionless, increase for increasing
$v_{\rm esc}$ and can depend only on the square of the velocities. Hence 
$f=f(v_{\rm esc}^2/v_{\rm rel}^2)$, and a more careful analysis shows
that indeed $f\propto v_{\rm esc}^2/v_{\rm rel}^2$. Combining all
factors, the capture rate follows as
\be  
 R_C = A n_X \:\frac{v_{\rm esc}^2}{v_{\rm rel}}\:\sum_i N_i\sigma_{iX} 
\ee
with $A\ap 2$ for the Sun. To obtain a numerical estimate,
we neglect helium and metals and set $N=M_\odot/m_H$. Using the rotation 
velocity of the Sun around the Galactic center as 
relative velocity $v_{\rm rel}\ap 220\,$km/h, 
$v_{\rm esc}=(GM_\odot/R_\odot)^{1/2}\ap 617\,$km/s and
$n_X=\rho_{\rm loc}/m_X$ with $\rho_{\rm loc}\ap 0.3\,$GeV/cm$^3$, it 
follows
\be  
 R_C \ap  10^{21} {\rm s}^{-1} \;
               \left(\frac{\sigma_{iX}}{10^{-42}\rm cm^2} \right) \: 
               \left(\frac{100\, {\rm GeV}}{m_X} \right) \,.
\ee

Next we estimate the annihilation rate $\Gamma_A$. Generically, the
assumption that the WIMP is a thermal relic fixes also the
annihilation cross section at present to 
$\langle\sigma_{\rm ann}v\rangle\sim 3\times 10^{-26}$cm$^3$/s.
(This could be avoided if s-wave annihilation is suppressed by
symmetry reasons.) Since the captured WIMPs orbit many times
in-between interactions, they can be characterized by a global
temperature $T$ and their density follows a barometric formula,
\be
 n_X(r) = n_0\exp(-m_X\phi(r)/T)
\ee  
determined by the local gravitational potential $\phi(r)$.  Hence
\be
 \Gamma_A = \langle\sigma_{\rm ann}v\rangle \;4\pi\int\d r\, r^2
 n_X^2(r) \,.
\ee
Determining first $R_A=\Gamma_A/N_X^2$, we see that this rate is fixed
by $\langle\sigma_{\rm ann}v\rangle$ and the ratio $V_2/V_1^2$ of
``effective volumes'' defined as 
\be \label{V_alpha}
 V_\alpha = 4\pi \int_0^{R_\odot}\d r\, r^2 \exp(-\alpha m_X\phi(r)/T) \,.
\ee
To obtain a numerical estimate, we use that WIMPs for $m_X\gg m_H$
are concentrated in the center of the Sun. Thus we approximate
the density by $\rho(r)\ap  \rho(0)\ap 150\,$g/cm$^3$, set $T$ 
equal to the central temperature of the Sun, 
$T=T_c\ap 1.4\times 10^7\,$K, and use 
$R_\odot\to\infty$.
Then the integrals (\ref{V_alpha}) can be performed and one obtains
\be \label{V_alpha2}
  V_\alpha =  6.5\times 10^{28} {\rm cm}^3
 \left( \frac{100\,{\rm GeV}}{\alpha\, m_X} \right)^{3/2}
\ee
and thus
\be
 R_A = \langle\sigma_{\rm ann}v\rangle \frac{V_2}{V_1^2}=
6.5\times 10^{28} 
 \left( \frac{100\,{\rm GeV}}{m_X} \right)^{3/2} \,.
\ee
Now we can determine the equilibration time as $\tau=(R_C R_A)^{-1/2}\ap
1.4\times 10^{16}\,$s. Since $\tau$ is much smaller than the age of
the Sun, we can use $\Gamma_A=R_C$. 
Denoting with $f_\nu$ the number of neutrinos produced per
annihilation, we obtain as our final result for the neutrino flux
$\phi_\nu$ from WIMP annihilations in the Sun
\be
 \phi = \frac{f_\nu R_C}{4\pi d^2} \sim 5\times 10^{-8} 
 {\rm cm^{-2}\, s^{-1} \, sr^{-1} }
\ee
at the Earth distance $d=1\,$AU. Comparing this flux with expected neutrino 
intensities and the sensitivity of neutrino telescopes discussed in 
Sec.~5.5 makes it clear that the detection of WIMP annihilations is
as challenging as the one of astrophysical neutrinos. The chances for 
success depend strongly on the WIMP mass and its annihilation spectrum.
Only if the energy of the neutrinos is sufficiently high, 
$E_\nu\gsim 100\,$GeV, the direction of the
produced muon provides a reasonable signature.

\section{Summary of possible and suggested signals for DM}

\begin{figure}
\begin{center}
\includegraphics[width=0.45\textwidth]{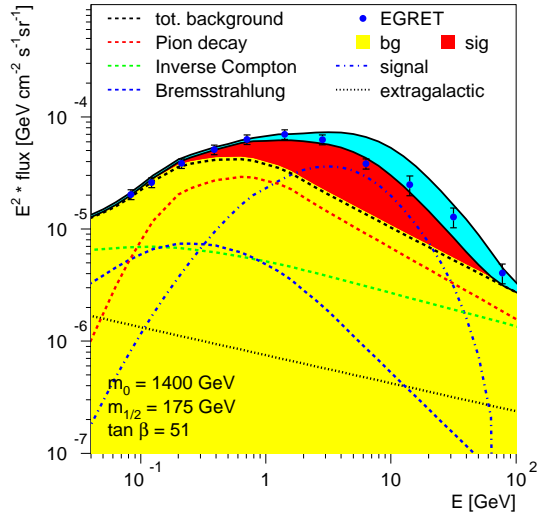}
\end{center}
\caption{\label{boer4}
EGRET data on diffuse photon background together photons expected from
CR interactions (yellow) and a putative DM signal (red) according 
Ref.~\cite{deBoer:2004xt}.} 
\end{figure}

The different annihilation channels and the DM distribution offers
several different  possible signals to detect DM annihilations:
\begin{itemize}
\item All particles from DM annihilations in the Sun or the Earth are
  absorbed except neutrinos. Apart from a small flux of neutrinos produced 
  by cosmic ray interactions in the solar atmosphere, no other neutrinos
  with energy $E\gsim$~GeV are expected from the Sun. 
  The direction provides however
  only for $E_\nu\gsim 100\,$GeV a reasonable signature.
\sitem Cold DM has small velocities and thus $XX\to2\gamma$ produces a
  sharp photon line at $E_\gamma=m_X$. This is a smoking-gun for DM,
  but since the DM particle is most likely electrically neutral, the
  lowest order contributions to these processes are box 
  diagrams, and the branching ratio is therefore very small. 
\sitem 
The main production channel for photons are decays of neutral pions. 
The main problem here
  is to separate a potential DM contribution from diffuse
  astrophysical Galactic and
  extragalactic backgrounds, 
  cf.~Fig.~\ref{boer4}. More promising might be to look at nearby
  dwarf galaxies that have with a larger DM fraction. 
\sitem Antiprotons could become visible below $\sim 1\:$GeV, but without any
specific signatures. Thus the identification of DM as source of these 
antiprotons requires again a precise modeling of antiproton flux produced by 
cosmic rays. 
\end{itemize}
Various astrophysical observations have been suggested
as signals for DM annihilations:
\begin{itemize}
\sitem The Integral satellite sees a strong positron annihilation line
  from the Galactic bulge. An explanation by DM annihilations into
  electron-positron pairs, $\bar XX\to e^+e^-$, requires
  that the electron-positron pair gets only little kinetic
  energy. Hence the DM particle should be very light, 
  $m_e<m_X\lsim 10\:$MeV -- which is not excluded but difficult to achieve. 
\sitem
The ``EGRET excess'', i.e.\ the red region in Fig.~\ref{boer4}, is a 
possible surplus of diffuse gamma-rays compared to the predictions in 
the simplest models for the propagation of cosmic rays. This potential excess 
has been explained both by DM annihilations in the Milky Way and as the
sum of DM annihilation in other galaxies. 
\sitem
The ``WMAP haze'', a potential excess of synchrotron radiation from the 
Galactic center in the WMAP data has been explained as synchrotron radiation
of electron produced in DM annihilations. 
\sitem
HESS, an atmospheric Cherenkov telescope, has observed TeV $\gamma$-rays from
the Galactic center. The flux extends however power-law like up to the
sensitivity limit ($\sim 30\:$TeV) and is thus difficult to combine
with the upper limit on $M$ for a thermal relic. 
\end{itemize}
For references to the original literature and a brief critical review 
see Ref.~\cite{susy}.